\documentstyle[12pt]{article}

\newcommand{\alt}{\mathrel{\raisebox{-.6ex}{$\stackrel{\textstyle<}{\sim}$}}}
\newcommand{\agt}{\mathrel{\raisebox{-.6ex}{$\stackrel{\textstyle>}{\sim}$}}}

\def\decayright#1{\kern#1em\raise1.1ex\hbox{$|$}\kern-.5em\rightarrow}

\catcode`@=11
\newcount\@tempcntc
\def\@citex[#1]#2{\if@filesw\immediate\write\@auxout{\string\citation{#2}}\fi
  \@tempcnta\z@\@tempcntb\m@ne\def\@citea{}\@cite{\@for\@citeb:=#2\do
    {\@ifundefined
       {b@\@citeb}{\@citeo\@tempcntb\m@ne\@citea\def\@citea{,}{\bf ?}\@warning
       {Citation `\@citeb' on page \thepage \space undefined}}%
    {\setbox\z@\hbox{\global\@tempcntc0\csname b@\@citeb\endcsname\relax}%
     \ifnum\@tempcntc=\z@ \@citeo\@tempcntb\m@ne
       \@citea\def\@citea{,}\hbox{\csname b@\@citeb\endcsname}%
     \else
      \advance\@tempcntb\@ne
      \ifnum\@tempcntb=\@tempcntc
      \else\advance\@tempcntb\m@ne\@citeo
      \@tempcnta\@tempcntc\@tempcntb\@tempcntc\fi\fi}}\@citeo}{#1}}
\def\@citeo{\ifnum\@tempcnta>\@tempcntb\else\@citea\def\@citea{,}%
  \ifnum\@tempcnta=\@tempcntb\the\@tempcnta\else
   {\advance\@tempcnta\@ne\ifnum\@tempcnta=\@tempcntb \else \def\@citea{--}\fi
    \advance\@tempcnta\m@ne\the\@tempcnta\@citea\the\@tempcntb}\fi\fi}
\catcode`@=12

\begin{document}

\font\fortssbx=cmssbx10 scaled \magstep2
\hbox to \hsize{
\hskip.5in \raise.1in\hbox{\fortssbx University of Wisconsin - Madison}
\hfill$\vcenter{\hbox{\bf MAD/PH/847}
            \hbox{August 1994}}$ }

\vspace{.5in}

\begin{center}
{\Large\bf Particle Astrophysics\\[.1cm]
 with High Energy Neutrinos}\\[.5cm]
{\bf Thomas~K.~Gaisser$^a$, Francis~Halzen$^b$ and Todor~Stanev$^a$}\\[.2cm]
{\it
$^a$Bartol Research Institute, University of Delaware, Newark, DE 19716, USA\\
$^b$Department of Physics, University of Wisconsin, Madison, WI 53706, USA\\}
\end{center}

\vspace{1in}

\begin{abstract}

\noindent
The topic of this review is the particle astrophysics of high energy
neutrinos. High energy is defined as $E_{\nu} > 100$~MeV. Main topics
include:

\noindent
-- atmospheric neutrinos and muons from $\pi$, $K$ and charm decay. They
probe uncharted territory in neutrino oscillations and constitute
both the background and calibration of high energy neutrino
telescopes,

\noindent
-- sources of high energy neutrino beams: the galactic plane, the sun,
X-ray binaries, supernova remnants and interactions of extra-galactic
cosmic rays with background photons,

\noindent
-- an extensive review of the mechanisms by which active galaxies may
produce high energy particle beams,

\noindent
-- high energy neutrino signatures of cold dark matter and,

\noindent
-- a brief review of detection techniques (water and ice Cherenkov
detectors, surface detectors, radio- and acoustic detectors,
horizontal airshower arrays) and the instruments under construction.
\end{abstract}

\thispagestyle{empty}

\newpage

\renewcommand{\Large}{\large}

\section{Introduction}

The scope of this paper is neutrino astronomy
for $E_\nu > 100$~MeV. Our main interest
is neutrinos from energetic astrophysical sources such as
binary stars and accreting black holes in Active
Galactic Nuclei (AGN).  We will also discuss
atmospheric neutrinos at some length because
they are the only neutrinos with $E>100$~MeV
that have yet been detected.  Atmospheric
neutrinos are both background and calibration beam
for high energy neutrino astronomy.  They are of interest in
their own right because they probe an uncharted range of neutrino
oscillation parameter space.  For stellar collapse neutrinos
and solar neutrinos we refer the reader to the recent review
of Totsuka~\cite{Totsuka92} and, for solar neutrinos to the
reviews of Bahcall {\it et al.}~\cite{Bahcall92,Bahbook} and
Turck-Chieze {\it et al.} \cite{Turck}.

In his classic review of cosmic ray showers in 1960 \cite{Greisen},
Greisen ends with a discussion of the prospects for
gamma ray and neutrino astronomy at very high energy.  He
notes that ``Since photons and neutrinos propagate
in straight lines, success in their detection
will open up broad new areas of astronomy.''  He discussed
the relation between photons and neutrinos from decay
of pions produced in the interstellar medium or near a
source.  He described a detector very much like the
present water Cherenkov detectors (complete with veto
shield against entering muons), and he estimated the rate of
interactions of atmospheric neutrinos in 3 kilotons
of sensitive volume to be 500 events per year.

The idea of detecting neutrinos by looking for neutrino-induced
upward or horizontal muons was suggested by
Markov \& Zheleznykh \cite{MarkZhel} at about the same time.
The process is
\begin{equation}
\nu_\mu\,+\,N\;\rightarrow\;\mu\,+\,{\rm anything},
\end{equation} where
$N$ is a nucleon in the material surrounding the detector.
The muon range increases with energy.  This
extends the effective target volume and makes it possible to see
neutrino-induced muons with detectors of moderate size.
Two groups (Kolar Gold Fields \cite{Achar,Menon} and
Case-Wittwatersrand \cite{Reines1,Reines2})
reported the first observations of atmospheric neutrinos
with the detection of horizontal muons in detectors so
deep that the muons could not have been produced in the
atmosphere.

        Atmospheric neutrinos are of current interest, despite their
long history and apparently mundane origin, because of the anomalous
flavor ratio observed for neutrino interactions in the large proton
decay detectors, Kamiokande \cite{KAM92a} and IMB \cite{IMB92a}.
The essential point is that, because of their large volume, these
detectors can measure interactions of neutrinos inside the detector.
They need not depend on the large external target mass provided by
the
long range of energetic muons produced in charged current
interactions of
muon-type neutrinos.  They can therefore study both $\nu_e$ and
$\nu_\mu$ interactions.  In all, more than a thousand atmospheric
neutrino events have now been measured by the various underground
experiments\cite{KAM92a,IMB92a,KGF,Frejus,Nusex,Soudan2}
The anomaly is that the observed ratio
of events produced by electron neutrinos to those from
muon neutrinos is significantly larger than expected.

Several new detectors designed specifically for high energy
neutrino astronomy are about to come into operation.
The Baikal experiment \cite{BAI} has already reported
some muon measurements \cite{Spier}.  The DUMAND \cite{DUM}
and AMANDA \cite{BAR,LOW} are being partially deployed
at present, and other detectors, such as NESTOR \cite{RES}
are in advanced prototype stages.  A major stimulus
for this activity is the prospect that Active Galactic
Nuclei (AGN) may  be prolific particle accelerators
and beam dumps, and therefore intense sources of
high energy neutrinos.

We have divided our review into three major sections:
({\em a}) atmospheric neutrinos, ({\em b}) possible sources of high
energy neutrinos of extraterrestrial origin and ({\em c})
neutrino detection.  We include some comments about high energy
gamma ray astronomy relevant to possible neutrino
sources at the beginning of part~({\em b}).
We begin with a brief treatment of neutrino production
in cosmic ray cascades, which is relevant both for
atmospheric and astrophysical neutrinos.

\section{Neutrino production}

Unlike
the typical monoenergetic beam produced by a machine, cosmic
accelerators produce power law spectra of ions at high
energy,
\begin{equation}
\phi_p\propto E^{-(\gamma+1)} \,. \label{phi_p}
\end{equation}
The observed high energy cosmic ray spectrum at Earth is
characterized by $\gamma\sim 1.7$.  In general, a cosmic
accelerator in which the dominant mechanism is first order
diffusive shock acceleration (first order Fermi mechanism), will
produce a spectrum with $\gamma \sim 1 +\epsilon$, where
$\epsilon$ is a small number.  The observed spectrum is
thought to be steeper than the accelerated spectrum because
of the energy dependence of the cosmic ray diffusion in the
galaxy.  The simplest way to understand this is to
think of the observer as inside a volume of ``containment''
from which the characteristic escape time decreases with
energy
\begin{equation}
\tau(E)\propto E^{-\delta} \,. \label{tau(E)}
\end{equation}
If $Q(E)$ is the rate of production of cosmic rays per
unit volume, then the observed cosmic ray density will be
\begin{equation}
\rho_{\rm CR}(E)\sim Q(E)\times \tau(E)\propto E^{-(2+\epsilon
+\delta)} \,.
\label{rho_CR}
\end{equation}
For $1\le E\le 100$~GeV, a value of $\delta\sim0.6$
can be inferred from observed ratios of secondary
cosmic ray nuclei (e.g. Li, Be, B) to their
progenitors (e.g. carbon and oxygen) \cite{Engelmann}.


Production of secondary particles ($S$)
is related to the spectrum of accelerated primaries ($P$) by
\begin{equation}
{{\rm d}P_S\over {\rm d}E_S} = {\Delta\over\lambda_P}
\int_{E_S}^\infty\,{{\rm d}n_{PS}(E_S,E_P)\over
{\rm d}E_S}\phi_P(E_P){\rm d}E_P \,,  \label{dP/dE}
\end{equation}
where $\Delta/\lambda_P$ is the probability of interaction
in traversing a small amount ($\Delta$) of target.  If the
distribution of secondaries depends only on the ratio of
energies, $x=E_S/E_P$, then the integral in Eq.~(\ref{dP/dE}) becomes
\begin{equation}
\phi_P(E_S)
\int_0^1 x^{\gamma-1}F_{PS}(x)\,{\rm d}x
\equiv \phi_P(E_S)\,Z_{PS} \,,  \label{phi_P}
\end{equation}
where
\begin{equation}
F_{PS}={1\over\sigma}\int{\rm d}^2p_T\,
E_S\,{{\rm d}\sigma_{PS}\over {\rm d}^3p }\approx E_S{{\rm
d}n_{PS}\over
{\rm d}E_S} \,. \label{F_PS}  
\end{equation}
For $\gamma > 1,\;F(0)$ does not contribute to the integral,
and the scaling approximation made here is an adequate
approximation for rough estimates.
This treatment generalizes readily to decay chains, such
as $p\rightarrow \pi^\pm \rightarrow \mu^\pm$, etc., and
it can be applied to cascades in galactic and stellar environments
as well as in the Earth's atmosphere.
In case of a thick target in which the primary beam is fully
attenuated, the production spectrum of secondaries is given
by Eq.~(\ref{dP/dE}) with the replacement
\begin{equation}
{\Delta\over\lambda_P}\rightarrow {\Lambda_P\over \lambda_P} \,,
\label{Delta/lambda}   
\end{equation}
where $\Lambda_P$ is the attenuation length of the primary.

In general, the flux of neutrinos
from decay of pions is given by~\cite{TKGbook}
\begin{equation}
 {dN_\nu \over dE_\nu} = {N_0(E_\nu) \over 1-Z_{NN}}
 \times \left \{ { A_{\pi\nu} \over 1+B_{\pi\nu}\,cos\vartheta
\,E_\nu/\epsilon_\pi} + (\cdots) \right \} \,, \label{dN/dE}
\end{equation}
where $A_{\pi\nu} = Z_{N\pi}(1-r_\pi)^\gamma/(\gamma+1)$,
$r_\pi=\left({m_\mu/ m_\pi}\right)^2$ and $B_{\pi\nu}$ is a
constant that depends on nucleon and pion attenuation lengths.
The first term inside the curly brackets
represents neutrinos from decay of pions.
The energy $\epsilon_\pi$ is a characteristic energy
that reflects the competition between
decay and interaction in the medium.  For cascade
development in the Earth's atmosphere $\epsilon_\pi\sim115$~GeV;
it is larger for more tenuous media, such as the atmosphere of the
Sun.

The $(\cdots)$ represents the contributions of other mesons, with
$\epsilon_{K^\pm}\sim 850$~GeV and $\epsilon_{D^\pm}\sim 4\times
10^7$~GeV
in the Earth's atmosphere.  Each term also contains the appropriate
branching ratio, e.g.\ 0.635 for $K^\pm\rightarrow\mu+\nu_\mu$.
For $E_\nu\ll\epsilon_i$, all parent mesons decay,
and the neutrino spectrum
is parallel to the primary nucleon flux.
For  $E_\nu\gg \epsilon_i$
the neutrino spectrum
steepens by one power of $E_\nu$.
In the atmosphere, muons with
$E_\mu \gg \mu\,c^2\times 15\;{\rm km}/c\tau_\mu\sim 2$~GeV
reach the surface and stop before they decay.

In the
energy range important for contained events ($0.1<E_\nu<2$~GeV)
decay in flight of atmospheric muons is the dominant source
of $\nu_e$, and an important source of $\nu_\mu$.
At much  higher energy, the dominant source of atmospheric
$\nu_e$ is

\begin{equation}
K_L^0\rightarrow\pi\,e\,\nu_e \,,  \label{mu}
\end{equation}
The relative contributions of the various decay modes
to lepton spectra in the
Earth's atmosphere are discussed in detail by
Lipari~\cite{Paolo92}.
If we consider neutrino production in astrophysical settings, with
typical matter densities of 10$^{10}$~atoms/cm$^3$, pions and kaons
will
always decay and the neutrino spectrum will follow the
nucleon spectrum also at very high energy.

In addition, muon decay will continue to be an important source of
neutrinos at high energy as well as low.
Thus we can define three types of neutrino spectra
that can be produced by cosmic rays:

\begin{enumerate}

\item Atmospheric neutrinos, which follow the incident cosmic ray
spectrum
with $\gamma \sim2.7$ up to $\sim$100~GeV and steepen toward
$\gamma\sim3.7$
at higher energy. The  position of the bend increases with increasing
zenith angle, which generates a characteristic angular dependence of
the
atmospheric neutrino spectrum described by the angular factor in
Eq.~(\ref{dN/dE}). Electron neutrinos, that come mostly from muon
decays have a spectrum with one power of $E$ steeper.

\item Neutrinos produced by galactic cosmic rays in interactions with
interstellar gas. These extraterrestrial neutrinos should follow
the cosmic ray spectrum up to the highest energies, since all
interaction
products, including muons, decay.

\item Neutrinos produced by cosmic rays {\it at their acceleration
sites}
and following the hard ($\gamma \sim 2.0$--2.2) cosmic ray source
spectra,
which are not yet affected by the energy-dependent escape from the
Galaxy.

\end{enumerate}
These three types of neutrino fluxes are illustrated schematically
in Fig.~1.

Atmospheric neutrinos have been detected and studied
extensively. Diffuse galactic neutrinos should exist with intensities
comparable to the diffuse galactic
gamma ray background~\cite{Bloemen}. They should be
detected by a future generation of detectors
of sufficient size to see neutrinos at a rate of several per
10$^5$~m$^2$ per year~\cite{GSH}. These two fluxes can be used for
calibration
of high energy neutrino telescopes, which have as their principal
goal
the search for high energy neutrinos from energetic astrophysical
systems.

The existence of neutrinos associated with cosmic ray sources
is more problematic.  It requires substantial acceleration
in compact sources with sufficient local gas to act as a
beam dump.  The {\em possibility} (not certainty) of such
point sources is suggested by the fact that the standard model
of cosmic ray acceleration by supernova blast waves
in the diffuse interstellar medium cannot accelerate
particles to the highest observed energies.  An alternative
for the higher energy cosmic radiation is acceleration
in compact sources.  The argument goes as follows.

First order Fermi acceleration at supernova blast
shocks offers a very attractive model for a galactic acceleration
mechanism, providing about the right power and spectral shape.
Shock acceleration takes time, however, because
the energy gain occurs gradually as a particle diffuses back and
forth across the shock front.  The finite lifetime
of the shock thus limits the maximum energy per particle
that can be achieved at a particular supernova.  The
acceleration rate is
\begin{equation}
{{\rm d}E\over{\rm d}t}\;\simeq K\,{u^2\over c}\,Z\,e\,B,\;
{\rm so}\;\;\;E_{\rm max}\, <\, {u\over c}\,Z\,e\,B\,L \,,
\label{Emax}
\end{equation}
where $u$ is the shock velocity, $Z\,e$
the charge of the particle being accelerated and $B$
the ambient magnetic field.  The numerical constant $K\sim 0.1$
depends on the details of diffusion in the vicinity of the shock.
The crucial length scale in  Eq.~(\ref{Emax})
is given by
$L\sim u\,T$, where $T\sim 1000$~yrs for the free expansion phase
of a supernova. Using this kind of argument, Lagage
\&~Cesarsky~\cite{Lagage}
show that, in its simplest version (shock velocity parallel to
magnetic field direction, $B = B_{\rm ISM}\sim 3\mu$Gauss)
$E_{\rm max}$ can only reach energies
$\alt 10^{14}$\,eV${}\times Z$ for an accelerated nucleus. That
leaves a large gap of some three orders of magnitude that cannot be
explained by the ``standard model'' of cosmic ray origin.
To reach a higher energy one has
to increase significantly $B$ and/or $L$.

One possibility is to use the much higher magnetic fields associated
with
some energetic astrophysical systems. There is no shortage of such
objects in the Galaxy, the
most obvious being neutron stars and black holes, as well as young
supernova remnants.  Some examples of possible acceleration sites
will be discussed in Section~6 below.  Any such compact region
with active particle acceleration would be a likely site for
production of high energy neutrinos and photons through interactions
of the accelerated particles with the ambient gas and radiation
fields.

We emphasize that this is not the only possibility.
Some argue \cite{Peters,Axford}
that explaining the higher energy cosmic rays by a new
source is unnatural because it requires fine tuning
to produce a smooth spectrum where cosmic rays
from the second source join onto those from the first.

There are several ways
to extend the basic supernova mechanism to higher energies.
One possibility \cite{VolkBier} takes advantage of the fact that
some supernovas explode into the stellar wind of a progenitor
star rather than the interstellar medium.  If the progenitor
wind carries a high enough magnetic field, then higher
top energies can be achieved (see Eq.~\ref{Emax}).
Other possibilities involve
a configuration in which the magnetic field is quasi-perpendicular
to the shock normal \cite{Jokipii} or the interaction of high energy
cosmic rays with expanding shocks of several supernovas in an
active region \cite{Axford}.  Mechanisms such as these would
not be correlated with point sources of high energy gamma rays
and neutrinos.

Perhaps the most exciting possibility at present is the
suggestion \cite{Protheroe,KazEll}
that particle acceleration plays a central role
inside AGN and that interactions of these high energy particles
with dense photon fields and gas in the central regions of
AGN  \cite{Begelman,Stecketal,MannBier,SzaboPro92} will lead to
production of neutrinos of very high energies.  This possibility
is the subject of \S 6.

Figure~1 illustrates the window of opportunity for
high energy neutrino 
astronomy.   The steepening at high energy of the spectrum of
atmospheric neutrinos, which dominate the total neutrino
flux at low energy, allows the possibility of
reasonable signal/background ratios, shown
schematically by the shaded area on Fig.~1.
The small angle between the parent
neutrino and the secondary lepton in charged current interactions at
high energy allows for accurate source location and enhanced
signal/background ratio in the case of point sources.
The exact position
of the crossover from atmospheric to astrophysical neutrinos,
depends on luminosity, distribution and distances of potential
sources and on energy response and, in the case of point sources,
angular resolution of the detectors.  These will be discussed
for a number of sources in Sections~6--8 below.

\section{Atmospheric neutrinos}
 The atmospheric cascade that produces the cosmic ray neutrino
beam is depicted in Eq.~(\ref{cascades}):
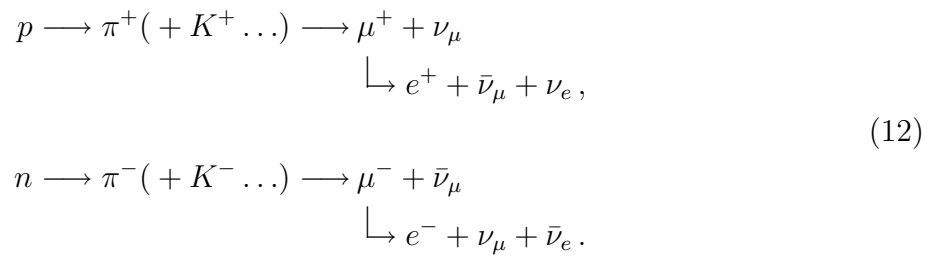
\begin{eqnarray}
p \longrightarrow \pi^+({}+K^+\dots) \longrightarrow
&&\hskip-1.5em \mu^+ + \nu_\mu \nonumber\\
        &&\hskip-1.5em\decayright{.2} e^+ + \bar\nu_\mu + \nu_e
\,,\nonumber \\
\label{cascades}\\
n \longrightarrow \pi^-({}+K^-\dots) \longrightarrow
&&\hskip-1.5em \mu^- + \bar\nu_\mu \nonumber \\
         && \hskip-1.5em\decayright{.2} e^- + \nu_\mu + \bar\nu_e \,.
\nonumber
\end{eqnarray}
Protons also produce negative mesons and neutrons produce positive
mesons,
but the same charge processes are slightly favored by the steep
spectra
and the excess of same-charge mesons in the forward fragmentation
region.

Although analytic expressions like Eq.~(\ref{dN/dE})
are qualitatively correct, more detailed
calculations are needed for a precise evaluation of the atmospheric
neutrino flux.  Several complications must be accounted for:

\begin{itemize}

\item
The primary cosmic ray spectrum is not a simple power law, especially
${}\alt 10$~GeV and ${}\agt 100$~TeV.
Moreover, it depends on location and direction (because of the
geomagnetic cutoff) and on the epoch of the solar cycle.

\item
Muon energy loss, decay and polarization must be accounted for.

\item
The inclusive cross sections do not have exactly scale-invariant
forms.  Furthermore, nuclei, as well as nucleons, are involved in
the collisions.

\end{itemize}

\subsection{Contained Events}

Contained events are those neutrino interactions that originate
within the detector's fiducial volume and whose interaction products
are
all contained within that volume.  Most such events have
lepton energies in the GeV range or less.  In this case, most
muons decay, and all the complications listed above come into play.

Qualitative expectations for the neutrino ratios follow
from simple kinematics of the $\pi\rightarrow\mu\rightarrow e$ decay
chain.  Because of the asymmetry of the
$\pi\rightarrow\mu+\nu_\mu$ decay mode,
each of the two neutrinos from muon decay has about
the same energy as the neutrino from pion decay.  Thus in a given
energy range
\begin{equation}
{\nu_e\over\nu_\mu}\,\approx {1\over 2}, \;\;
{\bar{\nu}_\mu\over \nu_\mu}\,\approx 1, \;\;{\rm and}\;\;
{\bar{\nu_e}\over \nu_e}\,\approx\,{\mu^-\over\mu^+}\,<\,1.
\label{R_nu}
\end{equation}
The excess of $\nu_e$ to $\bar{\nu}_e$ is a consequence of the excess
of protons to neutrons in the incident cosmic ray beam.

Several detailed calculations \cite{GBarr89,Lee90,Honda90,Bugaev89}
of the $\sim$GeV neutrino fluxes agree with each other in
finding a value for the neutrino flavor ratio within 5\% of each
other.  Attention is focussed on the flavor ratio because
most of the sources of uncertainty in the calculation of neutrino
fluxes  cancel in calculating the ratio.
(These uncertainties include normalization of primary spectrum,
treatment of
geomagnetic effects and parameterization of pion production in
collisions of cosmic ray protons and helium with nuclei of the
atmosphere.) Specifically, the neutrino flux calculations give

\begin{equation}
R_{e/\mu}\;=\;{\nu_e\,+\,{1\over3}\bar{\nu}_e\over
               \nu_\mu\,+\,{1\over3}\bar{\nu}_\mu}\;=\;0.49\pm 0.01
\end{equation}
for $0.1<E_\nu< 1$~GeV \cite{roysoc}.
The fact that this expectation is apparently significantly
violated~\cite{KAM92a,IMB92a}
is largely responsible for the great interest in atmospheric
neutrinos.

The water Cherenkov detectors \cite{KAM92a,IMB92a} in fact see
as many \cite{KAM92a} or more \cite{IMB92a} electrons than muons.
Part of the difference between expectation and observation
is a consequence of the fact that the acceptance of the
detectors spans a larger energy range for electrons
than for muons.  Even after accounting for flavor-dependence
of the acceptance, however, a significant discrepancy remains.

Detailed simulations of the detector
response, including acceptance effects, lead to the result that
the ratio of ratios for charged leptons from interactions
of $\nu$ and $\bar{\nu}$ is
\begin{equation}
{ (\mu/e)_{data}\over(\mu/e)_{sim}}
\;=\;0.60\,\pm0.06\,\pm0.05
\end{equation}
for Kamiokande with 6.2 kT-yrs of data (389 contained,
single-ring events) \cite{Kajita}.
The corresponding IMB result is

\begin{equation}
{ (\mu/e)_{data}\over(\mu/e)_{sim}}
\;=\;0.54\,\pm0.03\,\pm0.05
\end{equation}
for 7.7 kT-yrs
of data (507 contained, single-ring events \cite{IMB92a}).
The water detectors thus show at least
a $4\sigma$ discrepancy between
observation and expectation for the $\nu_\mu/\nu_e$ ratio.

Measurements with tracking calorimeters give mixed results, and
the statistical uncertainties are significantly larger than for
the water detectors.
NUSEX (0.74 kT-yrs, \cite{Nusex}) and Frejus (1.56 kT-yrs,
\cite{Frejus})
are both consistent with expectation. In contrast, Soudan 2 (1.0
kT-yrs,
\cite{Soudan2}) finds a ratio of ratios similar to Kamiokande.
The number of events is relatively low in all three experiments, and
the systematic effects are also different.

In view of the complexity of the analysis involved, it is of interest
to ask to what extent the data from the two water detectors are
consistent
with each other.  Beier {\em et al.} \cite{Beier} compared the
IMB and Kamiokande data at stage when Kamiokande had 4.92 kT years
of data \cite{KAM92a} and IMB had 3.4 kT years \cite{IMB91}.
They found the data to be fully consistent
between the two experiments.

Table~\ref{cont_t} shows a comparison based on the more complete
data sets, 7.7 kT years for IMB \cite{IMB92a} and 6.2 kT years
for Kamiokande \cite{Kajita}.  The comparison is made by converting
Kamiokande data to IMB.  Three factors are involved:\\
1) The larger exposure of IMB (7.7/6.2 = 1.24).\\
2) Higher
muon threshold at IMB--the muon threshold
is 300 MeV/c for IMB as compared to 200 MeV/c for Kamiokande.
Using the momentum spectra of Ref.~\cite{Oconnell},
we estimate that this reduces the Kamiokande muon rate by 0.79.\\
3) The lower geomagnetic cutoffs at IMB.
Using the cutoff effects from Ref.~\cite{GBarr89},
this increases the muon rate by a factor 1.23 and the electron
rate by 1.34.  The larger correction factor for electrons
is a consequence of the fact that the electron
threshold (100 MeV/c for both experiments) is
lower than the muon threshold, and the geomagnetic filter
affects low energies more than high.\\
These three correction factors are applied successively to the
Kamiokande
column to obtain the converted Kamiokande numbers that may be
compared directly with the IMB column.

\smallskip

\begin{table}[h]
\begin{center}
\caption{ \label{cont_t} Comparison of contained event rates.}\smallskip
\begin{tabular}{lccccc}
\hline\hline
\vrule width0pt depth5pt height 14pt& Kamiokande & exposure & $\mu$-threshold &
geomagnetic & IMB
\\[.1cm]\hline
muons & 191 & $\rightarrow$237 & $\rightarrow$187 & $\rightarrow$230
& 182 \\
electrons & 198 & $\rightarrow$246 & $\rightarrow$246
&$\rightarrow$330 & 325\\
total & 389 & $\rightarrow$483 & $\rightarrow$433 &
$\rightarrow$560
& 507 \\ \hline\hline
\end{tabular}
\end{center}
\end{table}

\smallskip

After the conversion, the Kamiokande muon/electron ratio (230/330)
appears
somewhat higher than for IMB (182/325).  Another consideration that
must
be borne in mind is that the second half of the IMB data was taken
during a period of maximum solar modulation.  The event rate at
IMB is expected to be somewhat lower during maximum
solar modulation, whereas Kamiokande, with its higher geomagnetic
cutoff is much less sensitive to modulation \cite{GBarr89}.
This effect should be the same for $\nu_\mu$ and $\nu_e$, however.
Recently, Beier \& Frank \cite{geneb} have pointed out that the
momentum
spectra of the electrons in the two experiments are also
somewhat different.
Beam tests of the response of water Cherenkov detectors
to electrons and muons will soon be carried out at
KEK.  This should help to resolve questions about the
efficiency for discrimination between neutrino flavors
as a systematic error in this type of detector.

Another possible source of systematic
error that has been pointed to is the cross section
for charged current interactions of neutrinos in
nuclei.  The Fermi gas model has been used for calculation
of the spectra of the produced charged leptons
by both Kamiokande \cite{Takita} and IMB \cite{Casper}.
Recent calculations by Engel {\em et al.} \cite{Vogel}
include several effects that go beyond the Fermi gas model.
They find no significant shift
in the spectra of electrons relative to muons which
would distort the inferred $\nu_e/\nu_\mu$ ratio.
In addition, there is some direct confirmation of
the Fermi gas model for $E_\nu> 400$~MeV in data
discussed in Ref.~\cite{Mann}.  Another experiment \cite{LosAlamos},
which appears to show an anomalous result for
the muon spectrum in $\nu_\mu +{\rm carbon}\rightarrow
\mu+\ldots$, is in any case below the energy range
of interest here ($p_\mu>200$~MeV/c for Kamiokande
and $p_\mu>300$~MeV/c for IMB).

Assuming that the problem with the contained events
reflects an intrinsic property of particle physics (rather
than a lack of understanding of detector response
or neutrino cross sections), one needs to know whether
there are too few $\nu_\mu$ or too many $\nu_e$.
For example, \cite{Honda92}
with the calculation of Refs.~\cite{GBarr89,Honda90},
an interpretation in terms of neutrino oscillations can
be explained by $\nu_\mu$ disappearance (e.g.
$\nu_\mu\leftrightarrow\nu_\tau$ oscillations),
but not by $\nu_\mu\leftrightarrow\nu_e$, which would
increase the predicted flux of $\nu_e$. On the other hand,
comparison of the data to the calculations of
Refs.~\cite{Lee90,Bugaev89}
(which are some 30\% lower than those of
Refs.~\cite{GBarr89,Honda90})
prefers $\nu_\mu\leftrightarrow\nu_e$ in order to boost up
the predicted $\nu_e$ flux as well as lower the $\nu_\mu$ flux.
Fogli {\em et al.} \cite{Fogli} have recently reviewed the limits
of various neutrino oscillation scenarios that could explain the
atmospheric neutrino anomaly.

A more exotic explanation is the suggestion that there is an
excess of electrons due to proton decay in the mode
$p\rightarrow e\,\nu\,\bar{\nu}$ \cite{TMann}.  This
interpretation requires a calculated atmospheric
flux with a low normalization (so the muon rate is
correctly predicted).  The atmospheric
flux should also have a shape such that the deficit of
electrons from atmospheric $\nu_e$'s occurs preferentially
at low energy.  The deficit can then be filled in by the
characteristic
energy spectrum of a three-body proton decay.  The calculation
of Ref.~\cite{Bugaev89} has just these features.

A subset of the authors of Refs.~
\cite{GBarr89,Lee90,Honda90,Bugaev89}
is investigating the source(s) of the difference among the
normalization
and shapes of the calculations.  It appears that the main cause
of the characteristic shape and low normalization of the calculation
of Ref.~\cite{Bugaev89} is the parameterization of pion production
in collisions of protons with light nuclei \cite{TAUP}.

The production of $\pi^\pm$ with $E_\pi<2$~GeV is significantly
lower in the parameterization of Ref.~\cite{Bugaev89} than in
Refs.~\cite{GBarr89,Honda90}.

This uncertainty could be reduced by comparison to
measurement of muon fluxes at high altitude.  Existing
high altitude data \cite{Conversi,Bogomolov,Baradzei}
have large uncertainties, but they somewhat
favor the higher flux calculations.  A new set of
experiments \cite{MASS} should be able to fix the normalization
to perhaps 10\%.
If the contained neutrino anomaly is due to
$\nu_\mu\leftrightarrow\nu_\tau$ oscillations,
then the derived parameters suggest the effect should
also show up
as an apparent deficit in the neutrino-induced upward muon sample.
We return to this question in the next section.

\subsection{Upward Muons}
Neutrino-induced muons are of interest for two reasons:
first, they  extend the study of the atmospheric neutrino spectrum
to higher energy, and second,
they are the expected signal of high energy astrophysical neutrinos.
In this section we
first discuss the relation between a spectrum of neutrinos from any
source and the muon flux that it produces.  We then review the
current status of muons produced by atmospheric neutrinos
including the extent to which these measurements restrict the
neutrino oscillation interpretations of the contained event
anomaly.  In later sections of the paper we use the formulas of
this section to discuss possible signals of high energy astrophysical
neutrinos.

\subsubsection{Neutrino-induced muons}
The detection of neutrino interactions inside the detector volume
becomes more difficult for higher energy neutrinos because of the
steeply falling neutrino spectrum.
It is possible to enhance the effective volume
of the detector by looking for muons generated in
charged-current interactions of $\nu_\mu\,(\bar{\nu_\mu})$
in the rock below the detector. The effective detector volume is
then the product of the detector area and the muon range in rock
R$_\mu$.
TeV muons have a typical range of one kilometer in rock,
which leads to a significant increase in  effective detector volume.
The technique works only for muons entering the detector from below
or near the horizontal (upward going muons),
because the the background of downward atmospheric muons dominates
any neutrino-induced signal from above.

The average muon energy loss rate is
\begin{equation}
{\left\langle{\rm d}E\over{\rm d}X\right\rangle}\;=\;
-\alpha(E)\,-\,\beta(E)\times E,
\label{muloss}
\end{equation}
where $X$ is the thickness of material in g/cm$^2$.
The first term represents ionization losses, which
are approximately continuous, with $\alpha\sim 2$~MeV~g$^{-1}$cm$^2$.
The second term includes the catastrophic processes of
bremsstrahlung, pair production and nuclear interactions, for
which fluctuations play an essential role.  Here
$\beta\sim 4\times 10^{-6}$~g$^{-1}$cm$^2$.

The energy-dependent energy loss rates for each process
are tabulated in Ref.~\cite{Lohmann}.
The critical energy above which the radiative processes
dominate is
\begin{equation}
E_{\rm cr}\;=\;\alpha/\beta\;\approx\;500\;{\rm GeV}.
\end{equation}

To treat muon propagation properly when $E_\mu>E_{\rm cr}$ requires
a Monte Carlo calculation of the probability $P_{\rm surv}$ that
a muon of energy $E_\mu$ survives with energy $>E_\mu^{\rm min}$
after propagating a distance $X$ \cite{LS}.  The probability
that a neutrino of energy $E_\nu$ on a trajectory through a
detector produces a muon above threshold at
the detector is \cite{GG87,LS}
\begin{equation}
P_\nu(E_\nu,E_\mu^{\rm min})\,=\,N_A\,\int_0^{E_\nu}\,{\rm d}
E_\mu{{\rm d}\sigma_\nu\over{\rm d}E_\mu}(E_\mu,E_\nu)\,
R_{\rm eff}(E_\mu,E_\mu^{\rm min}),
\label{P_nu E}
\end{equation}
where
\begin{equation}
R_{\rm eff}\,=\,\int_0^\infty
\,{\rm d}X\,P_{\rm surv}(E_\mu,E_\mu^{\rm min},X).
\end{equation}

The flux of $\nu_\mu$-induced muons at the detector
is given by a convolution of the neutrino spectrum $\phi_\nu$
with the muon production probability (\ref{P_nu E}) as
\begin{equation}
\phi_\mu(E_\mu^{\rm min},\theta)\;=\;
\int_{E_\mu^{\rm min}}\,P_\nu(E_\nu,E_\mu^{\rm min})\,
\exp[-\sigma_{\rm tot}(E_\nu)\,N_A\,X(\theta)]\,\phi_\nu(E_\nu,\theta).
\label{N_mu}
\end{equation}
The exponential factor here accounts for absorption of neutrinos
along the chord of the Earth, $X(\theta)$.  Absorption becomes
important for $\sigma(E_\nu)\agt 10^{-33}$~cm$^2$
or $E_\nu\agt 10^7$~GeV.

The event rate is now simply calculated by multiplying Eq.~(\ref{N_mu}) with
the effective area of the detector.
One can now tabulate
$P_{\nu \to \mu}$ for a given muon energy threshold and fold it with
fluxes of neutrinos of different origin to calculate the expected
event
rate. Figure~2 shows $P_\nu$ for two values of muon
threshold
energy, 1 GeV and 1 TeV. The solid lines are
for $\nu$ and the dashed lines for $\bar{\nu}$.

There is some uncertainty in the calculation of the neutrino
cross sections for $E_\nu\gg10$~TeV because of the
required extrapolations of the structure functions to small
$x\ll 10^{-4}$.  We have used
the charged current cross section of Ref.~\cite{Renoquigg}.
For back-of-the-envelope calculations $P_\nu(E_\nu,0)$
can be approximated by two power laws (shown by the straight
lines for the 1 GeV case in Fig.~2):
\begin{equation}
\begin{array}{r@{\quad}l}
 P_{\nu\to\mu} \simeq 1.3\times10^{-6}\,E^{2.2}&
\hbox{for }E=10^{-3}\hbox{--1 TeV}\\
                \simeq 1.3\times10^{-6}\,E^{0.8}&
\hbox{for }E= \phantom01\hbox{--10$^3$ TeV}
\end{array}   \label{P_nutomu}
\end{equation}
where $E$ is in TeV.
The two energy regimes directly reflect the energy dependence of the
neutrino
cross section and the effective
muon range in Eq.~(\ref{P_nu E}).   $\sigma_\nu\sim E$ at low energy
followed
by a slower energy  dependence above 1~TeV reflecting the effect of
the $W$
propagator.  The muon range makes a transition from increasing
linearly
to constant behavior in a similar energy range.
\subsubsection{Atmospheric neutrinos above 1 GeV}

The atmospheric neutrino spectrum is quite steep at high energy,
approaching $E^{-(\gamma+2)}\approx E^{-3.7}$ for $E\gg 1$~TeV
(see Eq.~\ref{dN/dE}).  For this reason, despite the increase
of $P_\nu$ with energy, the contribution to the upward muon
flux from atmospheric neutrinos with $E>10$~TeV is small.  We
illustrate this in Fig.~3, which shows the
distribution of energies  of
atmospheric neutrinos
that gives rise to upward muons.  In this example,
``throughgoing'' is defined as $E_\mu>4$~GeV and
``stopping'' as $1<E_\mu<4$, and the fluxes are averaged
over all angles below the horizontal.  The response curve
for contained interactions is shown for comparison.

The upward going neutrino rate has been measured with significant
statistical accuracy by three experiments---IMB
\cite{IMBmu,IMBsing},
Baksan \cite{Baksan} and Kamiokande \cite{KAMmu}.
Since the experimental arrangements, biases and effective areas as a
function of the zenith angle are quite different, it is impossible
to compare these results exactly. One can, however, scale the
quoted experimental $E_\mu^{\rm min}$ to a common value of $3$~GeV,
which is the effective threshold at Kamiokande.\footnote{Kamiokande
does not use a sharp threshold of $E_\mu^{\rm min}=3$~GeV.
Rather, the definition of a throughgoing muon is any muon with a
projected trajectory $>7$~m inside the detector which actually
exits from the detector.}
The data are converted using Eq.~\ref{N_mu} for each angular bin.
A comparison between angular distribution of upward muons,
measured by the three experiments is shown in Fig.~4.
The quoted muon energy thresholds, measured rates averaged over
the upward hemisphere and the rates converted to a common
threshold are given in Table~\ref{upmu_t}.  The units are
10$^{-13}$cm$^{-2}$s$^{-1}$sr$^{-1}$.
The numbers in the last column of Table~\ref{upmu_t} are to be
compared with a calculation of the upward
muon flux due to atmospheric neutrinos.  Calculated
values  range from 1.95 to 2.36,
depending on the neutrino cross section and muon flux used for the
calculation \cite{Frati92}.  Given the experimental errors and
uncertainties in the input to the calculations, the observed
rates are consistent with expectation.

\begin{table}[h]
\begin{center}
\caption{ \label{upmu_t}
Measured upward muon rates ({\em data}) {\em shifted\/} to a common
threshold.}\smallskip\tabcolsep=2em
\begin{tabular}{lccc}
\hline\hline
\vrule width0pt depth7pt height 14pt
Experiment & $E_\mu^{\rm min}$(GeV) & {\em data} & {\em shifted}
\\\hline
Baksan \cite{Baksan} & $1.0$ & $2.77\pm0.17$ & 2.08 \\
IMB \cite{IMBmu,IMBsing} & $2.0$ & $2.26\pm0.11$ & $1.92$ \\
Kamioka \cite{KAMmu} & $3.0$ & $2.04\pm0.13$ & $2.04$ \\ \hline\hline
\end{tabular}
\end{center}
\end{table}

The question then arises whether the agreement between
calculation and measurement of upward muons is good
enough to rule  out
some possible explanations of the contained event anomaly
in terms of neutrino oscillations.  We first note that
the ``allowed region'' \cite{KAM92a} of parameter space for
explanation of
the contained event anomaly in terms of
$\nu_\mu\leftrightarrow\nu_\tau$
oscillations should lead to a significant depletion of
the neutrino-induced muon flux.

For example, for $\delta m^2 =8\times 10^{-3}$~eV$^2$,
and $L=10^4$~km (a typical propagation distance
for an upward neutrino that interacts below the detector)
the first node of
\begin{equation}\label{P_nu}
1\,-\,P_{\nu_\mu\rightarrow\nu_\mu}\;=\;
P_{\nu_\mu\rightarrow\nu_\tau}\\
\;=\;\sin^2\,2\theta\,\sin^2\left(
1.27\,\delta m^2\,{L_{km}\over E_{\rm GeV}}\right )
\end{equation}
is at $E = 65$~GeV.  This energy is in the middle of the
energy range important for generation of neutrino-induced
upward muons (see Fig.~3).  Thus, for large mixing angles
the upward muon flux would be significantly suppressed,
approaching the level of
$1\,-\,{1\over 2}\sin^2\,2\theta$ for $L\,\delta m^2\,/\,E >> \pi/2$.

Zenith angle dependence of the neutrino signal probes
oscillation lengths $L$ varying from $\sim$10 (downwards)
to 10$^4$ km (upward).
For small values of $E$ and large values of $L$ the value of
the second $\sin^2$ in Eq.~(\ref{P_nu}) will average to 1/2 and
the zenith angle dependence will not be observable. This is
the case for the contained events. Explicit variation
with zenith angle may be observable for higher
energy neutrinos if the contained event anomaly
is due to neutrino oscillations.\footnote{There is a hint of
such behavior in the recent Kamiokande preprint~\cite{KAM94}, received after
this manuscript was prepared.}

The major sources of
uncertainty in Eq.~(\ref{N_mu}) are the neutrino
flux and, to a lesser extent, the
charged current cross section in the relevant energy
range from $\sim$1 to ${\sim}10^4$~GeV.
Various calculated values of the neutrino
flux are listed in Table~\ref{upp_t} at three characteristic
energies.  In the most important region for
upward, throughgoing muons the calculations differ by
as much as 17\%.  Different standard parametrization
of the charged current cross sections also differ by as
much as 13\%.

\begin{table}[h]
\begin{center}
\caption{\label{upp_t} $\displaystyle{{\rm d}N\over {\rm d}\ln E_\nu}\;
(\nu_\mu+\bar{\nu}_\mu,\;{\rm cm}^{-2}\,{\rm s}^{-1}\,{\rm
sr}^{-1})$}  \smallskip \tabcolsep2em
\begin{tabular}{llll}
\hline \hline
        & \multicolumn{1}{c}{10 GeV} & \multicolumn{1}{c}{100 GeV}
&
\multicolumn{1}{c}{1000 GeV} \\ \hline
Volkova \cite{Volkova80} & $6.0\times10^{-4}$ & $6.1\times 10^{-6}$&
$4.5\times 10^{-8}$ \\
Mitsui \cite{Mitsui} &  6.3  &  6.2  &  4.1  \\
Butkevich \cite{Butkevich} & 7.3 & 6.9 & 4.2 \\
Bartol~\cite{AGLS92} & 6.9 & 7.2 & 4.7 \\
\hline \hline
\end{tabular}
\end{center}
\end{table}

Both IMB \cite{IMBmu}
and Kamiokande \cite{KAMmu} used low values of the cross section
\cite{EHLQ} and the neutrino flux \cite{Volkova80} as central
values for comparison with their data.  In Ref.~\cite{Frati92}
it was shown that, with this input the Kamiokande data on
throughgoing muons
rule out a region of parameter space for
$\nu_\mu\leftrightarrow\nu_\tau$
oscillations for $\sin^22\theta>0.4$ and $\delta
m^2\agt
5\times 10^{-3}$.  This is very similar to the excluded region
obtained by the IMB group starting from the same assumptions.

When a better representation of the cross section \cite{Owens}
and a higher neutrino flux \cite{Butkevich} are used, however,
the region excluded by upward, throughgoing muons
is much smaller.  In particular, much of the region ``allowed''
by the Kamiokande contained events \cite{KAM92a} is also allowed
by the Kamiokande measurement of upward muons.  In the further
numerical examples below the neutrino cross section and flux
used for calculations correspond to these higher
inputs \cite{Owens,Butkevich}.

Although consistent conclusions seem to emerge from
interpretation of the Kamiokande and IMB data on upward muons,
the same cannot be said for Baksan.  The Baksan group
still exclude most of the
Kamiokande ``allowed'' region even when they use the highest
neutrino flux calculation \cite{Butkevich}.
The Baksan limit on
$\nu_\mu\leftrightarrow\nu_\tau$ oscillations is based on upward
events near the vertical, specifically zenith angles
in the interval $-1<\cos\theta<-0.6$.

They observe 161 upward muons in this angular range during a live
time of 7.15 years.  With the
low neutrino flux \cite{Volkova80} they expect 142, as compared to
163 with the high neutrino flux \cite{Butkevich}.
In the comparable angular region
the Kamiokande measurement is $(1.42\pm0.18)\,\times\,10^{-13}$
upward muons per (cm$^2$~s~sr).  The calculated result with high flux
is 1.76 and with low flux 1.55.  The prediction of Kamiokande
with the high neutrino flux but assuming
$\nu_\mu\leftrightarrow\nu_\tau$
oscillations is 1.34 (instead of 1.76), which is completely
consistent with the experimental value of $1.42\pm0.18$.
This is for $\sin^22\theta=0.5$
and $10^{-2}<\delta m^2<10^{-1}$~eV$^2$.  The preliminary
result from MACRO ~\cite{MACROmu} is similar to the Kamiokande result:
they measure $74\pm 9\pm 8$ events as compared to a calculated
number for the same exposure of $101\pm 15$.

In summary, therefore,
we conclude that the present data on upward muons {\em do not} rule
out a $\nu_\mu\leftrightarrow\nu_\tau$ oscillation at a level
sufficient to explain the contained event anomaly.  A tenfold
increase in
statistics would help resolve the situation, particularly if the full
zenith angle range can be measured accurately.  For example,
with $\delta m^2 \sim 0.1$~eV$^2$ the flux would be significantly
suppressed near the horizontal as well as for $\cos\theta < -0.2$,
but for $\delta m^2 \sim 0.01$~eV$^2$ the near horizontal flux
would not be suppressed.

  The IMB group points out
that the fraction of upward muons that stop in the
detector is relatively insensitive to uncertainties in the
calculation
because the flux normalization cancels.  They find~\cite{IMBmu} that
the
measured fraction of stopping
muons rules out a portion of the ($\delta m^2, \sin^2\theta$) plane
for $3\times 10^{-4}<\delta m^2 < 10^{-2}$~eV$^2$ and large mixing
angle.
The constraint on the $\delta m^2$ parameter from the fraction
of stopping muons comes from the absence of
a distortion of the muon energy spectrum.  The relevant
neutrino energies are illustrated in Fig.~3. For example,
if
$\delta m^2\sim 10^{-3}$~eV$^2$ the transition probability
(\ref{P_nu}) is relatively large for
$E_\nu\sim 10$~GeV (and $L\sim10^4$~km) but negligible
for $E_\nu\sim 100$~GeV.  In this case one would have a significant
distortion of the energy spectrum of upward muons and hence an
anomaly
in the stopping fraction provided the mixing angle is
sufficiently large.  On the other hand, if
$\delta m^2\agt 10^{-2}$ then both the
high and low energy portions will be affected similarly and no
distortion of the stopping/throughgoing ratio would occur.

It should be mentioned that the IMB constraint from the
stopping fraction is based on
use of a single neutrino flux calculation~\cite{Volkova80}.
It therefore reflects a particular assumption about
the slope of the primary cosmic ray spectrum and other
factors that could affect the shape of the neutrino spectrum.
One example of such a factor is the uncertainty in the
production of kaons, because kaons contribute about
50\% of the throughgoing signal but only about 25\% of
the stopping muons.  Nevertheless, as
Table~\ref{upp_t} illustrates, the uncertainty in shape
is less than the uncertainty in normalization.

\subsection{Neutrinos and Muons from Charm production}

So far in discussing atmospheric neutrinos we have considered
only those which come from decay of pions, kaons and muons.
The (semi)-leptonic decay of charmed particles, produced in the
interactions
of cosmic rays in the atmosphere, is also a source of atmospheric
muons and
neutrinos---the ``prompt'' leptons.  These prompt leptons are also
described by Eq.~(\ref{dN/dE}), but with a much larger value of
$\epsilon_{\rm charm}\sim10^4$~TeV, which is a consequence
of the short lifetime of charmed particles.  Charmed particles
almost always decay before they interact in the atmosphere.

Thus, whereas the spectrum of conventional muons and neutrinos
becomes one power of energy steeper than the primary spectrum
for $E\agt1$~TeV, the prompt leptons continue with the same
power as the primary spectrum to much higher energy.
Prompt leptons eventually dominate the atmospheric spectra at
energies above 10--100~TeV as a result of their flatter energy
spectrum.
For this reason, an estimate of prompt neutrinos is important for
estimating the background for astrophysical neutrinos.
Prompt neutrinos and muons have a characteristic
isotropic angular
distribution, in contrast to the
characteristic $\sec\theta$ dependence of the decay products of
pions and kaons above $\sim1$~TeV (see Eq.~\ref{dN/dE}).


Searching for an
isotropic component of the atmospheric muon spectrum
with deep underground detectors is a traditional technique
to look for a prompt component and hence to estimate
the charm cross section.
Underground detectors exploit the depth-intensity
relations for muons  in order
to obtain different muon thresholds.
There is some weak evidence \cite{KGF2,BAKSAN} for
a prompt muon component at a level corresponding to
one prompt muon for every 1000 pions produced at the
same energy. Because of the small probability for
pion decay at high energy,
this would actually correspond to a rather large charm
cross section.
 The raw measurements are, however,
difficult to interpret~\cite{kokoulin}. This is not
surprising as  there are
many problems in detecting prompt muons with underground
experiments~\cite{HVZ}.

It is also difficult to predict accurately the flux of prompt leptons
because
the experimental status of charm production at laboratory energies is
rather
confusing. Especially large uncertainties are associated with the
production
of mesons and baryons in the Feynman $x_f \rightarrow 1$ region. As
is the case for
strange particles, the (forward) $x_f \rightarrow 1$ behavior is
expected to
vary strongly with the nature of the produced particle. Neutrino and
muon
fluxes are rather sensitive to the behavior of the cross section at
large
$x_f$ because of the steep parent cosmic ray spectrum. Theory cannot
come to
the rescue. Perturbative calculations are unreliable at low energies
because
of their sensitivity to the assumed quark mass and renormalization
scale.
Because charm is predominantly produced by the fusion of gluons at
high
energies, the cross section critically depends on the low-$x$
behavior of the
gluon structure function which is poorly or totally unknown depending
on the
energy\cite{LHC90}. A
calculation of the high-energy charm cross section is actually beyond
the
scope of perturbative QCD because it
requires the resummation of large logarithms of
$1/x$~\cite{collins}.

Prompt neutrino (or muon) fluxes corresponding to five
parametrizations
of the
high energy behavior of the charm production cross section
\cite {HVZ} are shown in Fig.~5
(solid lines).  For comparison, we also show
the conventional muon flux from the decay of pions and kaons.
The left dashed line is for vertical muons and the right for
horizontal.
The prompt curves represent
both neutrino and muon fluxes which are very nearly
identical for two-body decays of massive particles into light
leptons.
The prompt muon flux from charm decay
is independent of zenith angle up to $10^8$~GeV.

In contrast, the
conventional muon and neutrino fluxes have a strong dependence on
zenith
angle, with a ratio of vertical to horizontal fluxes approaching an
order of
magnitude at high energy.

Prompt production dominates above an energy whose value depends on
zenith angle and, of course, on the assumption for the high-energy
charm cross
section. This cross-over energy is lower for neutrinos than for muons
because
the conventional muon flux exceeds the conventional neutrino flux.
The five models for the hadroproduction of charm are discussed in
detail in
reference~\cite{HVZ}. As an extreme guess on the high side they
assume a charm
production cross section which is 10\% of the total inelastic cross
section,
$\sigma_{in}$~\cite{DataBook}. It behaves like $\log^2(s)$ at high
energies. The
model is inspired by the fact that a high energy gluon fragments 10\%
of the
time into charm particles, and its
applicability is limited to energies above a few
TeV since at lower energies it violates charm cross-section
measurements from
accelerator experiments~\cite{ammar}. As a lower limit they evaluated
the charm
cross section from leading order perturbation theory~\cite{Nason88}
using
relatively hard parton distributions.
Some interesting results have been obtained by
an X-ray chamber array measuring the vertical muon spectrum up to
50~TeV~\cite{Afanasieva}. These data rule out the highest
parametrization
shown in Fig.~5, which is not totally surprising as
it also overestimates the accelerator data at low energy.
We therefore consider the curve labelled {\em B} in
Fig.~5
to be the upper limit for the atmospheric background of neutrinos
above 100 TeV.

\subsection{Neutrinos and Muons in Horizontal Air Showers}

Cosmic rays with energy of order $100$~TeV and higher initiate air
showers which penetrate deeply
enough to be studied with  particle detectors at ground level.
The detected flux is a steeply falling function of zenith angle
because the depth of atmosphere traversed by a cascade reaching
the ground rises rapidly from 1030 to 36000~g~cm$^{-2}$ as
the zenith angle varies from zero to 90 degrees. Thus near the
horizon ($90^\circ$), close to a thousand radiation lengths of
matter separate the interaction from the detector, and the
configuration of the experiment is analogous to that of any
accelerator-based beam dump experiment. Most secondaries
such as pions and kaons are absorbed in the dump and only
penetrating particles, such as muons and neutrinos produced
in the initial interaction, reach the detector. High energy muons
will traverse the atmosphere and occasionally lose energy
by catastrophic photon bremsstrahlung. If the photon shower
is produced close to the detector it will be recorded and
is referred to as a horizontal air shower.  Because horizontal
air showers are a signature for penetrating particles in general,
they can also be used to search for cosmic neutrinos, so we
discuss them in some detail.
For comparable muon and neutrino spectra, muons
will dominate the horizontal air shower flux as a result of their
larger interaction cross section.  Therefore energetic atmospheric
muons constitute the main background for horizontal
air showers from cosmic neutrinos.

High energy muons produced in cosmic ray interactions in the
first layers
of the atmosphere, will radiate hard bremsstrahlung photons that
initiate
electromagnetic cascades. If the bremsstrahlung interaction occurs
close to
the particle detector the cascade will be registered. Neutrinos can
similarly
interact deep in the atmosphere and deposit a fraction $y$ of their
energy in
particles which produce an electromagnetic shower close to the
detector. The
horizontal shower rate $\phi_{sh}$ is determined by convolution of i)
the
flux of parent muons or neutrinos $\phi_{par}$ and ii) the
$y$-differential
interaction cross section of the parent particles integrated over the
atmospheric depth $t$ at a given zenith angle,
\begin{equation}
\phi_{sh}(N_e) = \int_0^{t_{max}} dt \int_0^1 {dy \over y}
\phi_{par}\left(E_{par}={E(N_e,t) \over y}\right)
{d \sigma \over dy} {dE \over dN_e} \label{phi_sh} \;,
\end{equation}
using the following notation for the differential and integral shower
size
spectra
\begin{eqnarray}
\phi_{sh}(N_e)&=&{ d^4N_{sh} \over dN_e~dt~dA~d \Omega }
\label{phi_sh N_e}
\;,\\
\Phi_{sh}(N_e)&=&\int_{N_e}^{\infty} dN_e^* \phi_{sh}(N_e^*) \;.
\label{Phi_sh}
\end{eqnarray}
The notation for the parent spectrum $\phi_{par}$ is the same. For a
given
shower size and depth of first interaction the energy of the shower
is fixed,
on average, by the depth development of the cascade. This is the
meaning of
$E(N_e,t)$ in Eq.~(\ref{phi_sh}). This development is somewhat
different for
purely electromagnetic, hadronic or mixed showers. As a rule of thumb
the
shower size at maximum is roughly the half the energy in GeV
units.

Calculation of the rates of horizontal cascades with fixed shower
size $N_e$
are discussed in detail in Refs.~\cite{horshow,HVZ}.
Both analytic and Monte Carlo calculations can be made \cite{GHVZ}.
Starting from the  horizontal rate as given in
Eq.~\ref{phi_sh}, the
number of observed showers above a given
size and angle is given by
\begin{equation}\label{horizontal}
N_{shower}(N_e>N_{0},\theta>\theta_{0})=T \;
\int_{N_{e0}}^{\infty} dN_e A(N_e)  \int^{\Omega_0}_0 d\Omega\;
\phi_{sh}(N_e,\theta)
\end{equation}
where $A(N_e)$ is the effective area of the array and $T$
is the observation time. The method described above has been
used to calculate the horizontal air  shower rates, shown
in Fig.~18, associated with the muon fluxes of
conventional
and charm origin from Fig.~5.

Horizontal shower sizes in the range $N_e=10^3$--$10^5$ have been
extensively
studied by the University of Tokyo~\cite{tokiodata}, and their
observed rate of showers with zenith angle $\theta > 70^\circ$ is
consistent with production by conventional atmospheric
muons~\cite{tokiobremss}.    The EASTOP group have also
reported very recently a measurement of horizontal air showers
with $\theta > 75^\circ$
that is consistent with a muonic origin \cite{EASTOP}.

For shower sizes above $10^5$ the AKENO group has published an upper
bound on
muon-poor, air showers with zenith angle greater than
$60^\circ$~\cite{Akenobound}. The bound is obtained by selection of
muon poor
showers in order to isolate purely electromagnetic showers initiated
by
bremsstrahlung photons from muons~\cite{muonpoor}. The high energy
muon fluxes
from charm decay are therefore bounded by this data which provide us
with
indirect, but relevant information on the charm production cross
section. This
is illustrated in Fig.~6. It is clear that the largest
charm cross sections
are ruled out by both the data of the University of Tokyo or by the
AKENO bound.

It is interesting to point out that the differential muon spectrum
from prompt
decays at energy $E_{\mu}$ reflects $pp$ interactions of average lab
energy
roughly a factor 10 (8) times the muon energy, for a 2.7 (3) spectral
index~\cite{ZasMP,GHVZ}. The measurements for largest
$E_{\mu}$~\cite{Afanasieva} thus
correspond to an average proton lab energy around
$500$~TeV. Measurements of horizontal shower of sizes above $10^5$
correspond
to a primary photon energy of around $100$~TeV, which has been on
average
produced by a primary $p$-air collision of lab energy of order
$1.7$~PeV just
beyond the reach of TEVATRON. The AKENO bound extends the average
energy reach
for prompt production even further, to close to $100$~PeV of
laboratory
energy, although the information should be taken with care. Firstly,
there is
some inconsistency between the rate of horizontal showers measured by
the
university of Tokyo and the bound~\cite{Akenobound}. Secondly, the
measurement
relies on selecting muon poor showers at zenith angles above
$60^\circ$, and
contamination from ordinary cosmic ray showers can be a problem. The
already
difficult selection of muon poor showers is further complicated by
the
presence of the original muon that gave rise to the bremsstrahlung
photon. For
a more detailed discussion of the relevance of this data to the charm
cross
section we refer the reader to reference~\cite{GHVZ}.

\section{Gamma ray astronomy}

Many authors have discussed candidate point sources of
${}\agt{}$TeV neutrinos, especially in connection
with reported observations of air showers from point sources.
Possible sources include accreting X-ray
binaries~\cite{Eichler,Berezinsky,GS85}, compact binary
systems with interacting winds~\cite{HG}, a neutron star engulfed by
a giant companion~\cite{Berezinsky,BerGaleotti} and
young supernova remnants~\cite{BerPri,Sato77,SS77,GS87}.

There are two approaches to computing the signal expected from
such sources.  The first starts from observations (or limits)
on photon signals from candidate sources.
If the photons are products of decay of neutral pions, then
one expects a comparable flux of $\nu_\mu$.  Photons
can also be produced by electrons through synchrotron
radiation, bremsstrahlung and inverse Compton scattering.
To be as certain as possible that one is dealing with
$\pi^0$ $\gamma$-rays it is therefore desirable to look
in an energy range above that accessible to electrons, which
are typically limited to energies in the TeV range by ambient
magnetic fields.  We
use limits of observations on photons in the 100~TeV range,
i.e.\ from air shower experiments.  Typical limits for steady
emission from point sources are
in the range~\cite{Cygnus,CASA,SPASE,JANZOS,Tibet,TWJC}
\begin{equation}
{{\rm d}N_\gamma\over{\rm d}\ln E_\gamma}=E_\gamma\,
\phi_\gamma\,
 \alt 10^{-13}\;{\rm cm}^{-2}\,{\rm s}^{-1}
\label{dN_gamma}
\end{equation}
for ${E_\gamma}\sim 100$~TeV.  Limits on certain Northern hemisphere
sources obtained using the muon rejection technique are somewhat
lower, approaching $10^{-14}$~cm$^{-2}$s$^{-1}$ \cite{CASApoint}.

To find the implied limit on the corresponding neutrino flux requires
knowing the relation between photons and neutrinos at production
and the fraction of the produced photons that is absorbed in the
source.  The latter is highly uncertain.  For a spectrum of photons
from $\pi^0$-decay of the form
\begin{equation}
\phi_\gamma\;=\;{\rm Const}\times E_\gamma^{-\alpha}
\label{dN/dE_gamma}
\end{equation}
the corresponding spectrum of neutrinos from decay of $\pi^\pm$ is
\begin{equation}
\phi_\nu\;=\;{\rm Const}\times (1-r_\pi)^{\alpha-1}
\times{1\over 1-A_\gamma}\label{phi_nu},
\end{equation}
where $A_\gamma$ is the fraction of photons absorbed at the source
(which may in general be energy-dependent).
For $\alpha \agt 2$ the kinematic factor is
approximately $1\over 2$.  When $\nu_\mu$ from
decay of muons are added, the summed flux of  $\nu_\mu+\bar{\nu}_\mu$
is approximately equal to the photon flux at production \cite{TKGbook}.

We can convolve the neutrino flux (\ref{phi_nu}) corresponding
to the limit (\ref{dN_gamma}) with the charged current
neutrino cross section and muon range (see Eq.~\ref{N_mu})
to obtain a limit on the upward muon rate.
For spectral index $\gamma$ in the range $1.1$ to $1.3$, we find
\begin{equation}
{\rm Flux}(\uparrow\mu)\alt
{2\;{\rm events}\;(E_\mu>1\;{\rm TeV})
\over 10^5\rm\,m^2\,yr}\times{1\over(1- A_\gamma)}
\;.
\label{Flux}
\end{equation}
The total number of upward muon events with $E_\mu>1$~GeV is only a
factor of two larger for such a flat spectrum.

A source producing at this limit and having
$A_\gamma\agt 0.98$, i.e. a factor 50 enhancement of the neutrino
relative to
the photon flux, would be detectable in DUMAND in the sense of giving
$\sim$20 events per year in $2\times 10^4$~m$^2$ with $E_\mu>1$~TeV.
(The exact number depends on ``details'' such as the location
of the source relative to the detector, which determines
the fraction of the time it is sufficiently below the horizon
to produce a signal.)

The photon absorption factor $A_\gamma$ could be much
larger (e.g.\ in the case of the neutron star swallowed by
a giant star), but this would be at the expense of requiring
still greater power at the source.\footnote{It is shown
in Ref.~\cite{BerLearned} that the power of a hidden source
cannot be increased indefinitely without making the
object so hot and bright as to violate observations.}
This leads to the
other approach to estimating likely neutrino fluxes from
various sources.  It is straightforward to
calculate the power in accelerated protons required
to give a detectable signal of neutrino-induced
muons independent of any model of
photon reabsorption.  The result depends on
the distance to the source, the assumed spectral index
and the fraction of the accelerated proton beam that interacts.
Estimates~\cite{BerLearned,Gaisser90}
show that a power of about $10^{39}$ to $10^{40}$~erg/s of
accelerated
protons is required for a source at the distance of the
Galactic radius to produce a detectable signal in DUMAND,
assuming a fully absorbed proton beam.
This could be a young supernova remnant or a young pulsar.
A system accelerating particles with a power of $10^{38}$~erg/s
(e.g.\ an X-ray binary accreting at the
Eddington limit for a solar mass star with a large efficiency
for converting accretion energy into high energy particles)
would have to be relatively nearby ($\sim$1~kpc) to be detectable.

Recently the Whipple collaboration reported the observation of TeV
(10$^{12}$\,eV) photons from the giant elliptical galaxy
Markarian~421~\cite{Whmkn}, an observation which might be directly
relevant to
our quest for sources of high energy neutrinos. With a signal in
excess of
6~standard deviations, this is the first convincing observation of
TeV gamma
rays from outside our Galaxy. That a distant source like
Markarian~421 can be
observed at all implies that its luminosity exceeds that of galactic
cosmic
accelerators such as the Crab, the only source observed by the same
instrument
with comparable statistical significance, by close to 10 orders of
magnitude.
The Whipple observation implies a Mkn~421 photon luminosity in excess
of
$10^{43}$ ergs per second. It is interesting that these sources have
roughly
the same flux of energy per logarithmic energy interval in the TeV
region as in the GeV region.

Why Markarian 421? Whipple obviously zoomed in on the Compton
Observatory
catalogue of active galaxies (AGN) known to emit GeV photons.
Markarian, at a
distance of barely over 100~Mpc, is the closest blazar on the list.
Stecker
et al.~\cite{Steckerabs} recently pointed out that TeV gamma rays
are efficiently absorbed on infra-red starlight, anticipating that
TeV
astronomers will have a hard time observing 3C279 at a redshift of
0.54.
Production of $e^+e^-$ pairs by TeV gamma rays interacting with IR
background photons is the origin of the absorption. The absorption
is,
however, minimal for Mkn~421 with $z=0.03$, a distance close enough
to see
through the IR~fog.

This observation was not totally unanticipated. Many
theorists~\cite{HENA}
argue that blazars such as Mkn~421 may be powerful cosmic
accelerators
producing beams of very high energy photons and neutrinos.
Acceleration of
particles is by shocks in the jets which are a characteristic feature
of these
radio-loud active galaxies. Many arguments have been given for the
acceleration
of protons as well as electrons. Inevitably beams of gamma rays and
neutrinos
from the decay of pions appear along the jets. The pions are
photoproduced by
accelerated protons on the dense target of optical and UV~photons in
the
galaxy. The latter are the product of synchrotron radiation by
electrons
accelerated along with the protons. There are of course no neutrinos
without
proton acceleration. The arguments that protons are indeed
accelerated in
AGN are rather compelling. They provide a ``natural'' mechanism for
i) the energy transfer from the central engine over
distances as large  as
1~parsec, ii) the heating of the dusty disc over distances of several
hundred
parsecs and iii) the near-infrared cut-off of the synchrotron
emission in the
jet. Protons, unlike electrons, efficiently transfer energy in the
presence of the magnetic field in the jet. A detailed case for proton
acceleration in active galaxies is made in reference~\cite{bierkofu}.

Other possible models for the emission of gamma radiation up
to 1 TeV from Mkn 421 involves inverse Compton scattering by
accelerated electrons, rather than $\pi^0$ $\gamma$-rays
\cite{SchlickDerm,Zdzolik,SikBegRees}. Such models therefore do
not predict a corresponding flux of high energy neutrinos.

\section{Guaranteed Sources of High Energy Neutrinos: the Galactic
Plane and the Sun}

By their very existence, high-energy cosmic rays do guarantee the
existence of
a definite source of high energy cosmic neutrinos~\cite{newXIII}.
Cosmic
rays interact with the interstellar gas in our galaxy and are
therefore
inevitably accompanied by a flux of diffuse photons and neutrinos
which are
the decay products of the pions produced in these interactions.
A rough estimate of the diffuse fluxes of gamma rays and neutrinos
from the galactic disk can be obtained by convoluting
the observed cosmic ray flux with interstellar gas with a nominal
density of
1 particle
per cm$^3$. The target material is concentrated in the disk of the
galaxy and
so will be the secondary photon flux. Its observation would reveal
``point
sources'' associated with molecular clouds and the spiral arm of the
galaxy.

An estimate of the expected fluxes at TeV and PeV energy can be
easily
performed under the assumption of a constant cosmic ray density in
the Galaxy.
Imagine a concentration of matter of density $\rho$ and linear
dimension $R$.
For example, the flux at Earth of photons  generated by pions produced in
cosmic ray interactions with this matter is given by
%
\begin{equation}
\Phi_{\gamma(\nu)} = \Phi_{CR}\, f_A \left[ \sigma_{\rm inel} \over
m_N \right] [\rho R]
{ 2 Z_{N \pi^0} \over \gamma + 1} \; , \label{Phi_gamma}
\end{equation}
where
${\Phi}_{CR}\approx 1.8\,E^{-2.7}$~cm$^{-2}$sr$^{-1}$s$^{-1}$GeV$^{-1}$
is the cosmic ray intensity,
${\sigma}_{\rm inel}$
is the total inelastic $pp$ cross section, $m_N$ is the nucleon mass
and
$\rho R$ is the column density of the source. The quantity
$Z_{N\pi}=1/\sigma\,\int dx\,x^\gamma\,d\sigma/dx$ is the
spectrum-weighted moment for production of pions by nucleons with
differential
energy spectrum $E^{-(\gamma+1)}$, and $f_A$ ($\simeq 1.22$) is a
correction
factor to account for the fact that some primaries and targets are
nuclei~\cite{Schaefer}.  As noted in the previous section,
the differential spectrum
of $\nu+\mu + \bar{\nu}_\mu$ is very nearly equal to the that of
photons after accounting for muon decay.

In a detailed calculation, as recently performed by Berezinsky et
al.~\cite{newXIV}, one must explicitly account for the energy
dependence of
the inelastic cross-section, of the particle physics parameters and
of the
spectral index $\gamma$. Nevertheless, Eq.~(\ref{Phi_gamma}), which
neglects
these energy
dependences, can be used to make adequate first order estimates of
the $\gamma$-ray and neutrino fluxes.
Equation~(\ref{Phi_gamma}) states
that the
photon to cosmic ray ratio is directly proportional to the linear
matter
density $\rho R$. The
ratio is of order $6\times 10^{-5}$ for a column density of
0.1~grams/cm$^2$.
Maps of the galactic linear column density can therefore be directly
translated
into photon or neutrino fluxes with the assumption that the cosmic
ray density
in the Galaxy is constant at its local value. The predicted flux is
of order
$10^{-5}$ of the cosmic ray flux in the PeV energy range, with the
different
estimates varying within ``a factor".
Observation of a photon flux at this level has turned out to be
challenging.  The best experimental limits \cite{CASAMIA} are still
an order of magnitude higher than expectation \cite{newXIV}.

It is clear that a roughly equal diffuse neutrino flux is produced by
the decay of charged pion secondaries in the same collisions.
For example, from Eq.~(\ref{Phi_gamma}),  and
assuming a threshold of $E_\mu>1$~TeV, we can estimate the
number of neutrino events from within 10~degrees of the disc as
5~events per
year for a 10$^5$\,m$^2$ detector at the South Pole which views
1.1~steradian
of the outer Galaxy with an average density of 0.013~grams/cm$^2$.
This would
be increased to 15~events per year if the spectral index in the outer
galaxy
were indeed as small as 1.4, as suggested by an
analysis~\cite{Bloemen} of
the GeV $\gamma$-ray results of the COS-B satellite.
Another interesting example is the direction of Orion,
a molecular cloud with a column density of
$0.04$~g/cm$^2$ and an angular width of $0.07$~sr \cite{COMPTEL}.
For this case we estimate 0.3 neutrino-induced muons per
year in a $10^5 \,\rm m^2$ detector in this angular bin.
There are several gas
concentrations of similar or smaller density in the galaxy.
These numbers account for the fact that neutrinos are
produced by the decay of muons as well as pions.
Although these rates
are significantly below the atmospheric background, the source
is guaranteed and the event rate might be significantly higher if
there are
regions of the galaxy where the spectrum is flatter than the local
spectrum.

The above estimates assume a cosmic ray intensity that is constant
throughout the disk of the Galaxy and equal to that
measured at Earth. A recent COMPTEL~\cite{COMPTEL} observation of
3 to 7 MeV fluxes from the region of Orion may suggest, however, that
the cosmic ray energy density could be significantly higher in that
region. The observed $\gamma$-ray line intensities (from excited
$^{12}$C and $^{16}$O nuclei) would correspond to cosmic ray energy
density of more than 50 eV cm$^{-3}$, a factor of 100 higher than
in the vicinity of the Earth. Although the lines are generated
by cosmic ray nuclei of kinetic energy around 10 MeV, which are
subject to strong solar modulation inside the solar system, it is
quite possible that the cosmic ray density in a wider energy range
is significantly higher in this active star-formation region.
The limits derived from the COS-B data\cite{Bloe84}
allow a cosmic ray intensity and respectively neutrino fluxes
higher by factor of 5 in the region of Orion.

 The other guaranteed extraterrestrial source of high energy
neutrinos is
the Sun. The production process is exactly the same as for
atmospheric
neutrinos---cosmic ray interactions in the solar atmosphere.
Neutrino
production is enhanced because the atmosphere of the Sun is much
more tenuous---the scaleheight of the chromosphere is $\sim$115 km,
compared with 6.3 km for the upper atmosphere. As a result
$\epsilon_\pi$
is higher by a factor of $\sim$20, as is the energy where the slope
of the
neutrino spectrum increases. The difference is even larger for cosmic
rays
that enter the Sun at large angles and never reach atmospheric
densities
higher than 10$^{-7}$ g/cm$^3$. A detailed calculation of the
neutrino
production by cosmic rays in the solar
atmosphere~\cite{SeckSG} shows  a
neutrino spectrum higher than the angle averaged
atmospheric flux by  a factor
of $\sim$2 at 10 GeV and a factor of $\sim$3 at 1000 GeV.

The decisive factor for the observability of this neutrino source
is the small solid angle (6.8$\times$10$^{-5}$ sr) of the Sun.
Although the rate of the neutrino induced upward going muons is
higher
than the atmospheric emission from the same solid angle by a factor
of $\sim$5, the rate of muons of energy above 10 GeV in a 10$^5$
m$^2$
detector is only 5 per year. Taking into account the diffusion of the
cosmic rays in the solar wind, which decreases significantly the
value of the flux for energies below one TeV, cuts this event rate
further
by a factor of 3. Folded with a realistic angular resolution of 1
degree,
observation of such an event rate also requires a 1 km$^2$ detector.

\section{Possible Galactic Neutrino Sources}

In \S2 we noted that the galactic cosmic radiation above
$\sim$100~TeV
might be accelerated in compact sources, and that, if so, these would
be likely point sources of photons and neutrinos.  In \S4 it was
shown
that present limits on $\sim$100~TeV gamma rays from potential point
sources make it unlikely that there will be prolific galactic
point sources of neutrinos.  Before discussing several possible
types of point sources in more detail, we give an example to set
the scale for what may be the maximum reasonable expectation for
neutrinos from galactic point sources.  We consider a
two-component model of the cosmic radiation as described in \S2
in which the low energy component steepens around $100$~TeV
and a high energy component dominates at much higher energies.
For this illustration we assume that the
high energy component is produced in compact sources scattered
in the disk of the Galaxy.

We start by estimating the power that would be needed
to supply such a high energy component.  In a two-component
picture \cite{tkgkofu} the cosmic ray energy density in energy range
between 100 and 1000 TeV is about $10^{-15}$~erg/cm$^3$, of which
about half would be from the high energy component.
%
%
%
The local energy density in this component is thus estimated as
\begin{equation}
\rho_E =
5\times10^{-16}\rm erg\,cm^{-3}\;. \label{rho_E}
\end{equation}
Assuming this is typical of the energy density elsewhere in the
Galaxy,
the luminosity of the galaxy in such particles is then
\begin{equation}
\epsilon\times{\cal L}_p =
{V_{\rm gal}\,\rho_E\over\tau} = 1.5\times10^{38}\rm\,erg\;s^{-1}
\;, \label{L_p}
\end{equation}
where $V_{\rm gal}$ is the volume of the galaxy and $\tau$ the
confinement or
lifetime of PeV cosmic rays in the galaxy. The values of these
parameters
are very uncertain and also depend on the model of propagation and
escape of cosmic rays from the Galaxy.
The numerical result in Eq.~(\ref{L_p}) is obtained for
$V_{\rm gal}=5\times10^{66}\rm\,cm^3$ (the volume of the galactic
disk)
and  $\tau =5\times10^5$~year as an estimate of the time cosmic rays
spend in the disk.  The parameter
$\epsilon < 1$ in Eq.~(\ref{L_p})  is
the fraction of the accelerator power required
just for the decade above 100 TeV.  If we assume that these
accelerators produce a hard spectrum with equal energy per
logarithmic interval, then the estimate of the total power
needed to maintain the steady observed  PeV cosmic ray flux
is ${\cal L}_p\sim10^{39}$~erg/sec.  This source will resupply the
galaxy and compensate for the loss of cosmic rays resulting from
their limited
confinement time~\cite{AMHNat}.

Let us next assume that a comparable amount of energy is absorbed
by collisions of protons with gas near the high energy accelerators.
Then the total power in neutrinos is related to the
neutrino flux at Earth from all compact sources by
\begin{equation}
{\cal L}_p\,D_{p\to\nu_\mu}\,D_\nu=\sum_i(4\pi d_i^2)\,
\int dE\,E\,{dN_{i,\nu}\over dE}
\label{L_p D}
\end{equation}
Here $D_{p\to\nu_\mu}\ (\simeq0.3)$ is the fraction of energy in an
accelerated proton spectrum $\propto E^{-2}$ that goes into $\nu_\mu
(\bar{\nu}_\mu)$,
$D_\nu\ (\simeq1)$ describes the fraction of neutrinos that
escape from the sources and
$d_i$ is the distance to the $i$th source.  If there are $n$ such
sources distributed throughout the galactic plane, then an estimate
of the distance to the nearest source is $d\sim R_g/\sqrt{n}$ where
$R_g\sim 10$~kpc is a nominal radius of the disk in which the sources
are
concentrated.  If each source has a particle luminosity of ${\cal
L}_p/n$,
then number of sources cancels in the relation between the
total cosmic ray luminosity ${\cal L}_p$  and the neutrino luminosity
of a ``nearest neighbor'' source \cite{CowGais}.  One has
\begin{equation}\label{1source}
\int dE\,E\,{dN_{i,\nu}\over dE}\;=\;{
{\cal L}_p\,D_{p\to\pi}\,D_\nu\,/n\over 4\pi R_g^2\,/n},
\end{equation}
from which we estimate the neutrino flux from a nearby source as
%
\begin{equation}  E\,{{\rm d}N_\nu\over{\rm d}E}\;=\;
2\times 10^{-11}\,{100\;{\rm TeV}\over E}\,{\rm cm}^{-2}\,{\rm s}^{-1},
\label{dN_nu/dE}
\end{equation}
assuming an $E^{-2}$ spectrum.  Such a flux of high energy
$\nu_\mu+\bar{\nu}_\mu$ would give  some 300 events per year
in $10^5$~m$^2$.

This is a very high flux and should be considered an
extreme upper limit for a neutrino-induced signal
from a galactic point source.
To avoid the existing limits on photons from point sources
(\ref{dN_gamma}), one would need a factor $A_\gamma \agt 100$ absorption
of high energy photons in the source.
Absorption arguments depend strongly on the exact mechanism.
If photon absorption is due to interactions and cascading on the
ambient matter at source, the accelerated protons will also be
absorbed, which weakens the original motivation for this estimate.
It is, however, not only possible, but likely~\cite{ProthSAbs} that
the high energy $\gamma$-rays would be absorbed in
$\gamma\gamma \rightarrow e^+e^-$ collisions on the
strong radiation field at the source.
In this case the protons will not be absorbed since the
photoproduction
threshold is $(m_\pi/m_e)^2$ higher and the protons lose very little
energy in $p\gamma \rightarrow e^+e^-$ collisions.

The preceding argument is based on an assumed random distribution
of point sources in the disk of the galaxy.  Because of the
cancellation
of the number $n$ of sources in Eq.~(\ref{1source}), a similar
estimate can be made
of the neutrino flux from a single source at the
galactic center.  We note that the upper bound on
the $>100$~TeV gamma flux from the galactic center is \cite{JANZOS}
$\sim 2\times 10^{-13}$~cm$^{-2}$s$^{-1}$.  There could, however, be
significant absorption of a high energy photon
source from the galactic center.  Thus, as
pointed out by Berezinsky~\cite{XII}, {\em not} seeing high energy
neutrino
emission from the center of the galaxy would be an interesting
result.

We now look in somewhat more detail at two possible classes of
galactic point sources of neutrinos.

\subsection{X-ray Binary Systems.}

The interest in X-ray binaries was initiated by the
reports~\cite{SamStamm} of detection
of  UHE ($> 10^{14}$~eV) $\gamma$-rays from Cygnus X-3. Such
$\gamma$-rays
would most likely be produced in inelastic hadronic interactions, in
which case they would be accompanied by high energy neutrinos.
Current upper limits on steady emission from Cygnus X-3 are
about an order of magnitude lower than the level implied by
the original report, which referred to showers with energies
above $2\times 10^{15}$~eV.
Nevertheless, it is still interesting to consider X-ray binary
systems as potential accelerators of UHE cosmic rays and
to ask at what level one might expect accompanying neutrino fluxes.

X-ray binaries consist of a compact object (neutron
star or a black  hole)
and a non-compact companion star. Such systems are dynamically
complicated, involving mass transfer from the
companion onto the  compact
object through an accretion disk. Neutron stars are known to have very
strong (10$^{12}$~G)  surface
magnetic fields and sometimes millisecond periods.
Both the accretion  and
the magnetic dipole radiation are possible energy
sources. The  existence of high
magnetic fields and plasma flows creates the
environment necessary for  the
formation of strong shocks, and corresponding
particle acceleration.  The
companion star itself, the accretion flow,
or the heavy stellar winds  might
be targets for inelastic nucleon interactions and neutrino
production.

Calculations of the neutrino flux expected from Cygnus X-3 were done
independently by different authors~\cite{GaiStan}, and the results agree
to better than a factor of two
for similar assumptions about the input parameters and the configuration
of the accelerated beam and beam dump.
The total  upward going muon flux for a distance of 10 kpc
is 2--3$\times$10$^{-15}$ cm$^{-2}$s$^{-1}$, i.e.\ 50--100 upward
going muons
per 10$^5$~m$^2$ per year for a fully efficient detector.  Fluxes
at this level are well above the atmospheric background
for $E_\mu\agt 100$~GeV \cite{GG87}. Such a large flux corresponds
to a proton luminosity at the source of 2$\times$10$^{39}$~erg/s,
comparable to the generic estimate in the introduction to this
section.   The estimated flux is a factor $\sim$10 lower here, however,
because the models typically assume a 10\% duty cycle for the beam to
intercept the target mass (e.g. the companion star).

Models motivated by the original Cygnus X-3 observations in which
the high energy gamma rays emerge from source obviously cannot be
correct for Cygnus X-3 in view of the current limits from air
shower observations.  Emission of neutrinos in other models in
which the photons are absorbed can, however, be obtained by
scaling from these calculations by an assumed proton
luminosity and the distance of any potential source.

The crucial question then is the luminosity that might be
expected from such systems.
For accretion powered sources the luminosity is limited by
the Eddington luminosity
\begin{equation}
L_{Edd}=4\pi G M m_p/\sigma_T \; {\rm erg/s} \;, \label{L_Edd}
\end{equation}
which is the maximum X-ray luminosity that will not prevent
accretion.
Since the proton inelastic cross-section is lower than the Thomson
cross
section $\sigma_T$, technically the proton luminosity can exceed
$L_{Edd}$. On
the other hand $L_{Edd}$ can only be achieved at the surface of the
neutron
star and a realistic luminosity limit depends on the ratio of the
neutron star
radius  to the shock radius $R_{ns}/R_s$. Thus a reasonably
optimistic limit
for the proton luminosity will be
\begin{equation}
L_p^{max} = L_{Edd} \times (R_{ns}/R_s) \times
(\sigma^{inel}/\sigma_T) \;,
\label{L_p^max}
\end{equation}
i.e.\ of the order of or lower than $L_{Edd}=1.4\times10^{38}\times
M/M_\odot$~erg/s, corresponding to a rate of upward TeV muons
$<50$ events per year in a $10^5$~m$^2$ detector for a source
at 10 kpc.

Another potential source of energy in an X-ray binary is
pulsar rotation.
 Harding \& Gaisser~\cite{HG} have studied proton acceleration at
X-ray
binaries powered by the pulsar through a pulsar wind shock.
An absolute upper bound on the energy is the power released
by magnetic dipole radiation,
\begin{equation}
L_d =  4 \times 10^{43} B_{12} P_{ms}^{-4} {\rm erg/s} \;,
\label{L_d}
\end{equation}
where $B_{12}$ is the pulsar surface magnetic field strength in
10$^{12}$~G
and $P_{ms}$ is the pulsar period in milliseconds. Discussing
different
X-ray systems, however, they end up with a maximum proton
acceleration
luminosity of $6\times10^{38}$~erg/s for Cygnus X-3 with a pulsar
period of 12.8~ms \cite{Turver}.
This is still factor of 2 smaller than  the
luminosities used in the Cygnus X-3 estimates above and illustrates
that an X-ray
binary has to put almost all of its energy in high energy protons to
be
detectable in neutrinos. For a more modest X-ray binary at the
galactic center
that accelerates 1/10 $L_{Edd}$ in high-energy protons the actual
upward muon
rate will be $\sim$3 events per 10$^5$\,m$^2$ per year.

\subsection{Young Supernova Remnants}

Young supernova remnants are another candidate for production of
observable
neutrino fluxes \cite{BerPril,Sato}.
If protons are accelerated inside a young
supernova remnant, they will interact with the material of the
expanding
shell and produce $\gamma$-rays and neutrinos until the particle
adiabatic
losses exceeds the collision loss.  In the approximation of
a uniform density shell of mass $M$ expanding with
velocity $v=10^9$~cm/s this occurs at
\begin{equation}
\tau_a\,=\,({{3 M c\, \sigma_{pp}} \over {4 \pi m_H v^3}})^{1/2}\,
        =\, 1.3 \times 10^7 ({{M} \over {M_\odot}})^{1/2}\; s.
\label{tau_a}
\end{equation}
The active time during which
the production is significant is of order 1~year. Two modifications
of this idea~\cite{GHS} were motivated by the explosion of SN1987A
and the
proliferation of detailed supernova models that followed.
If one accounts correctly
for the velocity distribution of the supernova shell, the $\gamma$ and
$\nu$ emission time increases by a factor of three. Also, if the
accelerated particles are contained within the shell as
it expands the pathlength for collisions will increase
and the duration of the signal will be extended for a period
estimated in Ref.~\cite{GHS} as $\sim 10$ years, with a gradual
decrease in intensity.  If the accelerated protons are not
confined in the shell, the duration of the signal would be
1--2 years \cite{Yamada}.

All these considerations concern the target for inelastic
interactions.
The other ingredient is the abundance of accelerated protons at this
stage of remnant evolution.
The pulsar wind model~\cite{GHS} utilizes the
pulsar spin down energy to create a shock inside the contact
discontinuity
of the shell. The proton luminosity  is bounded by the magnetic
dipole
luminosity  of the pulsar given in Eq.~(\ref{L_d}).  The efficiency
for producing a signal in such a model depends on the
efficiency for accelerating protons and on the degree of
mixing between the accelerated particles and the expanding
shell.  The latter depends on mixing the pulsar wind region
with the shell through Rayleigh-Taylor instabilities.

Although it is clear now that SN1987A does not contain
a strong pulsar, it is still of interest to discuss the
signal that could be expected from a
a young galactic supernova ($\sim$10~kpc) with a rapidly
spinning, strongly magnetized pulsar.
The answer is extremely sensitive to the magnetic field
and pulsar period assumed.  Both parameters enter into
the pulsar power and into the maximum energy.  For example,
for $P=10$~ms and $B_{\rm surface}=10^{12}$~Gauss and a
25\% efficiency for particle acceleration and interaction,
the model of Ref.~\cite{GHS} gives $10^{39}$~erg/sec
and $E_p^{\rm max}\approx 10^5$~TeV.  The corresponding
neutrino luminosity would be sufficient to produce a
signal of $\sim 100$ upward muons in $10^5$m$^2$
for several years.  For a longer pulsar period and/or
a smaller surface magnetic field, both the maximum
energy and the available power rapidly decrease.

Berezinsky \& Ptuskin~\cite{BerPtus} argued that acceleration
at the supernova blast wave could also produce an observable
signal, even though in this case the accelerated particles
are not deep inside the expanding shell.  When a supernova
expands into the surrounding medium it drives a blast wave
ahead.  There is also a reverse shock in the supernova
ejecta.  Particle acceleration occurs at both shocks, with
the accelerated particles injected into the
respective downstream regions,
which are contained between the two shocks.
The kinetic energy of the expanding shell is huge
(of order $10^{51}$~erg/s) but the rate at which it is
dissipated is limited by the rate at which matter is swept
up by the expanding shell.   Thus the luminosity from accelerated
particles in this region is quite
sensitive (quadratically~\cite{BerPtus}) to the mass loss rate
of the progenitor star, which was relatively low for SN1987A.
For what is considered a ``typical'' mass loss rate of
$10^{-5}\,M_\odot$~\cite{BerPtus}, the estimated neutrino
flux for a supernova at 10~kpc corresponds to several hundred
upward muons with $E_\mu>100 GeV$ in the first 100 days \cite{BerPtus}.
The rate falls off slightly faster than 1/t.

The big disadvantage of young supernova remnants as potential
neutrino sources is, of course, that supernova explosions
are rare events. The one that  we
were lucky to observe, SN1987A, was not only quite distant,
in the  LMC,
but also shows no signs of pulsar activity at a level above
$\sim$10$^{37}$\,erg/s.

\section{Possible Extragalactic Sources}

Active galactic nuclei are the most luminous objects in the
Universe and have long been recognized as possible sources
of high energy signals \cite{BlaBla}.
These first estimates were mostly based on the total AGN power
and number density. More recent calculations \cite{BierStrit,SikBeg}
developed the idea
in two important ways. They first identified the potential importance
of hadrons (especially neutrons) for transporting energy in
active galactic nuclei. Secondly, shock
acceleration models were at least crudely incorporated into the
AGN models, and the photoproduction process was shown to be the
most important one for proton energy loss. This led to estimates
of the maximum proton energy achievable in acceleration at
AGN shocks and to the prediction of high energy neutrino fluxes.

Active galactic nuclei have luminosities ranging from 10$^{42}$
to 10$^{48}$ erg/s, which corresponds to black hole masses
from 10$^4$ to 10$^{10}$ $M_\odot$ \cite{mjrees} on the natural
assumption
that they are powered by Eddington limited accretion onto a black
hole.
AGN's have generally flat emission spectra
with a luminosity up to $\sim$3$\times$10$^{46}$ erg/s per
decade of energy. In the IR band a steady dust emission is observed,
most probably coming from a large region far away
from the core. The main thermal feature is the UV bump,
which is variable on a timescale of days and weeks\cite{UVvar}.

Its energy  source is either X-ray heating\cite{UV_X}
or viscous heating of the accretion disk\cite{UVvar},
either of which would be  closely related to the
central engine. X-rays  have a hard, nonthermal
spectrum, variable on even shorter timescales\cite{Xvar},
which often cuts off at several MeV.
AGN's have been extensively studied at radio frequencies,
where the most general identification is as radio-{\it loud}
or radio-{\it quiet}, depending on the fraction of energy in
the radio portion of the spectrum \cite{Sanders}.
Roughly 10\% of all observed AGN's are classified as
radio-loud \cite{fraction}. Blandford \cite{Blandford}
suggests that radio-loud AGN's have rapidly spinning black
holes and therefore also strong jets. The UV bump is not always
easy to see in radioloud AGN's.

Two possible sources within AGN's of intense, high
energy neutrino  fluxes
have been identified.  The first is associated with the central
engine and the second with production in jets associated with
blazars, which are radio-load AGN's in which the observer is
illuminated by the beam of a jet.  We first discuss  central
emission.

\subsection{Generic AGN}
To introduce most of the parameters important for the production of
neutrinos,  we  briefly describe the spherical
accretion model used in most of the calculations
of the neutrino production in central regions of
AGN's \cite{SikBeg,Stecketal,BegSik,SzaboPro92}.
Some of the limitations of this model will be mention
in \S7.4 below.  The model is based on work
performed by Kazanas, Protheroe and Ellison \cite{Protheroe,KazEll}.
They assume that close to the black hole the accretion flow
becomes spherical and a shock is formed where the ram pressure
of the accretion flow is balanced by radiation pressure near the
black hole.  The shock radius  is parameterized by
$R = x_1 \times R_S$, where
$R_S$ is the gravitational (Schwarzschild) radius of
the black hole, and $x_1$ is estimated to be in the range $10$ to
$100$ \cite{SzaboPro92}.
The continuous emission is dominated by the ultraviolet and
X-ray radiation, which
are assumed to emanate from inside the radius enclosed
by the shock.
Since the region inside the shock is optically thick, the radiation
density at the shock can be estimated
from the surface brightness of the AGN.  This leads to the relation
\begin{equation}\label{radiation}
U_{rad} \simeq L \times (\pi R^2 c)^{-1}
\end{equation}
between luminosity and radiation density in the central region.
Since $R=x_1\times R_S\propto L_{\rm Eddington}$, it follows
from Eq.~(\ref{radiation}) that $U_{rad}\propto L^{-1}$.  Numerically,
\begin{equation}\label{numeric}
U_{rad}\sim 2\times 10^6 {\rm erg/s}\times {1\over L_{45}}
\times \left({30\over x_1}\right)^2,
\end{equation}
where $L_{45}$ is the luminosity divided by $10^{45}$~erg/s.
The radiation energy density also defines the magnetic
field value $B$
at the shock under the assumption of equipartition
of the radiation  and
magnetic energy.  For the numerical example above
$B\sim 7000\;{\rm Gauss} \times (L_{45})^{-{1\over 2}}
\times {30\over x_1}$.

Acceleration of protons is assumed to occur by the first order
diffusive Fermi mechanism at the shock, resulting in an
$E^{-2}$ differential spectrum that extends up to $E_{\rm max}$.
Energy loss processes occur during acceleration,
including $p\gamma\rightarrow N\pi$ and $p\gamma\rightarrow
p+e^++e^-$
in the dense radiation fields as well as $pp$ collisions in the gas.
All three processes contribute an energetic electromagnetic
component, either through $\pi^0\rightarrow\gamma\gamma$ or
by production of electrons.  Both photo-meson production and
$pp$ collisions also give rise to neutrinos via the
$\pi^\pm\rightarrow\mu^\pm\rightarrow e^\pm$ decay chain.
In the astrophysical environment all unstable particles
(except quasi-stable neutrons) decay
practically without energy loss.
An important detail is that photoproduction of charged
pions by protons is dominated by the $n\pi^+$ channel
\cite{Stecker68}.

Although high energy neutrinos escape directly from the core,
the electromagnetic component does not.  The
core is optically thick to photons with energies greater
than $\sim 5$~MeV.  All $\gamma$-rays generated in the
dense photon field immediately lose energy in
$\gamma\gamma\rightarrow e^+e^-$ collisions.  Inverse
Compton/pair-production cascades downscatter all electrons and
photons to X-ray and lower energies. The essential
ingredient of these models is that the observed X-ray spectrum
is produced as the end product of the electromagnetic cascades
initiated by high energy photons and electrons produced by
the accelerated protons.  Thus, estimates of expected
neutrino fluxes from  individual AGN's are normalized through the
model to their observed X-ray luminosities.

The proton density at the shock, $n_p(R)$, can be estimated from the
accretion rate needed to support the black hole luminosity, and
from the radius and accretion velocity at the shock.  It is
\begin{equation}
n_p \simeq 1.3 \times 10^8 x_1^{1/2} R^{-1.5} L^{1/2} Q^{-1} {\rm
cm}^3,
\end{equation}
where $Q$ is the efficiency for converting accretion power
into accelerated particles at the shock.
Such proton densities are not only
a good injection source for proton acceleration, but also
a possible target for $pp$ interactions.

The proton energy loss, however, is dominated at high energy
by the photoproduction
process $p\gamma \rightarrow n\pi^+ (p\pi^0)$ simply
because the target photon density $n_{ph}$ is much higher
than $n_p$. For thermal radiation with temperature T$^\circ$K
 the density ratio would be
\begin{equation}
{{n_p} \over {n_{ph}}}\, \simeq \,
2.5 \times 10^{-13}\, x_1^{3/2}\, T\,  Q^{-1}.
\end{equation}
The high cross-section pair production process
($p\gamma \rightarrow pe^+e^-$) is relatively unimportant
because of the low proton energy loss per collision.
The thermal radiation  corresponds
to photon energies in the range $1$ to $40$~eV.

The relative importance of the different energy-loss
mechanisms  depends in
detail on the energy-dependence of the various cross sections
and on the intensity and spectral shape of the target
radiation  field.  The
detailed calculations \cite{Stecketal,SzaboPro92,SikBeg,BegSik} use
numerical and/or Monte Carlo techniques to follow the production,
propagation and cascading of the secondary particles inside the
central region and to determine the fluxes of neutrinos, nucleons
and X-rays that emerge.  Without going into such detail, it is
still possible to describe in a semi-quantitative way the basic
results,
especially the shape and upper limit of the neutrino spectra.
To do this, we make use of the approximate form of the radiation
field given by Stecker {\em et al.} \cite{Stecketal}.

The minimum energy of a target photon for
photoproduction by a proton of energy $E_p$ is
\begin{equation}\label{Ecritical}
\epsilon\;\approx {\Delta^2 - m_p^2\over 2 E_p}
\end{equation}
where $\Delta = 1.232$~GeV is the mass of the $(3,3)$ resonance.
The collision length of a nucleon against photoproduction
is \begin{equation}\label{ell}
\ell^{-1}\;=\;\int\,\sigma(\epsilon)\,n(\epsilon) {\rm d}\epsilon,
\end{equation}
where $n(\epsilon)$ is the number density of photons (differential
in energy).  Using a resonance approximation for the
cross section  gives
\begin{equation}\label{ellapprox}
\ell^{-1}\;\approx\;{\pi\Gamma\Delta\over\Delta^2-m_p^2}\,\epsilon\,
n(\epsilon)\times \sigma_{\rm peak},
\end{equation}
where $\Gamma\approx 115$~MeV is the width of the $\Delta$
resonance and $\sigma_{\rm peak}\approx 5\times 10^{-28}$~cm$^2$.
Since $R\propto L$ and $n(\epsilon)\propto U_{\rm rad} \propto
L^{-1}$,
the ratio $R/\ell$ is independent of luminosity in the model.
{}From the spectrum of Ref.~\cite{Stecketal} one finds that
$R/\ell<1$ above the UV bump, i.e. for $\epsilon>40$~eV.
{}From Eq.~\ref{Ecritical} this corresponds to
$E_{\rm crit}\approx 8\times 10^6$~GeV.
Thus nucleons with energy less than $\sim 10^{16}$~eV can escape
from the central region ($r<R$) {\em if they propagate
rectilinearly}.

Nucleons that escape no longer contribute to the production of
secondary photons and neutrinos.  This has little effect on the
predicted down-scattering into the X-ray region since most of
the energy has already been dumped by nucleons with
higher energy (provided the accelerated spectrum extends
to $E_{\rm max}\gg 10^{16}$~eV).  The assumption made about
propagation
of protons does, however, have a crucial effect on the predicted
neutrino spectrum.  If, as assumed by Stecker {\em et al.}
\cite{Stecketal},
protons travel in straight lines inside the central region, then
the neutrino spectrum will follow the proton spectrum only
down to an energy roughly
\begin{equation}
\langle{E_\nu\over E_p}\rangle\times 10^{16}\;{\rm eV}\sim 5\times
10^5\;
{\rm GeV}.
\end{equation}
At lower energy the neutrino spectrum d$N_\nu/{\rm d}E_\nu$ will be
constant, reflecting the flat momentum distribution of neutrinos
produced in a $p\gamma$ collision.  If, as seems more likely,
protons remain confined in the central region by the same turbulent
magnetic fields necessary for the diffusive shock acceleration to
work, then the neutrino spectrum will follow the proton spectrum
down to much lower energy.  Both Refs.~\cite{SzaboPro92} and \cite{BegSik}
assume that protons will be confined to the central region.

Figure~7 illustrates the  difference the
assumption of  proton
confinement makes. It shows the model neutrino spectra
($\nu_\mu+\bar{\nu}_\mu$) for the extragalactic source 3C273.
The thin lines show several of the models of Protheroe and
Szabo \cite{SzaboPro92}, who performed their calculation for
$x_1$ values from 10 to 100 and two different photon target
spectra.  The thick line represents the model
of Stecker {\it et al.}\cite{Stecketal}.  For both sources
the neutrino spectrum continues to follow the $E^{-2}$
proton spectrum down to low energy in the calculation where the
protons remain confined in the central region.
Because of the large neutrino flux
in the important region around 1 TeV, the models of
Szabo and  Protheroe
generate significantly more upward going muons than
predicted by Stecker et al. \cite{Stecketal}.

The slight dip in the spectra of Ref.~\cite{SzaboPro92}
around $10^4$--$10^5$ GeV is caused by proton energy loss to
$e^+ e^-$ pair production, which dominates proton
energy losses for proton energies between $\sim$30
and $\sim$3000~TeV. This feature is much more prominent
in the calculation of Ref.~\cite{BegSik}
than in Ref.~\cite{SzaboPro92}. As a consequence of
the larger relative contribution of pair production,
the predicted neutrino-induced signal of Ref.~\cite{BegSik}
is somewhat smaller
than that of Ref.~(\cite{SzaboPro92}), as shown with a
dash line on Fig.~7.
Sikora and Begelman \cite{BegSik} give only the spectral
shape for a generic source. We have normalized their
neutrino spectrum to the 3C273 luminosity.
The exact position of the maximum
neutrino energy is thus uncertain, because it depends on the
parameters of the particular source.

Neutrons are not confined by magnetic scattering in the inner
region.  Thus neutrons with $E<E_{\rm crit}$ escape from
the central region inside R provided they do not decay first.
For the relevant range of parameters, neutrons with energy
above a TeV will usually escape.  These energetic neutrons
decay at distances ranging from $\sim$0.01 to $\sim$100 parsec,
and their decay products can have a profound effect on energy
transport in AGN's, for example driving winds \cite{SBR} and
producing radio emission \cite{Kazanas} far from the core.

As far as neutrinos are concerned, the escape of neutrons
from the core has an interesting consequence for the
shape of the spectrum
of electron antineutrinos.  The dominant channels
for photoproduction of charged mesons by nucleons are
$p\gamma\rightarrow n\pi^+$ and $n\gamma\rightarrow p\pi^-$.
The kinematics of the
$\pi^+\rightarrow\mu^+\rightarrow e$ decay chain is such that
the flux of $\nu_\mu$ from pion decay is approximately
equal to the flux of $\bar{\nu}_\mu$ from muon decay, and
vice versa for decay of $\pi^-$.  Thus from protons
one has roughly equal fluxes of
$\nu_\mu$, $\bar{\nu}_\mu$ and $\nu_e$ from the $\pi^+$.
The neutron chain leads to $\bar{\nu}_e$ instead of $\nu_e$.
Thus for $E>E_{\rm crit}$, when both neutrons and protons interact
inside the core region, the flux of $\bar{\nu}_e$ is nearly
equal to the flux of $\nu_e$ (only slightly suppressed by the
small energy loss of the nucleon in $p\gamma\rightarrow n\pi^+$).
At lower energies, the neutrons escape before interacting.
One then gets $\bar{\nu}_e$ from $n\rightarrow p\,e^-\,\bar{\nu}_e$,
with, however, a strong kinematic suppression because of the
very small energy transfer to the leptons in $\beta$-decay
of the neutron.  For $\bar{\nu}_e$ from neutron $\beta$-decay, the
the spectrum of $\bar{\nu}_e$ is a factor $\sim 5\times 10^{-4}$
lower than the parent neutron spectrum, as compared to a factor
of about $5\times 10^{-2}$ when $E>E_{\rm crit}$ and the
process $n+\gamma\rightarrow \pi^-\rightarrow\mu^-\rightarrow
\bar{\nu}_e$ can occur.

As it turns out, the photoproduction in the UV bump also limits
the maximum proton acceleration energy $E_p^{max}$,
and hence the maximum neutrino energy.
This differs from the situation in a more diffuse environment,
such as acceleration by a supernova blast wave expanding
into the interstellar medium.  In that case the upper limit
is determined by the characteristic lifetime of the shock.
(See the discussion of Eq.~11 above.)  $E_p^{max}$ is roughly
proportional to $L^{1\over 2}$, reaching a value of
10$^{17}$ eV for  $L=10^{45}$ erg/s, with at least
a factor of two uncertainty \cite{SzaboPro92}.

This maximum energy  can be estimated
by comparing the acceleration rate (\ref{Emax})
to the energy loss rate, $K_{\rm inel}E_p\,c/\ell$.
The acceleration rate in this case is
\begin{equation}\label{Eloss}
{{\rm d}E\over {\rm d}t}\sim 0.1 {u^2\over c} e B
\approx  2\times 10^5\;{\rm GeV\,s}^{-1}
\times{30\over x_1}\times\left( L_{45}\right )^{-{1\over 2}},
\end{equation}
where we have used the equipartition estimate of the magnetic
field from Eq.~(\ref{numeric}).
Since (see Eq.~\ref{ell}) $\ell^{-1}\propto n(\epsilon)\propto L^{-1}$,
we estimate
\begin{equation}
E_p^{\rm max}\;\propto\;L^{1\over 2}.
\end{equation}
Using once again the radiation spectrum
of Stecker {\em et al.}, one finds that the energy loss rate
becomes comparable to the acceleration rate at the peak of
the UV bump, where $E_p\approx 3\times 10^8$~GeV.
For $E_p^{\rm max}$ above $\sim$10$^{19}$~eV,
which in this model can be achieved only in AGN
with total luminosity $\simeq$10$^{48}$~erg/s, proton
synchrotron radiation becomes
the most important energy loss channel.

The results of the calculation of Szabo \& Protheroe
\cite{SzaboPro92} can be summarized
by the following approximate formula \cite{ProthSta}, which
gives the neutrino flux at Earth from an AGN of given X-ray
flux and $E_p^{max}$ in [cm$^2$.s.TeV]$^{-1}$
\begin{equation}
F_\nu{E_\nu}\, \simeq \, 0.25 F_X \exp(-20 E_\nu/E_p^{max}) \times
E_\nu^{-2},
\end{equation}
where $F_X$ is the 2--10 KeV X-ray flux (erg cm$^{-2}$s$^{-1}$)
and $E_\nu$ is the neutrino energy in TeV.

The generic AGN model, described above, is a first order
approximation of the physical processes that may take place in
active galactic nuclei. The assumption of spherical accretion
could be an adequate representation of the accretion flow inside
a thick accretion disk.

An attempt to use a different geometry in the
center of AGN's was made in the model due to Nellen,
Mannheim \& Biermann \cite{Nellen}.
In this model it is assumed that protons are accelerated
somewhere near the central region, perhaps in the bases
of the jets, and that both X-rays and neutrinos are generated
in collisions of the accelerated protons in the inner
regions of the disk.  This model is less specific
than those of Refs.~\cite{SzaboPro92,Stecketal,BegSik}, and
the main production process is assumed to be quite different.
Nevertheless, the predicted neutrino fluxes are rather similar
to those of Refs.~\cite{SzaboPro92} and \cite{BegSik}.  This
is because the intensity is normalized to the X-ray luminosity
and the protons are assumed to be confined to the central
region until they lose all their energy through collisions.

 Individual radioquiet AGN will be difficult to detect,
although the atmospheric background in a 1$^\circ$ radius around
the source is extremely small,
$\sim$$1.6 \times 10^{-6}$ m$^{-2}$yr$^{-1}$ muons above 1 TeV.
Even with optimistic luminosities, the number of such events from
individual sources is less than 2--3 yr$^{-1}$ in a $10^5$ m$^2$
detector.  For example, the estimated rates from the models of
Ref.~\cite{SzaboPro92} are $\sim$1~per $10^5$~m$^2$yrs for
3C273 and $\sim3$ in the same units for NGC4151.

\subsection{Diffuse AGN neutrino flux}

 In their pioneering paper Stecker {\it et al}~\cite{Stecketal}
integrated the neutrino fluxes from single
generic AGN's to obtain a diffuse flux of neutrinos
from all cosmological AGN. The integration has to account
for the AGN density and luminosity distribution,
as well as for the neutrino adiabatic energy loss due to
the expansion of the Universe. This procedure is identical to
the integration used to calculate the value of the diffuse
X-ray background. In fact, it uses the AGN
luminosity function derived from X-ray observations \cite{Macc,Mori}
and assumes that the neutrinos and the X-rays have a common source.

  The AGN luminosity function as a function of the redshift can be
expressed as
\begin{equation}
\rho(L_X, z)\;=\; R_0^3 {{g(z)} \over {f(z)}} \rho_0 \left( {{L_X}
\over {f(z)}} \right),
\label {xraylum}
\end{equation}
where $\rho_0$ comes from measurements of the AGN luminosity,
$R_0$ is the present scale size of the Universe and $g(z)$
and $f(z)$ describe the number density and luminosity evolution
of AGN in the co-moving volume. Any AGN induced background,
including X-ray and neutrino,  will then
have energy spectrum~\cite{SzaboPro92}
\begin{equation}
{{dI} \over {dE}}\;=\; {{1} \over {4 \pi}} {{c} \over {H_0}}
{{1} \over { E R^3_0}}
\int dL_X \int _0 ^{Z_{max}} dz \times
\rho(L_X, z) (1 + z)^{- \alpha}{{dL} \over {dE}}\{ E(1+z), L_X\},
\label {cosmol}
\end{equation}
where $L$ is the appropriate differential luminosity and
$\alpha$=5/2 for the Einstein-de Sitter cosmological model.

Figure 8 shows the current estimates of the isotropic
$\nu$ background ($\nu_\mu+\bar{\nu}_\mu$). The estimates of Szabo \&
Protheroe are made with different values of x$_1$ and photon
target spectra, and they are integrated using two independent
sets of luminosity functions \cite{Macc,Mori}. The resulting $\nu$
flux extends to very high energy, where it dominates the
atmospheric neutrino background by several orders of magnitude.
Because of the isotropic nature of the background flux,
its major feature is the extremely flat energy spectrum.
The thick line shows the corrected prediction of Stecker
{\it et al.} \cite{Stecketal}. While the $\nu$ spectra
are now in very reasonable agreement at the
higher energy end, the biggest difference occurs at energies
below 3$\times$10$^5$ GeV, where Stecker {\it et al} spectrum
becomes flat while the spectrum of Protheroe \& Szabo  follows the
primary proton spectrum, for the reason described in
the previous subsection. The difference reaches 2.5--3 orders of
magnitude at $E_\nu=10^4$ GeV, which makes a crucial difference in
the estimate of the $\nu_\mu$-induced upward muon signal.

Figure 9 shows the muon fluxes generated by the isotropic
neutrino background as in the bracketing high and low models
of Szabo \& Protheroe  \cite{SzaboPro92} and by Stecker
{\it et al.} \cite{Stecketal}. These have been calculated \cite{TKGnu92}
as described in \S~3.2.1  for comparison with the Frejus
measurement \cite{Meyer}, which gives a 90\% C.L. upper limit
of 2.3 events with $E_\mu\,<\,2$~TeV
for the range of zenith angles $-0.3<\cos\theta<0.3$  \cite{Meyer}.
The corresponding upper limit \cite{Rode} is shown in Fig.~9.
The muon flux generated by atmospheric neutrinos
averaged over the same angular interval is shown for
comparison. The AGN background dominates at muon energies
above 1 TeV.

Although the diffuse neutrino flux from AGN's
is isotropic, the produced muons
will be suppressed in the vertical direction (from below) by
an amount that depends on the relative importance of
high energy neutrinos in the spectrum.  The interaction length
of neutrinos in the Earth is less than an Earth radius
when $\sigma_\nu\sim 10^{-33}$~cm$^2$, i.e. for
$E_\mu\sim  10^7$~GeV \cite{Renoquigg}.
Accounting for absorption in the Earth,
the predicted TeV muon rate in a
downward looking detector with acceptance of $10^5$m$^2$sr
will be 160 to 800 {\it per year} for
Ref.~\cite{SzaboPro92} and $\sim$40 for Ref.~\cite{Stecketal}
over an atmospheric background of $\sim$140 events.
The higher range of predictions
of Ref.~\cite{SzaboPro92}, however, are already ruled out
by the Frejus limit.

So far we have discussed signals generated by muon neutrinos and
antineutrinos.  Electron neutrinos of sufficiently high
energy can generate air showers which could be observable
above the background or ordinary showers near the horizontal
because of the great penetrating power of neutrinos.
Limits on horizontal showers have been given by the Akeno
air shower experiment, as discussed in \S3.4 above.
The Fly's Eye detector has searched for upward-going
showers, which would be generated by electron neutrinos from
below that interact near enough to the surface so the resulting
electromagnetic cascade emerges from the ground before it is
absorbed.  For $E_\nu\gg10^7$~GeV, these events would be mostly
near the horizontal since the Earth would absorb the more vertical
high energy neutrinos.  The Fly's Eye limits~\cite{RMB} apply for
$E>10^8$~GeV and are discussed further below in connection with
cosmological neutrinos.

Electromagnetic cascades generated by charged
current interactions of electron neutrinos can
also be detected when they occur inside the volume of a
Cherenkov detector.  The rates are then simply
the convolution of flux, cross section and fiducial volume.
Figure~10 \cite{TKGVenice}
shows the rates predicted for the $\nu_e+\bar{\nu}_e$
spectra of Refs.~\cite{Stecketal,SzaboPro92}.
The dotted line shows the background
of atmospheric electron neutrinos.  Since the atmospheric
neutrino spectrum is steeper for $\nu_e$ than for $\nu_\mu$, the
flux of AGN $\nu_e$ crosses the atmospheric background at lower
energy than for $\nu_\mu$.  In the examples given in
Fig.~10, there are $\sim 10$ interactions per
1000 kt years of electron  neutrinos with
$E_\nu>1$~TeV in a typical model from Ref.~\cite{SzaboPro92} as
compared to $0.5$ in Ref.~\cite{Stecketal} and $0.2$ atmospheric
above the same energy threshold.  The rates plotted in
Fig.~10 are integrated
over all directions.  For $E_\nu\sim10^7$~GeV, absorption of upward
neutrinos by the Earth begins to be significant, suppressing the
quoted rates of $\nu_e$ from AGN slightly.

The spike at $6\times 10^6$~GeV in Fig.~10 represents the
interaction of $\bar{\nu}_e$ at the ``Glashow resonance''
\cite{Glashow}.
The resonance cross section
for $\bar{\nu}_e\,+\,e^-\,\rightarrow\,W^-\,\rightarrow
\bar{\nu}_e + e^-$ is \cite{HandM}
\begin{equation}
\label{glashow}
\sigma (\bar{\nu}_e e^- ) \;=\;
 {{G_F^2 s} \over {3 \pi}} \times
\left[{{M_W^4} \over {(s - M_W^2)^2 + \Gamma_W^2 M_W^2}}
\right]
\end{equation}
where $\Gamma_W \approx 2.1$~GeV is the width of the $W$ and
resonance occurs for $E_\nu = E_0 =  s/(2\,m_e)\approx
6.4\times10^6$~GeV. The peak cross section value is
\begin{equation}
\label{Glashowpeak}
\sigma(E_0)\;=\; {{ 1 } \over
{B_{W \rightarrow \bar{\nu}_e e^-}}}
{{ G_F^2 M_W^4 } \over {3 \pi \Gamma_W^2}}
\approx {\rm 9} \times {\rm 5.2} \times {\rm 10}^{-32} {\rm cm}^{2}
\approx {\rm 4.7} \times {\rm 10}^{-31} {\rm cm}^{2}
\end{equation}
for a total of nine (3 leptonic and 6 hadronic) W$^-$
decay channels.
Integrating Eq.~\ref{glashow} gives the rate per electron as
\begin{equation}
{\rm Rate}\;=\;{\pi\,\sigma(E_0)\,
\Gamma_W\,M_W\over 2\,m_e}\,\phi_{\bar{\nu}_e}
\left({M_W^2\over 2\,m_e}\right)\approx \phi_{\bar{\nu}_e}(E_0)\times
(2.4 \times 10^{-25}~{\rm GeV\,cm}^{2}).
\end{equation}

Detection of other exotic phenomena, such as
multi-W production \cite{Morris}, is also possible.

\subsection{AGN Jets}

The recent observations of GeV $\gamma$-rays from a large
number of extragalactic objects by the EGRET
instrument\cite{EGRET} on GRO has stimulated intense interest
in models of particle acceleration in jets with relativistic
bulk flow.  This is because most,
if not all, of the EGRET sources are radio-loud AGN, which are
thought to be AGN's viewed from a position illuminated by the
cone of a relativistic jet.  Jets carry a sizeable
fraction of the AGN luminosity. Moreover, the
apparent
luminosity to an observer looking at a small angle to the jet axis is
increased by a factor of up to $10^4$ for a jet Lorentz factor of 10.  This
is a consequence \cite{FKR} of the fact that $I(\nu)/\nu^3$ is a relativistic
invariant, so that $I(\nu) = \Gamma^3I^*(\nu\Gamma^{-1})$, where
$I^*$ is the photon intensity
(erg~s$^{-1}$cm$^{-2}$sr$^{-1}$Hz$^{-1}$)
seen by an observer moving with the gas in the jet and
$\Gamma$ is the Lorentz factor of the jet averaged over the cone of
the jet relative to the line of sight.

The interest intensified still further with the discovery of
$\sim$TeV photons from the nearby AGN Markarian 421 \cite{Whmkn}.
What is the origin of such high energy photons?

Proposed explanations can be divided into two classes.
The traditional approach to the production of very high energy
photons
is based on inverse Compton (IC) scattering of accelerated electrons on
a seed photon field. The photon field could be either external, i.e.
generated outside the electron acceleration region, or due to
the synchrotron radiation of the electrons
({\em synchrotron-self-Compton}).  Examples in this category
are Refs.~\cite{SchlickDerm,Zdzolik,SikBegRees}.

An alternative approach is that of Mannheim {\em et al.}
\cite{MannBier,BierMann,Mannbla,MannPRD}.  Following
the arguments of Ref.~\cite{BierStrit}, it is assumed
that protons also are  accelerated in the jets.
These protons lose there energy by collisions on the synchrotron
photons.  In the process they dump energy into photons, electrons
and neutrinos {\em via} production of $\pi^0$ and $\pi^+$.  (The jets
are sufficiently diffuse so that high energy neutrons escape and
production of $\pi^-$ is therefore greatly suppressed.)  The photons
and electrons are reprocessed, and cascade to form an $E^{-2}$ power
law
differential photon spectrum down to the energy below which the
accelerating region becomes transparent.  These photons which
originate from interactions of accelerated protons dominate
the high energy signal in this picture.  At low frequency
($\nu\alt 10^{15}$~Hz) synchrotron radiation
from the electrons dominates.

Both pictures have some difficulties to overcome.  For example,
the models that do not involve nucleons generally require a higher
bulk Lorentz factor of the jet.  The radiation target density has
to be high enough for IC scattering and, at the same time,
low enough for the generated $\gamma$-rays not to be absorbed
by $\gamma\gamma$ collisions. This is difficult to arrange for,
especially when IC scattering in the Klein-Nishina regime is the
relevant process. In addition, the energy densities in soft photons
(IR to X-rays) and $\gamma$-rays are comparable, which requires
that the two types of radiation are generated in different
locations~\cite{SikBegRees}. To prevent the electrons from losing
too much energy to synchrotron radiation,
the energy of the magnetic field in
the jet has to be of order 5\% of the radiation
density\cite{Zdzolik},
far from the 0{\it th} order assumption of equipartition.


There is a corresponding set of problems that the models
of hadronic origin have to overcome. Because of the smaller
rate of energy loss by protons, the jet Lorentz factor is
no longer a big limitation. The seed photon density, however,
has to be high enough for photoproduction to occur, and, as
a consequence,
$\gamma$-ray absorption in the source is a problem.
Similarly, there are several other free parameters of the
model, such as the ratio of power in protons and in electrons.
These are fixed from multi-frequency observations.
In Fig.~11 we compare results of quasi-simultaneous
observations of 3C273 in the optical, X-ray and $\gamma$-ray
bands~\cite{3CC} with the predictions of a synchrotron-self-Compton
model~\cite{SchlickDerm} (solid line) and a hadronic
model~\cite{MannPRD}.
The hadronic model  gives correctly the spectral
shape at X-ray and $\gamma$-ray energy. The turn-up in the
$>$10 GeV region can not be detected at Earth, because of the
absorption on the IR/optical background on propagation from
the distant source ($z = 0.158$ for 3C273).

The detection of neutrinos from blazars would confirm the
hadronic model, since there is no source of neutrinos in the
electromagnetic models.  In this connection, the uncertainty
in optical depth for photons in the source is particularly
problematic because it introduces extra model-dependence
in the relation between the observed photon flux and the
predicted neutrino flux.  This problem, of course, disappears
once the neutrinos are detected.  In the remainder of this section
we review the estimates of the predicted neutrino signals in
the hadronic model of production of high energy photons in
jets of AGN.

A calculation of the neutrino production in AGN jets was
first published by Biermann and Mannheim\cite{MannBier,BierStrit,BierMann}.
The produced neutrino flux reflects the physical conditions
in the AGN jet. Because of the low photon density
protons can achieve quite high energy at acceleration
($E_p^{max}\sim3\times 10^{10}$~GeV in the frame moving with the
bulk flow of gas in the jet (jet frame).

In addition both the acceleration and proton
interactions proceed in the jet frame, so the neutrinos
are boosted to high energy (blueshifted) with a Doppler
factor of order 10. In the case of 3C273 the flux
is shown by a dotted line in Fig.~7.

This model \cite{Mannbla} of the proton acceleration
and interactions envisions a bulk flow of magnetized plasma,
streaming from the base of the jet towards its end (the hot spot).
The acceleration occurs in a sheets of radial
dimension $R$ and thickness $D$ moving out
with the jet. In order to explain the variability of 3C279 on
timescale of a day the radiating sheet thickness should be
$D\,\sim 10^{15}$cm, much smaller than its radius
($\sim 10^{18}$  cm).
The acceleration of protons (and possibly nuclei)
proceeds {\em via} first order Fermi
acceleration in the frame of the flowing plasma.  As usual, the
acceleration rate is given by (compare Eq.~\ref{Eloss})
\begin{equation}\label{accel}
{1\over E}\,{{\rm d}E\over{\rm d}t}\propto {e\,B\,c\over E}.
\end{equation}
In this model,
protons collide with the synchrotron photons generated by the
accelerated electrons, and the photon spectrum is
approximated as a power law with
integral index $\gamma \approx -1$. 
The threshold photon energy for production of the $\Delta$
resonance obeys $E_\gamma({\rm threshold})\propto E_p^{-1}$,
so the density of photons at the resonance is
proportional
to the proton energy.  Since equipartition is assumed between
the energy in electrons and the magnetic field,
the normalization of the photon field is proportional to $B^2$.
Thus the loss rate for protons depends on magnetic field and proton
energy as
\begin{equation}\label{ploss}
(t_p)^{-1}\;\propto\;E_p\,B^2.
\end{equation}
Combining Eqs.~\ref{accel} and \ref{ploss}
using the numerical values of Ref.~\cite{BierStrit}, leads to an
estimate of the maximum proton energy in the jet frame of

\begin{equation}
E_p^{max,*}\le 2 \times 10^{10} B^{-1/2}\;{\rm GeV},
\end{equation}
where $B$ is the magnetic field strength in Gauss.

   The fact that the dominant energy loss process
for high energy  protons
in the hadronic model of AGN jets is photoproduction at
threshold in collisions on a power law
spectrum of photons (rather than a thermal spectrum) has an
interesting consequence also for the shape of the produced spectrum
of secondary pions.  If, as expected in first order shock
acceleration, the differential proton spectrum is $E^{-2}$, then
the differential pion production spectrum will be harder by one
power of energy, i.e. $E^{-1}$.  Occasionally energetic protons
also collide with thermal gas in the jet, which leads to an $E^{-2}$
spectrum of pions.  Thus the characteristic shape of the production
spectrum of pions is $E^{-2}$ at low energy, flattening to
$E^{-1}$ at high energy up to some characteristic maximum energy.

The spectrum of neutrinos from decay of $\pi^+$ and $\mu^+$ follows
the pion production spectrum, shifted down in energy by
appropriate kinematic factors.
All energies are boosted by the bulk Lorentz factor $\Gamma$
relative to an external observer.  The observed (boosted) maximum
neutrino energy is estimated by Mannheim \cite{Mannbla} as
$E_\nu^{\rm max}\sim 10^9$~GeV.  The boosted energy at which
the observed neutrino spectrum flattens from $E^{-2}$ to
$E^{-1}$ is in the range $10^5$ for 3C273 \cite{MannPRD}
to $10^7$  GeV
for Mkn 421 and 3C279 \cite{Mannbla}.
Photons are even more complicated than neutrinos
because their spectra at production
(from $\pi^0\rightarrow\gamma\gamma$ and from radiation by
electrons)
are reprocessed by pair cascading in the ambient
photon and magnetic fields.  The total photon luminosity
eventually emerges from the source at lower energy
(but boosted by
the bulk Lorentz factor).  Approximately equal amounts
of energy are carried by the four leptons that result from
the  decay chain
$$\pi^+\,\rightarrow\,\nu_\mu\,\mu^+\,\rightarrow
\,e^+\,\nu_e\,\bar{\nu}_\mu \; .$$ In addition,
$$p\gamma\,\rightarrow\,p\,\pi^0\;\approx\;
      2\times p\gamma\,\rightarrow\,n\,\pi^+\; $$
at the $\Delta$ resonance.  Thus 3/4 of the energy
lost to photoproduction ends up in the electromagnetic
cascade and 1/4 goes to neutrinos.  In addition, some of the
energy of the accelerated protons is lost to direct pair
production ($p\,+\,\gamma\rightarrow e^+\,e^-\,p$).  Thus
\begin{equation}\label{luminosity}
L_\nu\;\le\;{1\over 4} \,L_\gamma
\end{equation}

Equation~(\ref{luminosity}) can be used to relate an observed photon
spectrum to a predicted neutrino flux in the model.
The relation is further complicated by
attenuation of high energy
($\agt$~TeV) photons
during propagation from the source~\cite{Steckerabs,PrStabs}.

The spectra of
both $\gamma$-rays and neutrinos are
generated in $p\gamma$ and, to a lesser extent, $pp$ interactions.
For photoproduction the
energy carried by neutrinos is directly
related through kinematics
to the $\gamma$-ray luminosity as
\begin{equation}
L_\nu\, =\, \frac{3}{4} L_{\pi^+}\, =\,\frac{3}{13}L_\gamma,
\end{equation}
where $L_\gamma$ includes a contribution from $e^+e^-$ pairs.
The predicted flux for 3C273 \cite{MannPRD} is
shown in Fig.~7.
It would give a rate of upward muons with $E_\mu>1$~TeV of
$\sim$0.1 per year in a $10^5$~m$^2$ detector.

The simultaneous observation of the BLLac source Mkn 421 by
EGRET~\cite{EGRETmkn} and the Whipple observatory~\cite{Whmkn}
is especially valuable for the understanding of the physics of AGN
jets because of the large range of energy for detected photons.
The two observations define an energy spectrum with
$\gamma=2.06\pm0.04$ over more than four decades in
energy~\cite{Wspec}.
Mkn 421 is the closest source observed
by EGRET at a redshift of 0.031. This is significant because
of the absorption on propagation.
The exact energy dependence of the absorption feature is
uncertain because the magnitude and the energy spectra of the
IR and optical background(s) are not well known. Within a
factor of 2, however, 3 TeV $\gamma$-rays emitted at the
distance of Mkn 421 will already start being absorbed and will
show an apparent steepening of the spectrum independently of
the production spectrum.

One should appreciate that weakly interacting neutrinos will make
their way to
our detectors unattenuated by ambient matter in the source or by IR
light. So, while high energy photons are absorbed on intergalactic IR
photons
for AGNs much further than Mkn~421, neutrinos are not and sources
should be
detected with no counterpart in high energy photons.
Halzen and Vasquez~\cite{Hazquez}
scale the Mkn 421 $\gamma$-ray flux to neutrino flux, making a range
of
assumptions for the $\gamma$-ray absorption at source, expressed in
terms of the magnetic field value $B$ of the jet.
The VHE $\gamma$-ray flux from Mkn 421 is~\cite{Wspec}
\begin{equation}
\int_{1/2\,\rm TeV} dE \left[ E dN_\gamma\over dE \right] =
1.5\times 10^{-11} \, \rm cm^{-2} s^{-1} \;.  \label{int dE}
\end{equation}
It is assumed \cite{Wspec}
that the production spectrum (before attenuation)
is given by a power law with integral spectral index
$\gamma$.  The corresponding neutrino flux that would be expected if
the observed photon spectrum is to be understood in the hadronic
model is quite uncertain.  This is because the amount of
absorption in the source is not well determined.
What is needed is the
optical depth of the source, {\it i.e.}\ the jet.  Biermann
\cite{PLBEr} gives the following (admittedly model dependent) estimate:
\begin{equation}
        \tau_{\rm optical} = 2 \left[ B\over 1 \,\rm
Gauss\right]^{1/2}
        \left[ E_{\gamma}\over \rm 1\, TeV\right] \;.  \label{tau_o}
\end{equation}
The photon flux will be attenuated for energies above
which $\tau_{\rm optical}=1$. According
to (\ref{tau_o}) the optical depth of the source is unity for the
0.5~TeV photons observed by Whipple for a 1~G field.
The true value of the $B$-field in the jet is a guess
which ranges from $10^{-4}$ to  $10^4$~G.
The gamma ray flux can be computed inside the
source by correcting
the observed flux (\ref{int dE}) for absorption
in the jet. The  answer
depends critically on the magnitude of the
$B$-field. Once the unattenuated photon spectrum is established,
the neutrino flux is estimated simply by assuming
one neutrino per gamma ray, as appropriate for pion decay.

 Table~\ref{up muon} \cite{Wspec} shows the results for a range of
assumptions for  $B$
and the spectral index $\gamma$.
We conclude that in this scenario Mkn~421 should
produce a handful of upcoming muon events
per year in a generic
$10^5$~m$^2$ detector. Within the framework of the model
the magnetic field would be limited to $B\alt 1$~Gauss because
the observed
$\gamma$-ray spectrum extends above 1 TeV.
The situation could be different in other potential sources.

\begin{table}[h]
\caption{ \label{up muon}
Number of upcoming muons ($N$) per $ 10^5$~m$^2$ per y for the
different scenarios. ${\cal L_\gamma}$ is in $10^{43}$~erg/s. }
\begin{displaymath} \arraycolsep=1.5em
\begin{array}{c|c|c|c|c|c}
\hline
\hline
\vrule width0pt depth5pt height 14pt
B\mbox{(Gauss)} & \gamma & E_p^{max} & E_\gamma \; \mbox{for} \;
\tau_{opt}=1
& {\cal L_\gamma} & N \\
\hline
        & 1   &                     &                   & 30 & 2 \\
10^{-4} & 0.8 & 2 \times 10^{22}\rm eV & 50 \;  \mbox{TeV} & 500 & 11
\\
        & 0.4 &                     &                   & 10^6 & 450
\\
\hline
\vrule width0pt depth5pt height 14pt
1       & 1   & 2 \times 10^{20}\rm eV & 500 \; \mbox{GeV} & 200 & 13
\\
\hline
\hline
\end{array}
\end{displaymath}
\end{table}

 Stecker {\it et al.}~\cite{Steckhen} use the $\gamma$-ray
absorption on propagation to normalize the expectations from
other GRO sources. They find a flux of
neutrinos from the 3C273 jet sufficient to generate $\sim$0.1
muons above 1 TeV in 10$^5$ m$^2$ per year. The corresponding
flux from the 3C273 core is 40\% smaller. The quiescent state
of 3C279 would generate 5 muons, while the highest observed
$\gamma$-ray flux from 3C279 would correspond to 25 such muons.
The 3C279 core contribution is only 0.1 event.
These are the bracketing values for the expectations of
neutrino fluxes from active galactic nuclei, it the $\gamma$-rays
are indeed generated in  hadronic interactions by accelerated nuclei.

 Analogously to the background from generic, radio-quiet AGN,
one can integrate the emission of all AGN  jets to obtain
an isotropic ultra high energy neutrino background. Two estimates
of this background, due to Stecker\cite{Steckjet}
and Mannheim\cite{Mannbla}, are shown in Fig.~8.
Although the overall normalization of the isotropic neutrino
background from AGN jets is lower that that of generic AGN,
it extends to higher energy, and crosses it over at some very high
energy. The cross-over is explicit in the flux of Mannheim,
while Stecker {\it et al.} give only the slope ($\gamma$ = 2)
and the normalization.  In any case, the normalization is
somewhat uncertain because of the variability of the blazars.
The normalization of Mannheim's diffuse flux in Fig.~8
corresponds to an assumption of a 15\% duty factor for
blazars to be in a high state \cite{privateKarl}.

\subsection{AGN Neutrinos: Discussion and conclusions}

Although the spherical accretion model is
very useful for estimating the neutrino fluxes that
might be expected from cores of AGN, it is
subject to criticism from various points of view.
The model is only applicable
to accretion disks with thickness comparable to or exceeding
the dimension of the shock radius. It has to be constructed in
such a way that there is no leakage of the generated $\gamma$-rays
before their energy is downscattered to X-ray and longer
wavelengths. Any significant leakage would exceed the experimental
limits on diffuse extragalactic $\gamma$-rays. At the
same time, the radiation density cannot be so high as
to prevent the acceleration processes from occurring.
Some authors~\cite{BerLea}
estimate the source efficiency $Q$ to be lower than 1/3 and
ask if the conditions in the AGN nucleus are suitable for shock
formation at all. Others~\cite{MastKirk} show the danger of
overproducing background radiation through pair production and
synchrotron radiation, which could lead to shock instability and
drastically decrease $E_p^{max}$.  Some of these problems might be
avoided by placing the shocks in the bases of the AGN
jets~\cite{Nellen}.

 The calculations of the neutrino production in AGN jets are no
less difficult. To model correctly all the jet physics one
has to follow in some detail all the processes involved in
the frame of the relativistic plasma flow, including particle
acceleration and reacceleration at multiple shocks, $\gamma$-ray
production, multiplication and absorption in electromagnetic
cascades in a non stationary fashion. This is very complicated
problem that involves many free parameters. The simple scaling
of the neutrino fluxes with the $\gamma$-ray luminosity for
individual sources may not be exact, since the conditions at
the source are poorly known. The ratio of the magnetic to
radiation energy density, for example, which is essential for
the $\gamma$-ray absorption at the source, can vary within at least
one order of magnitude. The sources are also highly variable,
and many might have been observed during the peak of their activity.

The big question  is the fraction of the AGN luminosity
that goes through the nucleonic channel. Although it has been pointed
out~\cite{Begelman,BierStrit,SikBeg} that hadrons have
suitable interaction cross sections and are a natural vehicle for the
energy transport throughout the AGN disk, nucleons are not
strictly necessary for the solution of this problem. Since the
models of the non-nucleonic origin of the Mkn 421 $\gamma$-rays
are already struggling to extend the theory to $\gamma$-rays
above 1 TeV, a possible observation of, say, 10 TeV $\gamma$-rays
would be a confirmation of their $\pi^0$ origin.

This is hardly possible, however, because of the absorption
on the IR/optical background, even if the production spectrum
reaches much higher energy. There is only a slight chance
that~\cite{PrStabs},
for very low values of the extragalactic magnetic field, the
cascading on this background will flatten considerably the
spectrum observed in the GeV/TeV region. Such flattening would
reveal the extension of the production spectrum to much higher
energy and correspondingly confirm the $\pi^0$ origin of the
$\gamma$-ray flux.

The criticism above does not imply that the current predictions
are not reliable. They are results of the first generation
of research, which will become more exact in the near future. The
differences between various estimates reflects the uncertainties
of the calculations. Conclusions are that the expected
fluxes of source neutrinos are well below the sensitivity of
the currently active deep underground detectors with effective
area less than 1000 m$^2$. They are, however, tantalizingly close
to being detectable by the new generation of
detectors  especially designed for neutrino astronomy.

\section{Cosmological Neutrinos}

Another possible source of extremely energetic diffuse neutrinos
could be the interactions of ultra high energy cosmic rays on the
microwave background. The importance of such interactions was
noted by Greisen~\cite {Greicut} and independently by Zatsepin and
Kuzmin~\cite {ZatsKuz} soon after the discovery of the background
radiation.
These early papers stated the existence of an universal cut-off of
the cosmic ray spectrum due to photopion production. The question of
the production of neutrinos and $\gamma$-rays was developed later
in works by Wdowczyk {\it et al}~\cite{WTW},
Stecker~\cite {Steck3d}, Hill \& Scramm~\cite{HillSch},
Berezinsky and Grigorieva~\cite {BerezGri},
Halzen {\it et al}~\cite{HPSV} and others, and in a
recent paper of Yoshida \& Teshima~\cite {YoshTesh}, who perform a
detailed Monte Carlo calculation of the proton propagation in the
microwave background and the generation of neutrinos.

 The major source of energy loss is photoproduction, as described
in \S7.1. Here the target is the microwave background, with a density
of $\sim 400$~photons/cm$^{3}$ and an average energy
$\epsilon\simeq 7\times 10^{-4}$~eV, corresponding to the temperature
of the background radiation.
For cosmic rays exceeding
\begin{equation}
E_p \approx {\Delta^2 - m_p^2\over 2 (1 -\cos\theta) \epsilon}
\approx {5 \times 10^{20} \over (1 - \cos\theta)}\; {\rm eV} \; ,
\end{equation}
where $\theta$ is the angle between the proton and photon directions,
the photopion cross-section grows very rapidly to reach a maximum of
540 $\mu$b at the $\Delta^+$ resonance ($s = 1.52 \; {\rm GeV}^2$).
The $\Delta^+$ decays to $p \pi^0$ with probability of 2/3,
and to $n\pi^+$ with probability 1/3. Neutral pions give rise
to ultra-high-energy $\gamma$-ray fluxes,
and charged pions---to neutrino fluxes through the
decay channels of Eq.~(\ref{cascades}). Decay
kinematics is such that all three neutrinos take
approximately 1/4 of the parent pion energy. In
addition the neutrons also decay and produce a
small flux of  $\bar{\nu}_e$ at much lower energy.

 Because of the width of the thermal photon distribution, and the
isotropic
nature of the microwave background, there is some phase space for
photopion
production at proton energies as low as 10$^{19}$ eV. These are only
possible in head to head interactions on the high energy tail of the
microwave background spectrum (or on the infrared/optical
background).
Most of the proton energy loss in this energy range, however, is on
direct pair production ($p\gamma \rightarrow p e^+ e^-$),
which has a lower threshold but does
not contribute to the neutrino fluxes. Photopion production starts
dominating at energy above $3 \times 10^{19} eV$ and the
cross-section
reaches maximum at $\sim 5 \times 10^{20} eV$, where the proton mean
free
path $\lambda_p = (\sigma_{p \gamma} n_\gamma)^{-1}$ is
$\simeq 5 \times 10^{24}$ cm ($\sim$2 Mpc).
Since protons lose on the average 1/5 of their energy per interaction
the proton attenuation length $\Lambda_p$ comes to $\sim$ 10 Mpc, a
number that the exact calculation of Ref.~\cite {YoshTesh} shows is
reached for proton energies above 10$^{21}$ eV.

 The magnitude and intensity of the cosmological neutrino fluxes is
than determined by the maximum injection energy of the
ultra-high-energy cosmic rays and by the distribution of the
sources. If the sources are relatively close by, at distances
measured
in tens of Mpc, and the maximum injection energy is not much
greater than the highest observed cosmic ray energy (few $\times
10^{20}
eV$), the generated neutrino fluxes are negligible. If, however,
the highest energy cosmic rays are generated at many sources at large
redshift, then a large fraction of their injection energy would be
presently contained in $\gamma$-ray and neutrino fluxes. The most
important reason is that the energy density of the microwave
radiation, and the proton photopion production cross-section,
scales with $(1 + z)^4$. The effect is even stronger if the
source luminosity were increasing with $z$, i.e. cosmic ray sources
were more active at large redshifts---`bright phase' models.

 The neutrino flux is given by an integral identical to
Eq.~(\ref{cosmol}),
where $g(z)$ and $f(z)$ now correspond to number density and the
luminosity
function of the cosmic ray sources. In general, the cosmic ray
sources are defined by their injection spectra,
luminosity and cosmological evolution. The
normalization comes from the requirement that the ultra-high-energy
cosmic rays after propagation in the microwave background match the
observed spectra. The loss resulting in $\gamma$-ray fluxes,
downscattered on the microwave background, should not violate
the experimental limits on isotropic extragalactic
$\gamma$-rays~\cite{WW}.
Yoshida \& Teshima~\cite{YoshTesh} give the muon
and electron neutrinos separately for different injection models
characterized by the $z_{max}$ value, maximum injection energy
$E_{max}$
and different source evolution functions of the form
$\eta(z) = \eta_0(1 + z)^m$. Fig.~15 below shows these fluxes for
$E_{max}= 10^{22}eV$ and two extreme sets of evolution parameters:
$m=0$, $z_{max}=2$ (low) and $m=4$, $z_{max}=4$ (high). It is
important to
remember that such drastically different source evolution models
can fit equally well the observed cosmic ray spectrum.

  There are specific models that identify the sources of the
extragalactic cosmic rays. Rachen \& Biermann~\cite{RachBier} propose
that hot spots of Fanaroff-Riley class II radio galaxies, being
the largest and most powerful shock waves in the Universe, are
the dominant sources of cosmic rays of energy above 10$^{18}$ eV.
In this case $g(z)$ and $f(z)$ are the number and luminosity density
functions of FR-II galaxies, derived from radio observations at
particular radio frequencies.
They do not calculate the neutrino fluxes generated
in cosmic ray interactions of the microwave background, but since
their models match the observed
cosmic ray flux at energies around 10$^{18}$ eV, such a calculation
should be close to the results of Ref.~\cite{YoshTesh} for similar
source evolution functions, i.e. to be bracketed by the
extreme fluxes shown on Fig.~15.

  Independently of the specific model of the sources of the highest
energy cosmic rays, the associated neutrino fluxes can only dominate
the highest energy region, above $E_\nu$ = 10$^{17}$--10$^{18}$~eV.
This would only happen in the case that cosmic rays of energy
above 10$^{18}$ eV are indeed accelerated at numerous high redshift
astrophysical sources.
At lower energy the neutrino background is dominated by the
neutrinos generated in interactions at active galactic nuclei

 An intriguing possibility is
(see Refs.~\cite{OTW,HSW})  that
the highest energy cosmic rays are produced by energy loss of
superconducting cosmic strings. The strings lose energy in the form
of massive fermions ($M_F \propto 10^{15}$ GeV) that decay into
baryons and fermions. The spectrum is modified by interactions
on the 3K background but extends up to the Planck mass. This, plus
the contribution from strings at large red-shift,
increases the cosmic ray energy loss in the $>10^{19}$ eV range.
A neutrino flux a factor of 30 higher than the proton flux is
created. Only nucleons (not nuclei) can be
generated through this channel.

 More recently the neutrino emission from cosmic strings has
been discussed in the context of `cusp annihilation'
\cite{Bhatt,McGB,BhattR}.  The total energy of the string in the
region
of the cusp is released in the form of massive scalar and
gauge particles that decay rapidly into particle jets. Equal numbers
of
particles and antiparticles are generated by conservation of quantum
numbers. Neutrinos come mostly from the ordinary pion decay channels.
The shape of the jet fragmentation function is crucial for the
number of generated neutrinos and their energy spectrum. The
assumptions
used require an extrapolation of the observed jet fragmentation
function
up to the Planck scale. Another crucial parameter is $\mu$, the
string mass
per unit length, which defines to total luminosity of the string. For
$G \mu/c^2 \simeq 10^{-6}$, consistent with large-scale
structure formation and the observed anisotropy of the microwave
background, the generated neutrino fluxes are smaller than the
predictions
from ultra-high-energy proton propagation. Maximum neutrino fluxes
are obtained for $G \mu/c^2 \simeq 10^{-15}$.  In this case,
however, the cosmic strings would not have other
cosmological implications.

\section{Search for Dark Matter}

It is believed that most of our Universe is made of cold dark matter
particles. In the context of big bang cosmology,
these particles have
interactions of order the weak scale and masses of order
$M_W$, i.e. they are
WIMPs \cite{SeckelDM}.
{}From rotation curve measurements we also know
their density and average velocity in the galactic halo.
This information is the basis for estimating
the annihilation rate of WIMPS into
high energy neutrinos.

Galactic WIMPs, scattering off protons in the sun, lose energy. They
may fall below escape velocity and be gravitationally trapped.
Trapped dark matter  particles eventually come to equilibrium
temperature,
and therefore to rest at  the center of the sun. While the WIMP
density
builds up, their annihilation rate into lighter particles increases
until
equilibrium  is achieved where the annihilation rate equals half of
the
capture rate. The sun has thus  become a reservoir of WIMPs which
annihilate
into any open fermion, gauge  boson or Higgs channels. The leptonic
decays from annihilation channels such as
$b\bar b$ heavy quark pairs and $W^+W^-$ turn the sun into a source
of high energy neutrinos. Their energies are in the GeV
to TeV range, rather than in the keV to MeV range familiar
from its nuclear burning. These neutrinos can
be detected in deep underground experiments.

We will illustrate the power of neutrino telescopes as dark matter
detectors using as an example the search for a 500~GeV WIMP with a
mass outside  the reach of present accelerator and future LHC
experiments.
A quantitative estimate of the rate of high energy muons of WIMP
origin
triggering a detector can be made in 5 easy steps. An exact
quantitative calculation requires a complex code~\cite{HalzenDM}.

\smallskip
\noindent
{\it Step 1:} The halo neutralino flux $\phi_{\chi}$.
It is given by their number density  and average
velocity. The cold dark matter density implied by the observed
galactic rotation curves is $\rho_\chi$ = 0.4 GeV/cm$^3$. The
galactic halo is believed to be an isothermal sphere of WIMPs
with average velocity $v_\chi$ = 300 km/sec. The number density
is then
\begin{equation}
n_\chi = 8\times 10^{-4} \left[ 500{\rm\ GeV}\over m_\chi \right]\rm\
cm^{-3}
\end{equation}
and therefore
\begin{equation}
\phi_\chi = n_\chi v_\chi = 2\times 10^{4} \left[ 500{\rm\ GeV}
\over  m_\chi
\right] \rm\ cm^{-2}\, s^{-1} \,. \label{flux}
\end{equation}

\smallskip
\noindent
{\it Step 2:} Cross section $\sigma_{\rm sun}$ for
the capture of neutralinos by the sun.
The probability that a WIMP is captured is proportional
to the number
of target hydrogen nuclei in the sun (i.e.\ the solar mass divided by
the nucleon mass) and the WIMP-nucleon scattering cross section.
{}From dimensional analysis $\sigma(\chi N)\; \sim \;
\left(G_F m_N^2\right)^2/m_Z^2$
which we can envisage as the exchange of a neutral weak boson between
the WIMP and a quark in the nucleon. The main point is that the WIMP
is known to be weakly interacting. We obtain for the solar capture
cross section
\begin{equation}
\Sigma_{\rm sun} = n\sigma = {M_{\rm sun}\over m_N} \sigma(\chi N)
= \left[1.2\times 10^{57}\right] \left[10^{-41}\,\rm cm^2\right] \,.
\label{capture}
\end{equation}
\smallskip
\noindent
{\it Step 3:} Capture rate $N_{\rm cap}$ of neutralinos by the sun.
$N_{\rm cap}$ is determined by the WIMP flux (\ref{flux}) and the
sun's capture
cross section (\ref{capture}) obtained in the first 2 steps:
\begin{equation}
N_{\rm cap} = \phi_\chi \Sigma_{\rm sun} = 3\times 10^{20}\,\rm
s^{-1} \;\;{\rm for}\;m_\chi=500\;{\rm GeV}.
\end{equation}

\smallskip
\noindent
{\it Step 4:} Number of solar neutrinos of dark matter origin
One can check that the sun comes to a steady state where capture and
annihilation of WIMPs are in equilibrium. For a 500~GeV WIMP the
dominant annihilation rate is into weak bosons; each produces
muon-neutrinos
with a leptonic branching ratio which is roughly 10\%:
\begin{equation}
\chi\bar\chi \to WW \to \mu \nu_\mu \,. \label{branch}
\end{equation}
Therefore, as we get 2 $W$'s for each capture, the number of
neutrinos
generated in the sun is
\begin{equation}
N_\nu = {1\over 5} N_{\rm cap}
\end{equation}
and the corresponding neutrino flux at Earth is given by
\begin{equation}
 \phi_\nu = {N_\nu\over 4\pi d^2} = 2\times 10^{-8}\,\rm cm^{-2}
s^{-1} \,,
\label{nu flux}
\end{equation}
where the distance $d$ is 1 astronomical unit.

\smallskip\goodbreak
\noindent
 {\it Step 5:} Event rate in a high energy neutrino telescope.
For (\ref{branch}) the $W$-energy is approximately $m_{\chi}$ and the
neutrino energy half that by 2-body kinematics. The energy of the
detected muon is given by
\begin{equation}
 E_\mu \simeq {1\over2} E_\nu \simeq {1\over4}m_\chi \,.
\end{equation}
where we used the fact that, in this energy range, roughly half of
the neutrino energy is transferred to the muon.
 For the neutrino flux given by (\ref{nu flux}) we obtain
\begin{equation}
 {\rm \#\ events/year = 10}^5 \times \phi_\nu \times
\rho_{\rm H_2O}
\times
\sigma_{\nu\to\mu} \times R_\mu \simeq 100
\end{equation}
for a 10$^5$m$^2$ water cherenkov detector, where $R_\mu$ is the
muon range and $\phi_\nu \times \rho_{\rm H_2O} \times
\sigma_{\nu\to\mu}$ is the simple analog of Eq.~(\ref{P_nu E}).

This exercise illustrates that present high
energy neutrino telescopes (of area $\sim$10$^4$ m$^2$)
are powerful devices in  the search for dark matter and
supersymmetry.  They are complementary to present
and future accelerator searches in the sense that they are
naturally sensitive to heavier WIMP's because the underground high
energy neutrino detectors have been optimized to be sensitive
in the energy region where the neutrino interaction
cross section and the range of the muon are large. Also, for high
energy neutrinos the muon and neutrino are nicely aligned along
a direction  pointing back to the sun with good angular resolution.
In addition, in the estimate given above we neglected some
other decay channels that contribute neutrinos, and we did
not include the signal from annihilation of WIMP's
trapped in the center  of the Earth~\cite{GouldDM}.
Direct searches for dark matter
are clearly highly desirable.  To achieve comparable
sensitivity to these indirect searches requires a sensitivity
at the level of  0.05~events/kg\,day \cite{BottinoDM}.

An elegant way to extend the Standard Model is to make it
supersymmetric \cite{HaberDM}.
If supersymmetry is indeed Nature's extension of the
Standard Model it must produce new phenomena at or below the TeV
scale.  An attractive feature of supersymmetry is that it provides
cosmology with a natural dark matter candidate
in form of a stable, lightest
supersymmetric particle~\cite{SeckelDM}. There are a priori six
candidates:
the (s)neutrino, axi(o)n(o), gravitino and neutralino.
These are, in fact, the only  candidates
because supersymmetry completes the Standard Model all the way to the
GUT scale where its forces apparently unify. Because supersymmetry
logically completes the Standard Model with no other new physics
threshold up to the GUT-scale, it must supply the dark matter.
Here  we will focus on the neutralino, which,
along with the axion, is for various   reasons
the most attractive WIMP candidate~\cite{BerezinskyDM}.

The supersymmetric partners of
the photon, neutral weak boson and the two Higgs
particles form four  neutral
states, the lightest of which is the stable neutralino
\begin{equation}
\chi = z_{11} \tilde W_3 + z_{12} \tilde B + z_{13} \tilde H_1 +
z_{14} \tilde
H_2 \,.
\end{equation}
In the minimal supersymmetric model (MSSM)~\cite{HaberDM}
down- and  up-quarks
acquire mass by coupling to different Higgs particles,
usually  denoted by
$H_1$ and $H_2$, the lightest of which is required to
have a mass of  order the
$Z$-mass. Although the MSSM provides us with a
definite calculational framework, its parameters are many.
For the present discussion we  only have to
focus on the following terms in the MSSM lagrangian
\begin{equation}
L = \cdots \mu\tilde H_1 \tilde H_2 -{1\over2} M_1
\tilde B \tilde B -{1\over2}M_2 \tilde W_3 \tilde W_3
- {1\over\sqrt2} g v_1 \tilde H_1  \tilde
W_3 - {1\over\sqrt2} g v_2 \tilde H_2 \tilde W_3 + \cdots \,,
\end{equation}
which introduce the (unphysical) masses $M_1,\ M_2$ and $\mu$
associated with
the neutral gauge bosons and Higgs particles, respectively.
$M_1$ and  $M_2$ are
related by the Weinberg angle. The lagrangian
introduces two Higgs  vacuum
expectation values $v_{1,2}$; the coupling $g$ is
the known Standard  Model
SU(2) coupling. Although the parameter space of the
MSSM is more  complex, a
first discussion of dark matter uses just 3 parameters:
\begin{equation}
                \mu,\ M_2,\ {\rm and}\ \tan\beta=v_2/v_1 \,.
\end{equation}
Further parameters which can also be varied include the
masses of  top, Higgs, squarks, etc.

Neutralino masses less than a few tens of GeV have been excluded by
unsuccessful collider searches. For supersymmetry to resolve the
hierarchy problem of the Standard Model the masses of
supersymmetric
particles must be of order the weak scale and therefore, in practice,
at the
TeV scale or below. Also, if neutralinos have masses of order a few
TeV and
above, they overclose the Universe. Despite its rich parameter space
supersymmetry has therefore been framed inside a well defined
GeV--TeV mass window.

Assuming supersymmetry we can fill in some factors  in the
``back-of-the-envelope'' estimates in the previous section. In
supersymmetry, heavy WIMPs annihilate  preferentially
into weak  bosons. Other important annihilation
channels include \cite{DreesDM}
\begin{equation}
\chi + \bar\chi \to b + \bar b. \label{chi to b}
\end{equation}
Heavy quark decays dominate neutralino annihilation
below the  $WW$-threshold.
Also the dimensional estimate of  the neutralino-nucleon interaction
cross section $\sigma(\chi N)$ can be replaced by an explicit
calculation. It supports the dimensional estimate in the previous
section. $\sigma(\chi N)$ receives contributions from 2 classes of
diagrams: the exchange of Higgses and weak bosons, and the exchange
of squarks. The result is  often dominated by the large coherent
cross section associated with the  exchange of
the lightest Higgs particle $H_2$ and is of the form
\begin{equation}
\sigma = \alpha_H (G_F m_N^2)^2 {m_\chi^2\over(m_N + m_\chi)^2} \,
{m_Z^2\over
m_H^4}
\end{equation}
or, for large $m_{\chi}$
\begin{equation}
\sigma = \alpha_H \left(G_F m_N^2\right)^2 {m_Z^2\over m_H^4} \,.
\label{large}
\end{equation}
The proportionality parameter $\alpha_H$ is of order unity, but can
become as small as $10^{-2}$ in some regions of the MSSM parameter
space.
This is illustrated in Fig.~12 where the MSSM parameter
space is
parametrized  in terms of the unphysical masses $M$($\mu$) of
the unmixed wino(Higgsino).  (The ratio of the vacuum expectation
values associated with the two Higgs  particles
$v_2/v_1(=2)$ is here fixed to some arbitrary value.) The relation of
these parameters to the neutralino mass is shown in the figure.
The full  lines show fixed values of the neutralino mass $m_{\chi}$.
The lines labelled by  squares
trace fixed values of the ``coupling'' $\alpha_H$. The dashed area
indicates $M$, $\mu$ values which are excluded by cosmological
considerations. In standard big bang cosmology neutralinos with
the corresponding  parameters will overclose the Universe.

Note that for a given $\chi$ mass there  are two
possible states with the same $\alpha_H$ value. One of them will
preferentially annihilate into weak bosons, the other into fermions.
Therefore, their neutrino signature is provided by $W,Z$ decay and
semi-leptonic heavy quark decays, respectively. Fig.~12
illustrates
that, for heavy  neutralinos, which can only be searched for by the
indirect methods discussed here  and are
therefore of prime interest, any detector which can study dark matter
with $\alpha_H$ as small as 0.1 can exclude the bulk of the phase
space
currently available to MSSM dark matter candidates.

 Our main conclusions are summarized in Fig.~13
which exhibits, as a
function of the neutralino mass, the detector area required to
observe one event per  year. The detailed calculation confirms
our previous estimate of 100 events per 10$^5$~m$^2$
per year for a 500 GeV neutralino.
The two branches in this and the following figures correspond to the
two solutions for a fixed neutralino mass; see also Fig.~12.
Various
annihilation thresholds are clearly visible, most noticeable is the
threshold  associated with the $W,Z$ mass near 100~GeV.  The graphs
confirm that large detectors are required to
study the full neutralino mass range. It is clear from
Fig.~13,
however, that even detectors of more modest size significantly
extend the range explored by accelerators~\cite{MoriDM,LoSeccoDM}.
Neutralinos of 1~TeV mass are observable in a detector of area a few
times
$10^3\,\rm m^2$. The energy of the produced neutrinos is typically
``a fraction'' of the  neutralino mass, e.g.\ 1/2 for neutralino
annihilation into a $W$ followed by a  leptonic $e \nu$ decay.
For lower masses the event rates are small because the  detection
efficiency for low energy neutrinos is reduced. This mass range has,
however, already been excluded by accelerator experiments. For very
high
masses the number density of neutralinos, and therefore the event
rate,
becomes  small. This is not a problem as problematically large masses
are excluded by theoretical arguments as previously discussed.
The same results are shown in Fig.~14(a) as contours in the
$M, \mu$
plane which denote the neutrino detection area required for
observation
of  1~event per year. Clearly the $10^5\,\rm m^2$ contour covers the
parameter space.  The problematic large $\mu, M_2$-region does not
really represent a  problem as its parameters lead to values of the
matter density $\Omega$ exceeding  unity as shown in the accompanying
Fig.~14(b).

A realistic evaluation of the reach of an underground detector
requires more than counting events per year. Realistic simulations of
statistics and systematics must be done. Also a more complete mapping
of the MSSM parameter space is required. For those interested we
refer to Ref.~\cite{HalzenDM}.

\section{Event Rates in a Generic 0.1 km$^2$ Detector:
Synthesis}

 In the preceding sections we have attempted to summarize the new
limits that will be set and
the most likely observations that may me made by the next generation
of high energy neutrino telescopes. If past history is a guide,
however, the most important discoveries that occur when a new window
is opened may be completely unanticipated.

 We summarize some of the estimated event rates in
Table~\ref{events}.
The corresponding neutrino fluxes are presented on Figs.~15
and 16.
We remind the reader that a 0.1~km$^2$ detector is 2500 times larger
than IMB, 100 times MACRO or LVD, but only ``a factor'' larger than
many of the detectors under consideration or
construction~\cite{XVII},
e.g.\ AMANDA, BAIKAL, DUMAND and NESTOR. A list of operating and
proposed underground detectors having the capability to detect high
energy neutrinos is given in Table~\ref{detectors}.

 Table~\ref{events} gives the rates of upward going neutrino induced muons
of atmospheric and extraterrestrial origin. The absorption in the
Earth becomes important for the flatter extraterrestrial neutrino
fluxes and the event rates are given  both with and without
absorption. The event rates expected from astrophysical
neutrino sources are estimated with an account for absorption.

 Some of the event rates in Table~\ref{events} predicted for the
same type of source differ from each other by two orders of
magnitude.
This reflects the degree of uncertainty of our knowledge about
the conditions and the role of different physical processes for
the energetics of the source. The low event rates from the diffuse
AGN background come from the revised calculations of Stecker
{\it et al.}~\cite{Stecketal}, while the high rates reflect the
highest
neutrino background of Protheroe and Szabo~\cite{SzaboPro92}.
These highest rates are, however, in contradiction with limits set by
the
Frejus experiment~\cite{TKGnu92,Meyer,Rode}.
The highest rate of diffuse TeV muons
allowed by the Frejus limit is $\sim$200 per 10$^5$ m$^2$ per year.

\begin{table}[t]
\caption{  \label{events}}
\vspace{.25cm}\centering
\small
\begin{tabular}{ccc}
\hline\hline
\vrule width0pt height 14pt&
\multicolumn{2}{c}{EVENTS PER YEAR IN 0.1 KM$^2$}\\
\hline
\multicolumn{1}{l}{$\bullet$ ATMOSPHERIC (angle averaged, per steradian)}\\
 \underline{ muon energy} & \cite{Volkova80}
& \cite{AGLS92}  \\
$>1$ GeV\phantom0& 7800 & 8300  \\
$>1$ TeV\phantom0 & 129 & 104 \\
\hline
\multicolumn{1}{l}{$\bullet$ ATMOSPHERIC in 1$^\circ$ circle,
Ref.~\cite{AGLS92}}\\
 \underline {muon energy} & \underline{$ \cos\theta $ = 0.05}
& \underline{$\cos\theta$ = 0.95}  \\
$>1$ GeV\phantom0& 12.6  &  5.6  \\
$>1$ TeV\phantom0 & 0.21 &  0.05 \\
\hline
\multicolumn{1}{l}{$\bullet$ EXTRATERRESTRIAL FLUXES (angle averaged)}\\
\multicolumn{1}{l}{\hspace*{5mm}$\phi_\nu\, = \,{\rm 2.7}\times{\rm 10}^{-5}
(E_\nu/GeV)^{-1.7}$ cm$^{-2}$s$^{-1}$}\\
\underline{muon energy} & \underline{no abs.} &
\underline{with abs.} \\
$>1$ GeV\phantom0& 32.7 & 32.0 \\
$>1$ TeV\phantom0& 4.3 &  3.8 \\
\multicolumn{1}{l}{\hspace*{5mm}$ \phi_\nu = {\rm 4.0}\times {\rm 10}^{-8}
(E_\nu/GeV)^{-1}$ cm$^{-2}$s$^{-1}$}\\
\underline{muon energy} & \underline{no abs.} &
\underline{with abs.} \\
$>1$ GeV\phantom0& 8.8 & 6.6 \\
$>1$ TeV\phantom0& 5.0 &  3.3 \\
\hline
\multicolumn{1}{l}{$\bullet$ ASTROPHYSICAL DIFFUSE FLUXES (per steradian)}\\
\underline{muon energy}& \underline{plane of galaxy}& \underline{AGN}\\
$>1$ GeV\phantom0& 12--20 & 80--1600\\
$>1$ TeV\phantom0& 1.5--3.0& 40--800\\
also $\nu_e\;(6.3\,{\rm PeV})+e\to W^-$&& 0.3 per 1000 kton\\
\hline
\multicolumn{1}{l}{$\bullet$ ASTROPHYSICAL POINT SOURCES (E$_\mu >1$ TeV)}\\
\multicolumn{1}{l}{\quad Galactic source (Eq.~35)/100}&& 2.6\\
\multicolumn{1}{l}{\quad Extragalactic source (3C273)}&& 0.1--25\\
\hline
\multicolumn{1}{l}{$\bullet$ 500 GeV WIMPS from
\raise.3ex\hbox{$\bigodot$}}&&
 100\\
\hline\hline
\end{tabular}
\end{table}

 The estimated event rates from galactic sources come from the
considerations presented in \S6. The event rate generated by the
flux of Eq.~(35) is an extreme upper limit, which would be difficult
to reconcile with the
observational limits from VHE/UHE $\gamma$-ray observations.
As a conservative estimate we quote a rate corresponding to
a neutrino flux smaller by two orders of magnitude. The least
certain rate is the one expected for single AGN, given in
Table~\ref{events}
for the source 3C273. The smaller rate (0.1/yr) comes from the
estimates for emission from the jets by Mannheim~\cite{MannPRD} and
Stecker {\it et al}~\cite{Steckhen}. The highest event rate (25/yr)
actually comes to the neutrino flux corresponding to the `high state'
$\gamma$-ray flux of 3C279~\cite{Steckhen}. It is not likely that
such a high luminosity could be maintained at the source for periods
as long as a whole year. The atmospheric neutrino
background relevant for source searches is given in two directions:
close to the zenith and close to the horizon.

 Even the smallest predicted event rates for the diffuse AGN
background are easily detectable by a 10$^5$ m$^2$ neutrino
telescope. The expected ratio of signal to background for TeV muons
is from 0.3 to more than 2. The background, atmospheric neutrino
rate, is large and allows calibration and continuous monitoring
of the detector. The observations of the diffuse neutrino flux from
the plane of the galaxy are much more difficult, although
the likely concentration of the excess events in the direction
of the galactic plane should be of some help.

\begin{table}
\renewcommand{\thefootnote}{\fnsymbol{footnote}}
\setcounter{footnote}{0}
\caption{   \label{detectors}}
\def\undertext#1{\vtop{\hbox{\vphantom(#1}\kern3pt\hrule}}
\def\vstrut{\vrule width0pt height13pt depth4pt}
\def\pp{\phantom+}
\hfil
\vbox{\small
\halign{\vstrut #\hfil & \hfil#\hfil\tabskip1em & \hfil#\tabskip1em&
#\hfil\tabskip0em\cr
\multispan4 \vstrut \hfil
OPERATING DETECTORS WITH HIGH ENERGY NEUTRINO \hfil\cr
\noalign{\vskip-8pt}
\multispan4 \vstrut \hfil DETECTION CAPABILITY \hfil\cr
\noalign{\hrule}
\hfil \undertext{Detector}& \undertext{Location}&
\hidewidth\undertext{Area (m$^2$)}\,\footnotemark \hidewidth\hfil&
\hfil\undertext{Technique}\quad \cr
NUSEX& Mont Blanc& 10\pp& streamer tubes/Fe\cr
KGF& India& 20\pp& streamer tubes, very deep\cr
SOUDAN II& USA& 100\pp& drift tubes/concrete\cr
KAMIOKANDE& Japan& 120\pp& water Cherenkov\cr
BAKSAN& Caucasus& 250\pp& liquid scintillator tanks\cr
IMB& USA& 400\pp& water Cherenkov\cr
LVD& Gran Sasso& 300\rlap{\,\footnotemark}\pp
& liquid scintillator, streamer tubes\cr
MACRO& Gran Sasso& 850\pp& liquid scintillator, streamer tubes\cr
\noalign{\vskip6pt}
\multispan4 \vstrut \hfil FUTURE INITIATIVES (partial list)\hfil \cr
\noalign{\hrule}
\hfil \undertext{Detector}& \undertext{Location}&
\hidewidth\undertext{Area
(m$^2$)}\hidewidth\hfil&\hfil\undertext{Technique}\quad
 \cr
SNO& Canada& 600\pp& \quad D$_2$O\cr
SUPERKAMIOKANDE& Japan& 740\pp&  \quad water Cherenkov\cr
BAIKAL& Baikal& 2000\pp&  \quad water Cherenkov\cr
GRANDE type& USA, Italy, Japan& $\sim$30000\pp& \quad water
Cherenkov\cr
DUMAND& Hawaii& 20000+& \quad water Cherenkov\cr
AMANDA& South Pole& 20000+& \quad Cherenkov in deep ice\cr
RAMAND& Antarctica& $10^6\,\rm m^2\,$?& \quad microwave detection\cr
\noalign{\hrule}}}\hfill
\end{table}
\setcounter{footnote}{1}
\footnotetext{The total detector area for vertical upward upward
going neutrinos is given for all existing detectors. The effective
area for source searches depends on the detector and
source location and on the detector efficiency for different
zenith angles.}
\setcounter{footnote}{2}
\footnotetext{In operation}

 The atmospheric background in point source searches is
generally small. For energies of a TeV or more
the neutrino direction can be reconstructed to 1~degree or better.
We therefore expect less than one event per year in a 1$^\circ$ bin
from  the combined atmospheric and diffuse AGN backgrounds.
It is then quite likely that one or more sources will be discovered
by a 10$^5$ m$^2$ detector, {\em provided\/}
that hadronic processes play
an important role in the energetics of powerful astrophysical
objects.

With an even larger  1 km$^2$ detector we could then begin to study
neutrino sources in some detail. It may be possible not only to
count sources, but also to observe a multiplicity of sources
with enough statistics to begin extracting information from
energy spectra and temporal behaviour, particularly in comparison
with photon observations. Comparisons of the neutrino and
$\gamma$-ray
spectra contain information about the photon absorption at source
and during propagation to Earth, i.e.\ about important physical
properties of the sources and the intergalactic medium.
We should be able to observe episodic flux increases and maybe
even the periodicity of the neutrino emission from binary sources.
It is unlikely that such detailed observations can be carried out
with detectors smaller than 1~km$^2$.

Finally, we mention the possibility that such a neutrino telescope
can carry out Earth tomography \cite{DeRujGla}, employing
the attenuation of ultra high energy neutrinos, and making a direct
density profile of the Earth (seismic measurements only yield
velocity
profiles, and moreover give little information on the Earth's core).

Halzen \& Learned \cite{Learned} have presented
arguments such as these for doing neutrino
astronomy on the scale  of
1~kilometer. In order to achieve
large area it is
unfortunately necessary to
abandon the low MeV thresholds of detectors such as IMB and
Kamiokande. One
focuses on high energies where: i) neutrino cross sections are
large, ii) the
muon range is increased, iii) the angle between the muon and parent
neutrino
is less than 1~degree and, iv) the atmospheric neutrino background is
small.
The accelerator physicist's method for building a neutrino detector
uses
absorber, 3 chambers with $x,y$ wires with associated electronics
with a price
of $10^4$~US~dollars per m$^2$. Such a 1~km$^2$
detector would cost 10~billion dollars.
It is therefore a high priority to find
methods which are more cost-effective to be able
eventually to commission
neutrino telescopes with effective area of order 1~km$^2$.
Obviously the proven technique developed by IMB, Kamiokande and
others cannot be extrapolated to the 1~km scale. All present
proposals do however exploit the Cherenkov technique well-proven by
these experiments. The direction of the neutrino is inferred from
the muon direction which is reconstructed by mapping the Cherenkov
cone of the muon travelling through the detector. The arrival
times and amplitudes of the Cherenkov photons, recorded by a grid
of detectors, are used to reconstruct the track of the radiating
muon.

Detectors presently under construction have a nominal effective area
of $10^4$~m$^2$. Baikal is presently operating 36 optical modules and
the South Pole AMANDA experiment started operating 4 strings with
20 optical modules each in January 94. The first generation
telescopes~\cite{Learnedprime}
will consist of roughly 200 optical modules (OM). The experimental
advantages and challenges are  different for each experiment
and, in this sense, they nicely complement one another.
Briefly,

\begin{itemize}
\item AMANDA is operating in deep clear ice with an attenuation
length in excess of 60~m, which is similar to that of
the clearest water used in the Kamiokande and IMB detectors.
Although residual bubbles are found at
depth as large as 1~km, their density decreases rapidly with depth.
Ice at the South Pole should be bubble-free below 1100-1300~m as it
is in other polar regions~\cite{Goobar}. The ice provides a
convenient mechanical support for the detector. The immediate
advantage is that all electronics can be positioned at the surface.
Only the  optical modules are deployed into the deep ice.
Polar ice is a sterile medium  with a concentration of radioactive
elements reduced by more than $10^{-4}$ compared to sea or lake
water.
The low background results in an improved  sensitivity which allows
for the detection of high energy muons with very simple trigger
schemes which are implemented by off-the-shelf electronics.
Being  positioned under only 1~km of ice it is operating in a
high  cosmic ray muon  background.
The challenge is to reject the down-going muon background relative
to the up-coming neutrino-induced muons by a factor larger
than $10^6$. The group claims to be able to meet this challenge with
an up/down rejection which is at present similar to that of
the deeper detectors. The task is, of course, facilitated by the low
background noise. The polar environment is difficult as well, with
restricted access and one-shot deployment of photomultiplier
strings.
The technology has, however, been satisfactorily demonstrated
with the deployment of the first 4 strings. It is now clear
that the hot water drilling  technique can be used to deploy
OM's larger than the 8~inch photomultiplier tubes now
used to any depth in the 2.8~km deep ice cover.

\item BAIKAL shares the shallow depth with AMANDA, and has
half its  optical modules pointing up, half down. The depth
of the lake is 1.4~km , so the experiment cannot expand downwards
and will have to grow  horizontally. Optical backgrounds similar
in magnitude to ocean water have been discovered in Lake Baikal.
The Baikal group has been operating for one year an array
with 18 down-looking Quasar photomultiplier
(a Russian-made 15~inch tube) units in
April 1993, and may well count the first neutrinos in a natural
water Cherenkov  detector.

\goodbreak
\item  DUMAND will be positioned under 4.5~km of ocean water, below
most biological activity and well shielded from cosmic ray muon
backgrounds, which are a factor of 100 lower than for the shallower
detectors.  One  nuisance of the ocean is the background light
resulting from radioactive decays,  mostly K$^{40}$, plus some
bioluminescence, yielding a noise rate of 60~kHz in a single OM.
Deep ocean  water is, on the other hand very clear, with an
attenuation length of order 40~m in the blue. The deep ocean is
a difficult location for access and service.
Detection equipment must be built to high reliability  standards,
and the data must be transmitted to the shore station for processing.
It has required years to develop the necessary technology and learn
to work in an environment foreign to high-energy physics
experimentation.
The DUMAND group has successfully analysed data on cosmic ray muons
from the deployment of a test string~\cite{DUM}. The power and signal
cables from the detector location to shore (length 25~km) and the
junction box are already installed. The group will proceed with the
deployment of three strings in 1995.

\item NESTOR is similar to DUMAND, being placed in the deep ocean
(the Mediterranean), except for two critical differences.
Half of its  optical modules will point up as in BAIKAL. The angular
response of the detector is  being tuned to be much more
isotropic than either AMANDA or DUMAND, which  will give
it advantages in, for instance, the study of neutrino oscillations.
Secondly, NESTOR will have a higher density of photocathode (in some
substantial volume) than the other detectors, and will be able
to make local coincidences on lower energy events, even perhaps
down to the supernova energy range (tens  of MeV).

\end{itemize}

 Other detectors have been proposed for near surface lakes or
ponds (e.g.\break GRANDE, LENA, NET, PAN and the Blue Lake Project),
but at this time none is in construction~\cite{Learnedprime}.
These detectors all  would have the great advantage of accessibility
and ability for dual use as  extensive air shower detectors,
but suffer from the $10^{10}$--10$^{11}$ down-to-up ratio of
muons, and face great civil engineering costs (for water systems
and light-tight containers). Even if any of these are built it
would seem  that the costs may be too large to contemplate a
full, kilometer-scale detector.

In summary, there are four major experiments proceeding with
construction, each of which has different strengths and faces
different challenges.  For the construction of a 1~km scale
detector one can imagine any of the above detectors being
the basic building block for the ultimate 1~km$^3$
telescope. The redesigned AMANDA detector (with spacings
optimized to the attenuation length in excess of 60~m),
for example, consists of 5 strings on a circle with 60~meter
radius around a string at the center (referred to as a $1+5$
configuration). Each string contains 13 OMs separated by
15-20~m. Its effective volume is just below $10^7~m^3$. Imagine
AMANDA ``supermodules'' which are obtained by extending the basic
string length (and module count per string) by a factor close to 4.
Supermodules would then consist of $1+5$ strings with 51 OMs
separated by 20~meters on each string, for a total length
of 1~km. A 1~km scale detector then might consist of a $1+7+7$
configuration of supermodules, with the 7 supermodules
distributed on a circle of radius 250~m and 7 more on a circle of
500~m. The full detector then contains 4590 phototubes,  which is
less than the 7000 used in the SNO detector. Such a detector
(see Fig.~17) can be operated in a dual mode:
\begin{itemize}
\item it obviously consists of roughly $4\times15$ the presently
designed AMANDA array, leading to an effective volume of
$\sim6\times10^8$~m$^3$.  Importantly, the characteristics of the
detector, including threshold in the GeV-energy range, are the same
as those of the AMANDA array module.
\item the $1+7+7$ supermodule configuration, looked at as a whole,
instruments a 1~km$^3$ cylinder with diameter and height of 1000~m
with optical modules. High-energy muons will be superbly
reconstructed
as they can produce triggers in 2 or more of the supermodules spaced
by
large distance.  Reaching more than one supermodule (range of 250~m)
requires energy of 50~GeV. We note that this is the energy for which
a neutrino telescope has optimal sensitivity to a typical
$E^{-2}$ source (background falls with threshold energy, and until
about 1~TeV little signal is lost).
\end{itemize}
Alternate methods to reach the 1~km scale have been discussed by
Learned and Roberts~\cite{Roberts}.

How realistic are the construction costs for such a detector?
AMANDA's strings (with 10 OMs) cost \$150,000 including deployment.
By naive  scaling the final cost of the postulated $1+7+7$
array of supermodules is of order \$75 million, comparable to
that of Superkamiokande~\cite{Suzuki} (with $11{,}200 \times
20$~inch
photomultiplier tubes in a 40~m diameter by 40~m high stainless
steel tank in a deep  mine). It is clear that the naive estimate
makes several approximations over-  and underestimating the
actual cost.

In the next and final section of this paper we will briefly
review alternate ideas for making a kilometer-scale
neutrino detector in a cost-effective manner.

\section{Alternative Methods for Neutrino Detection}
\subsection{Radio Detection}
Over 30 years ago the suggestion was first made that radio
antennas might be able to detect microwave emission from
neutrino-induced cascades~\cite{XXIII}. This method can in
principle lead to the construction of relatively inexpensive
neutrino telescopes. Until recently only rough estimates have
been made of the power emitted~\cite{XXIV,XXV}. They indicated
that the detection threshold is quite high, so high that the
technique is probably insensitive to the atmospheric neutrinos.

That the radio signals emitted by neutrino-induced electromagnetic
cascades are even close to observability is the result of
interesting physics. According to the Frank-Tamm formula the power
radiated by a particle with charge $ze$ travelling a pathlength $l$
in a medium of refractive index $n$ is given by~\cite{XXVII}
\begin{equation}
{dW\over d\nu} = \left[{4\pi^2\hbar\over c}\,\alpha\right]z^2\nu
\left[1-{1\over\beta^2 n^2}\right]l\;, \label{dW/dnu}  
\end{equation}
where $\nu$ is the frequency and $\alpha=(137)^{-1}$. Naively
one might expect the power generated in a shower of $N$ charged
particles to be proportional to $N\langle l \rangle$. This
is not correct. If the emitted wavelength is large
compared to the physical dimensions of the shower, or equivalently,
the electric pulse generated by the shower is short compared
to the period of the waves observed, then the emission by the
shower particles is coherent and the power is of order
$(\Delta q\cdot N)^2\langle l\rangle$~\cite{XXIII,XXVIII}.
Here $\Delta q$ is the excess negative charge in the shower
\begin{equation}
\Delta q = {N(e^-)-N(e^+)\over N(e^-)+N(e^+)}\;, \label{Delta q}
\end{equation}
which enters in the coherent case because the electric fields
from opposite charges cancel for the same $\langle l\rangle$.
$\Delta q$ is positive mainly because Compton scattering of
shower photons on atomic electrons creates an excess of negative
charges in the shower. Coherence thus implies an enhancement
$(\Delta q)^2N$ which can compensate the loss in power associated
with the $\nu$ dependence of Eq.~(\ref{dW/dnu}). Roughly,
from visible light to GHz radiowaves the suppression associated
with the factor $\nu$ in the Frank-Tamm relation is a factor
$10^6$ which can be compensated by coherence because $N$ is of
order $10^6$ for PeV showers.

 The technique has been successfully tested in experiments
measuring radio emission by air showers observed in coincidence
with particle arrays~\cite{XXIX}. Whereas atmospheric fluctuations
make the systematics of the radioemission difficult to handle,
this is not a problem in denser material, like ice.
The physical dimension of the shower is reduced because the radiation
length is only 39~cm. The coherence is retained to higher
frequencies,
where more energy is available.
Determination of the precise threshold for observation in a medium
like ice depends on the details of the cascade and one has to perform
a real time numerical simulation of electromagnetic cascades in ice.
The calculations~\cite{XXVI} show that to a very good precision
the enhancement factor from coherence $(\Delta q)^2N$ is
proportional to the primary energy, and therefore the power of the
radioemission is proportional to the square of the cascade energy.

The critical parameter is the energy threshold for detection
which not only depends on the power generated but also on the
absorption in the ice and the background noise from the apparatus
and its environment. Absorption at 1~GHz depends critically on
temperature and therefore on the location of the experiment.
Determination of {\it in situ} background noise is a complex
problem. Experiments indicate that thermal noise of temperature 300K
represents an adequate guess of the background~\cite{XXX}. For this
assumption the amplitude of the noise spectrum rises linearly
with frequency, i.e.\ it exhibits the same dependence as the
signal below $\sim 1$~GHz~\cite{XXIX}.  In a detector of
bandwidth $\Delta \nu$ the noise varies as $\Delta \nu^{1/2}$,
and therefore the bandwidth enhances the signal to noise ratio
by the square root of the bandwidth. Neglecting absorption and
assuming $\Delta \nu=1$~GHz, a signal to noise ratio of unity is
achieved for a detection threshold linearly proportional to the
distance $r$ from the shower,
\begin{equation}
E_{\rm th}\,({\rm PeV}) \simeq 5\times10^{3}\, r\,({\rm m})\;.
\label{E_th}
\end{equation}
{\it E.g.}\ only neutrinos of 5~PeV and above can be sampled
in 1~km of ice. This is a very high threshold. Existing experiments
already set limits on high neutrino fluxes which imply extremely low
event rates above 5~PeV. This threshold estimate is based on a signal
to noise argument. The power in the signal is about an order of
magnitude smaller than the result quoted by Zeleznykh and
collaborators~\cite{XXIV}.

\subsection{Acoustic Detection}

High detection threshold may also be the main shortcoming of
the idea of acoustic detection of  neutrinos, originally
suggested by Bowen~\cite{XXXII}. Acoustic waves are produced
by the heating of the medium by the ionization loss of the
cascade particles. Shower direction can also be determined
by the timing and amplitude of the signal in different
detectors~\cite{XXXIII}. The beam pattern  is coherent
and reinforcing in a plane perpendicular to the cascade direction.
One can envisage a detector consisting  of pressure transducers
placed in an array. The optimal depth  is not as yet clear,
but it would have to be deep enough to avoid the problem of
acoustic wave scattering, and perhaps also to escape high
frequency surface noise.  The acoustic losses set the  size
scale  of the detector.

 Laboratory data~\cite{XXXIV} indicate that pure
ice has a sound velocity (of compressional waves) of
$v = 3200\,$m/s and a $Q$ of perhaps 1000 at low temperatures.
South Pole ice near the surface is below $-50^\circ$\,C.
With an attenuation length  of about 100~m, one can imagine
detectors in a lattice of  roughly  100~m spacing. Using
Bowen's~\cite{XXXII} figure of merit, we estimate a signal-to-noise
ratio of unity in a single detector at 100~m distance  from  a
6~PeV  cascade, with the signal-to-noise ratio scaling as the
square  of energy/distance.  The frequency maximum is about 20k~Hz.
Despite the dauntingly high detection threshold in comparison to the
optical technique, acoustic detection remains
interesting enough for further investigation for several reasons.
First, acoustic sensors are compact,  pressure  tolerant,
inexpensive (piezoelectric sensors are cheap compared to
photomultipliers), and  could be  installed relatively easily.
It does not seem outrageous to imagine a $30\times 30$
lattice of  strings  extending downwards for several kilometers
thus covering a volume of  the  order  of 30~billion tons.

\subsection{Horizontal Air Showers Revisited}

We already discussed in section 3.3 the implications of
horizontal air shower measurements for atmospheric muons
and neutrinos, especially the high energy component of
charm origin. We discuss here the use of horizontal
air showers to search for cosmic neutrinos.
Given a flux $\phi_\nu(E_\nu)$, the expected event
rate for neutrino-induced horizontal showers is calculated as
described in \S3.4, Eqs.~(\ref{phi_sh}) and (\ref{horizontal}).

%

 The atmospheric muon spectrum falls rapidly with energy
($\gamma  \sim 3.7$ for $\pi$, $K$ decay muons; $\sim$ 2.7
for muons from charm decay), while neutrino cross sections
rise with energy. For large enough fluxes of astrophysical
neutrinos, neutrino-induced showers can dominate at sufficiently
high energies. In addition to charge current $\nu_\mu
(\bar{\nu}_\mu)$
interactions, there will be a contribution from $\nu_e(\bar{\nu}_e)$.
For flat spectra and $\nu_{\mu}$ and $\nu_e$ fluxes of a
similar order of magnitude, the resonant $W^-$ production
dominates the rate of $\nu$-induced horizontal air showers
for shower sizes in the narrow range
$[2\times10^6$~to~$4\times10^6]$.

Horizontal shower measurements~\cite{tokiodata} have been compared to
expectations from diffuse AGN fluxes in Ref.~\cite{horshow},
where the potential of air shower arrays for detection of
diffuse neutrino fluxes was demonstrated.   Similar limits
have been obtained recently by the EASTOP Collaboration~\cite{EASTOP}.
Muon-poor showers
are selected in the Tokyo data~\cite{tokiodata},
so that the limit is stronger
for $\nu_e(\bar{\nu}_e)$ fluxes, which actually dominate
the neutrino-induced horizontal air shower signal.

Figure 18 shows the rate of (muon-poor) horizontal showers
associated with the neutrino emission by active galaxies.
Also shown is the background from atmospheric muons.
The signal-to-background increases with shower size and
exceeds unity for shower sizes exceeding $10^6$ for the
prediction of reference~\cite{SzaboPro92}. The figure
illustrates the kind of sensitivity air shower arrays
can reach in the detection of cosmic neutrinos. An improved
detection method would be to select all, not
only muon-poor, showers that are very close to the horizon.
This would avoid any contamination from cosmic rays without
rejecting hadron-like neutrino induced showers.
Unless prompt muon production by charm turns out to
be unexpectedly large, the larger of the range of
AGN isotropic backgrounds of  Szabo and
Protheroe~\cite{SzaboPro92} should be detectable by a
horizontal shower measurement with enough exposure
to detect showers with $N_e>10^6$. A flux peaked at energies
exceeding 1~PeV, will be strongly attenuated in the
earth and only detectable in upward going muons close
to horizontal angles. Under these circumstances, neutrino detection
via horizontal air showers may be a useful
alternative detection method.\\[5mm]
{\bf Acknowledgements.}
We are grateful for helpful discussions with many colleagues,
including Karl Mannheim, Ray Protheroe, David Seckel, Floyd Stecker
and Enrique Zas.
This research is supported in part by DOE Grants
DE-FG-91ER40626 (TKG \& TS) and
DE-AC02-76ER00881(FH). The work of FH is
also supported by the University of Wisconsin Research Committee
with funds granted by The Wisconsin Alumni Research Foundation.

\newpage
\section*{Figure Captions}

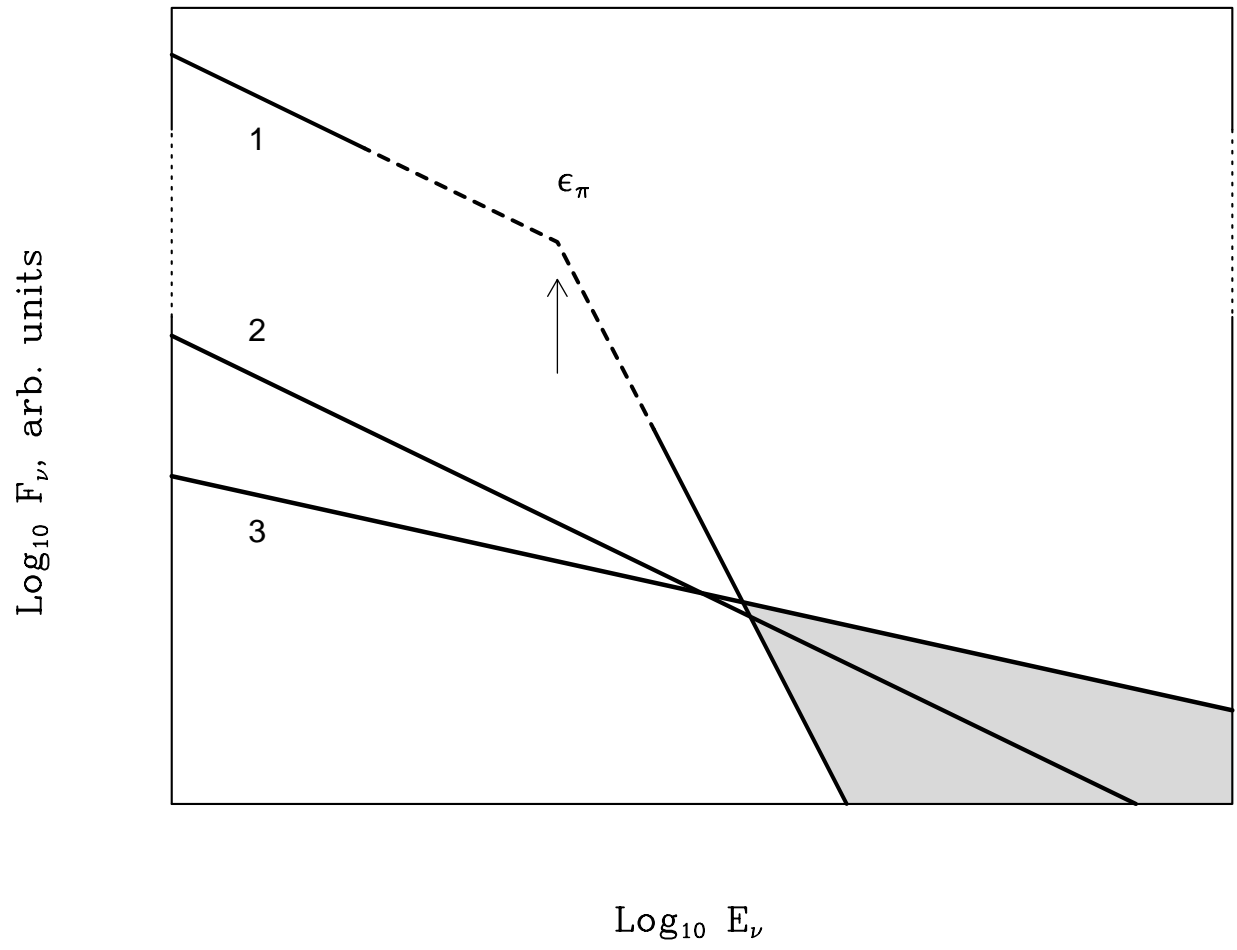
{\bf Fig.1.} A schematic presentation of neutrino fluxes
of different origin. (1)\ atmospheric neutrinos; (2)\ neutrinos
from cosmic ray interactions on galactic matter; (3)\ source
neutrinos.\\[4mm]
{\bf Fig.~2.} $P_\nu$ for two values of muon threshold
energy, 1 GeV and 1 TeV. The solid lines are for $\nu$ and
the dashed lines for $\bar{\nu}$. The dotted lines show the
power law approximations.\\[4mm]
{\bf Fig.~3.} Neutrino energies giving rise to
contained events, stopping and throughgoing muons.\\[4mm]
{\bf Fig.~4.} Upward going muon fluxes detected
by IMB \cite{IMBmu,IMBsing}(squares), Baksan \cite{Baksan}(stars) and
Kamiokande \cite{KAMmu}(circles), converted to a common
muon threshold energy of 3 GeV.\\[4mm]
{\bf Fig.~5.} Prompt neutrino ( or muon) fluxes
corresponding to five parametrizations of the
high energy behavior of the charm production cross section.
The vertical and horizontal fluxes of muons from the decay
of pions and kaons are shown with dash lines. See text and
Ref.~\cite{HVZ} for details.\\[4mm]
{\bf Fig.~6.} Rates of horizontal showers, initiated
by the muon fluxes shown in Fig.~5. See text.
The data points are from Ref.~\cite{tokiodata} and the dotted line
shows the shower rate expected from muons form $\pi$ and K
decay.\\[4mm]
{\bf Fig.~7.} Neutrino fluxes from 3C273 predicted
in different AGN models. Thick solid line shows the flux
of Stecker {\it et al}~\cite{Stecketal}; thin solid lines
show several models due to Szabo and Protheroe~\cite{SzaboPro92};
and the dash line is from Ref.~\cite{BegSik} (see text). These three
models are for neutrinos from the AGN nucleus. The dotted line shows
the calculation of Mannheim for the neutrino emission from the
AGN jet~\cite{MannPRD}.\\[4mm]
{\bf Fig.~8.} The isotropic neutrino background from
AGN. The thick line is the prediction of Stecker
{\it et al}~\cite{Stecketal} for generic AGN, and the thin lines
represent models of Szabo and Protheroe~\cite{SzaboPro92}.
The backgrounds from AGN jets are calculated by
Mannheim~\cite{Mannbla}
(dotted line) and Stecker {\it et al}~\cite{Steckjet}
(dash-dot line). The shaded
area shows the angle averaged flux of atmospheric neutrinos.\\[4mm]
{\bf Fig.~9.} Horizontal ($-0.3<\cos\theta<0.3$) muon
fluxes generated by  the isotropic neutrino background as in
the bracketing high and low models of Szabo \& Protheroe
\cite{SzaboPro92} (thin solid lines) and by Stecker {\it et al.}
\cite{Stecketal} (thick solid line).
The 90\% C.L. upper limit of the Frejus experiment \cite{Meyer}
is shown for muon threshold energy of 2 TeV.\\[4mm]
{\bf Fig.~10.} Differential interaction rate of
$\nu_e \;+\; \bar{\nu}_e$ as a function of $E_\nu$. Dotted line:
atmospheric; thin solid line: Ref.~\cite{SzaboPro92}; thick solid line:
Ref.~\cite{Stecketal}. The numbers by the curves are number of
interactions per 1000 kT per year, calculated for E$_\nu>$ 1 TeV.\\[4mm]
{\bf Fig.~11.} Comparison of quasi-simultaneous
observations of 3C273 in the optical, X-ray and $\gamma$-ray
bands~\cite{3CC} with the predictions of a synchrotron-self-Compton
model~\cite{SchlickDerm} (solid line) and a hadronic
model~\cite{MannPRD}.
See Ref.~\cite{3CC} for references to all experimental data.\\[4mm]
{\bf Fig.~12.} Contours in the $M,\mu$ plane of
constant $\alpha_{H}=1.0,\ 0.1,\ 0.01$ (boxes) and constant
neutralino mass $M_\chi=30, 100, 500$ and 1000~GeV (solid).
The shaded region is excluded by cosmological considerations.\\[4mm]
{\bf Fig.~13.} As a function of the neutralino
mass we show the telescope size required to be sensitive at the one
event per year level. We fix $\tan\beta=2,\ \alpha_{H}=0.1$. The
two branches correspond to the two solutions for fixed
$\alpha_{H}$.\\[4mm]
{\bf Fig.~14.} In the $M_2,\mu$ plane for
$M_{\tilde q}=\infty$, (a)~contours of constant detection rate
(events m$^{-2}$ yr$^{-1}$) and
(b)~regions of $\Omega_\chi h^2 > 1$ and $\Omega_\chi h^2 < 0.02$
which are ruled  out by cosmological considerations.\\[4mm]
{\bf Fig.~15.} Summary of isotropic neutrino fluxes
of energy above 1 GeV. (1)\ atmospheric neutrinos; (2)\
diffuse galactic neutrinos; (3)\ diffuse extragalactic
neutrinos---maximum and minimum predictions of Ref.~\cite{SzaboPro92};
(4)\ cosmological neutrinos---maximum and minimum predictions of
Ref.~\cite{YoshTesh}.\\[4mm]
{\bf Fig.~16.} Summary of source neutrino fluxes
of energy above 1 GeV. (1)\ neutrinos generated by cosmic rays
in the Sun~\cite{SeckSG}; (2)\ a galactic neutrino source;
(3)\ extragalactic neutrino source (3C273)~\cite{SzaboPro92};
(4)\ AGN jet emission (3C279 high state)~\cite{MannBier}.
The atmospheric
neutrino background within 1$^\circ$ is shown with a dash line.\\[4mm]
{\bf Fig.~17.} A possible configuration of a 1~km
neutrino detector, based on AMANDA-design supermodules.\\[4mm]
{\bf Fig.~18.} Rate of (muon-poor)
horizontal showers  associated with the neutrino
emission by active galaxies (thin lines are the
upper and lower bound of Ref.~\cite{SzaboPro92},
the thick line shows the background of
Ref.~\cite{Stecketal})
compared  to the background from
atmospheric muons (dotted line) and the experimental
results of Ref.~\cite{tokiodata}.
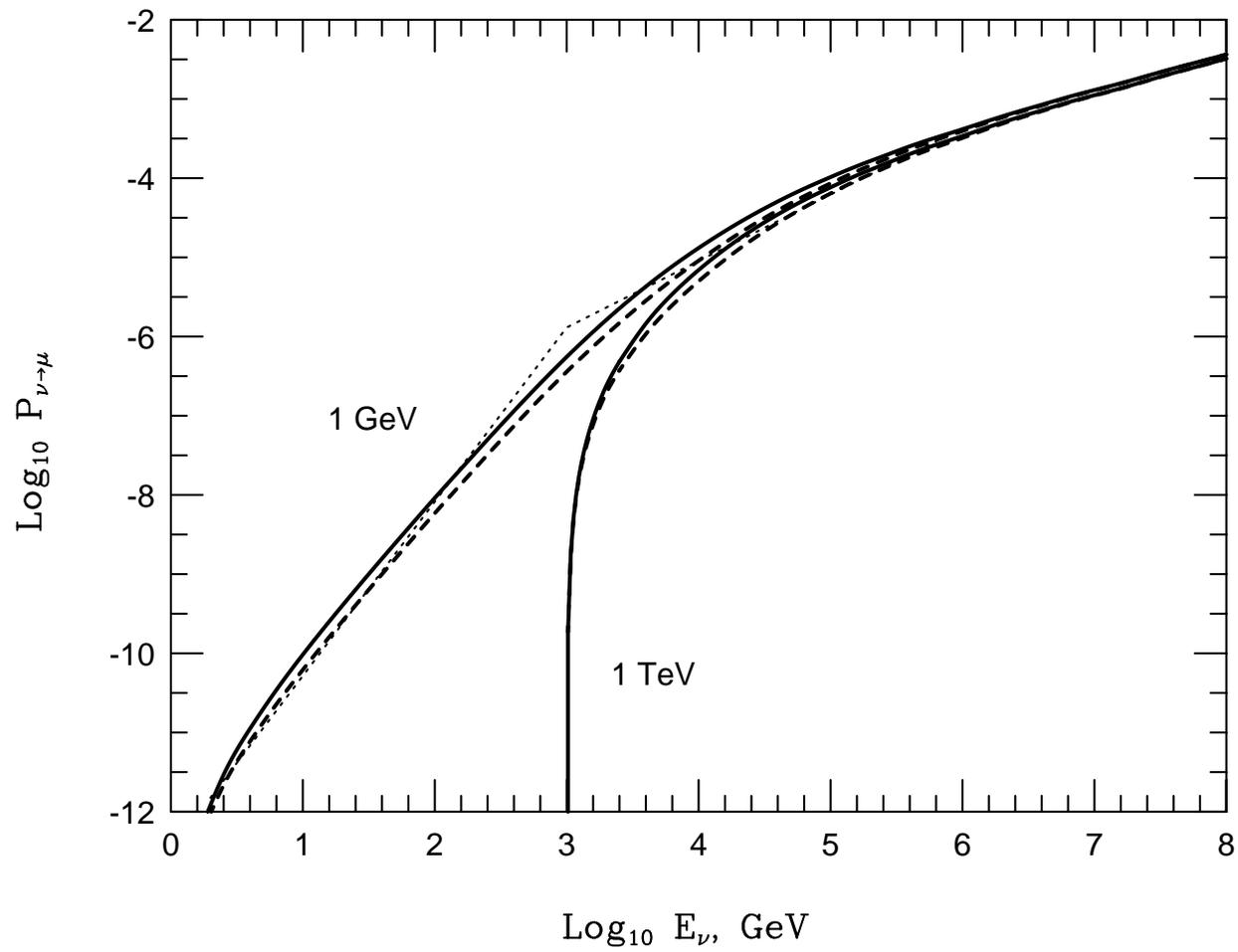
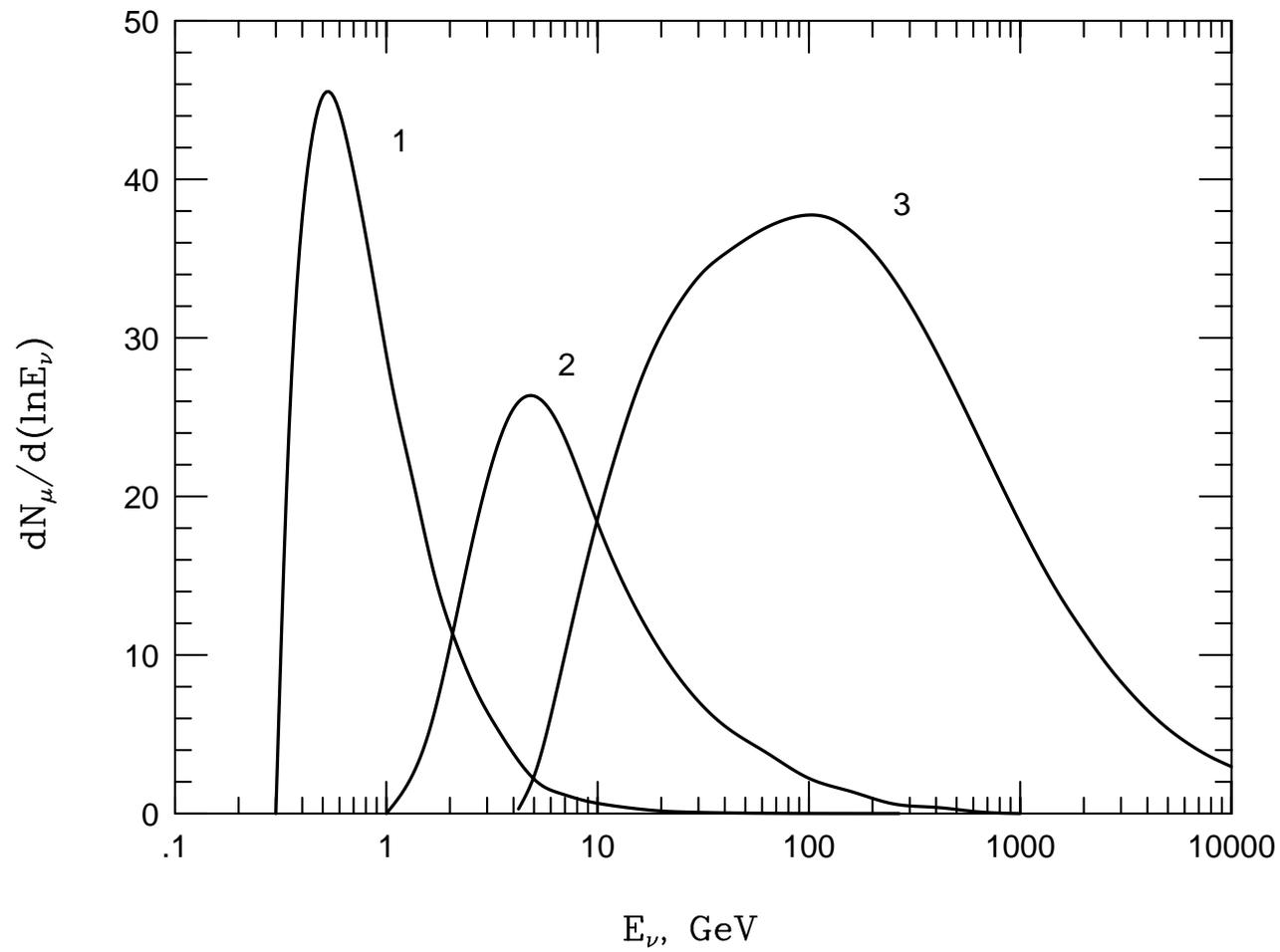
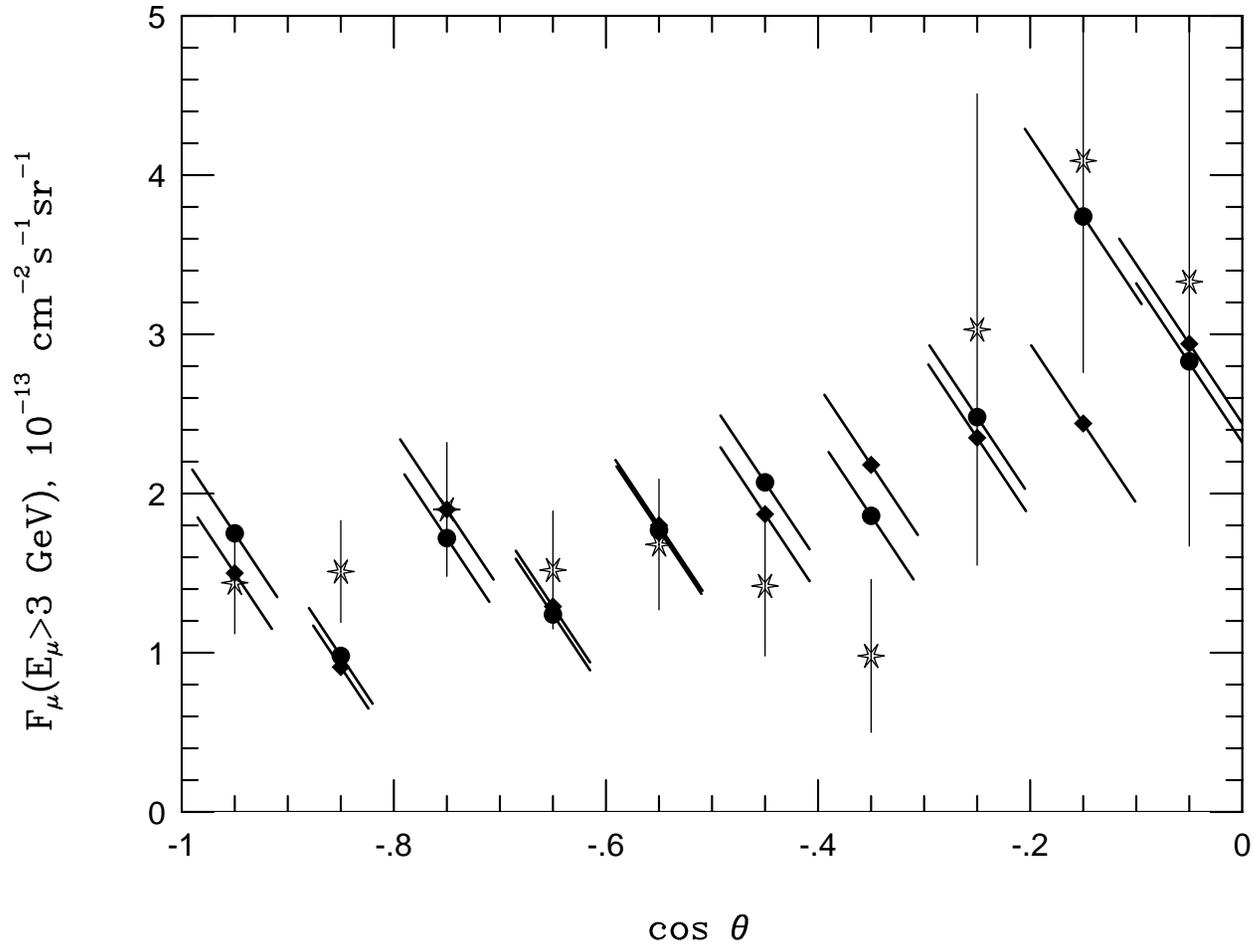
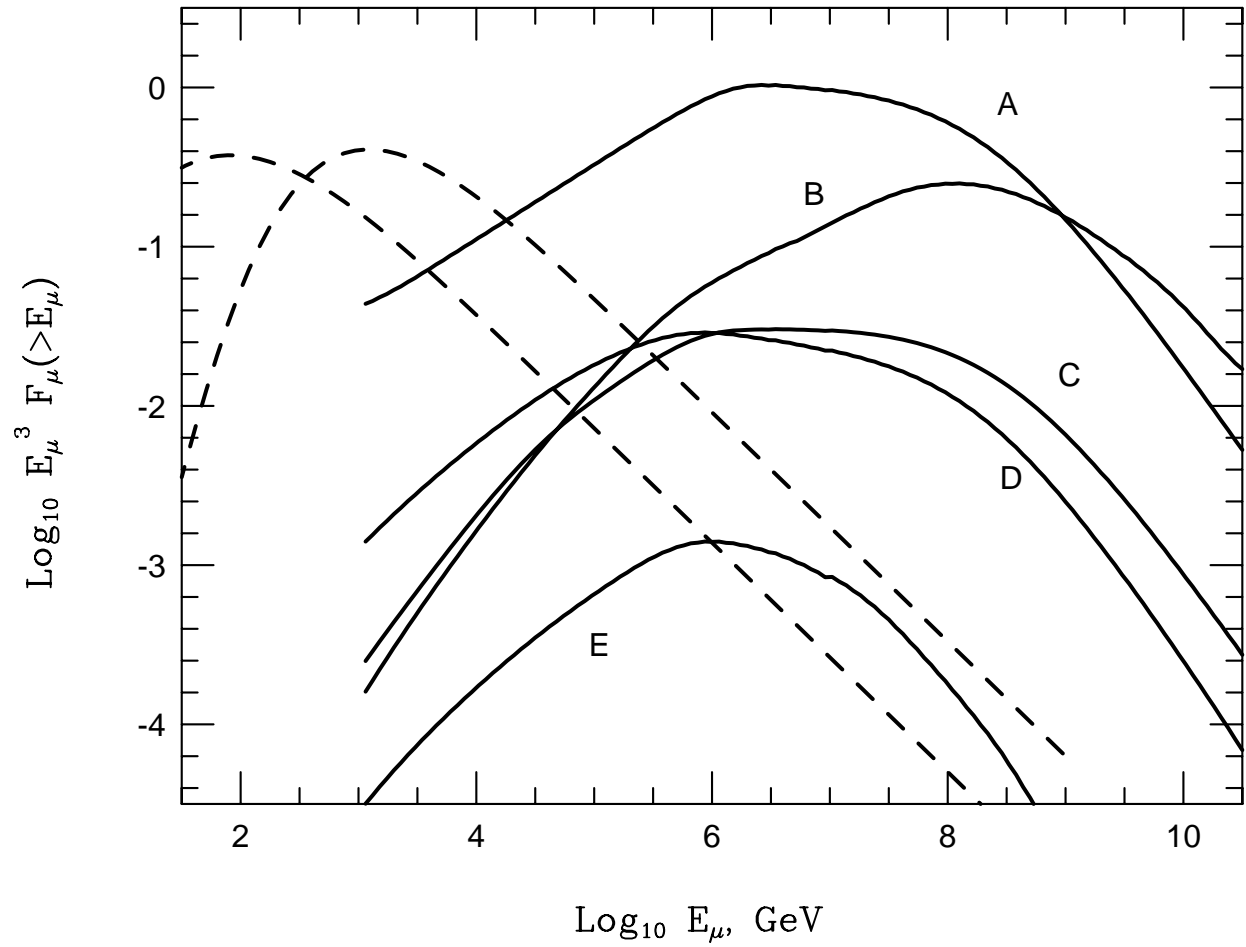
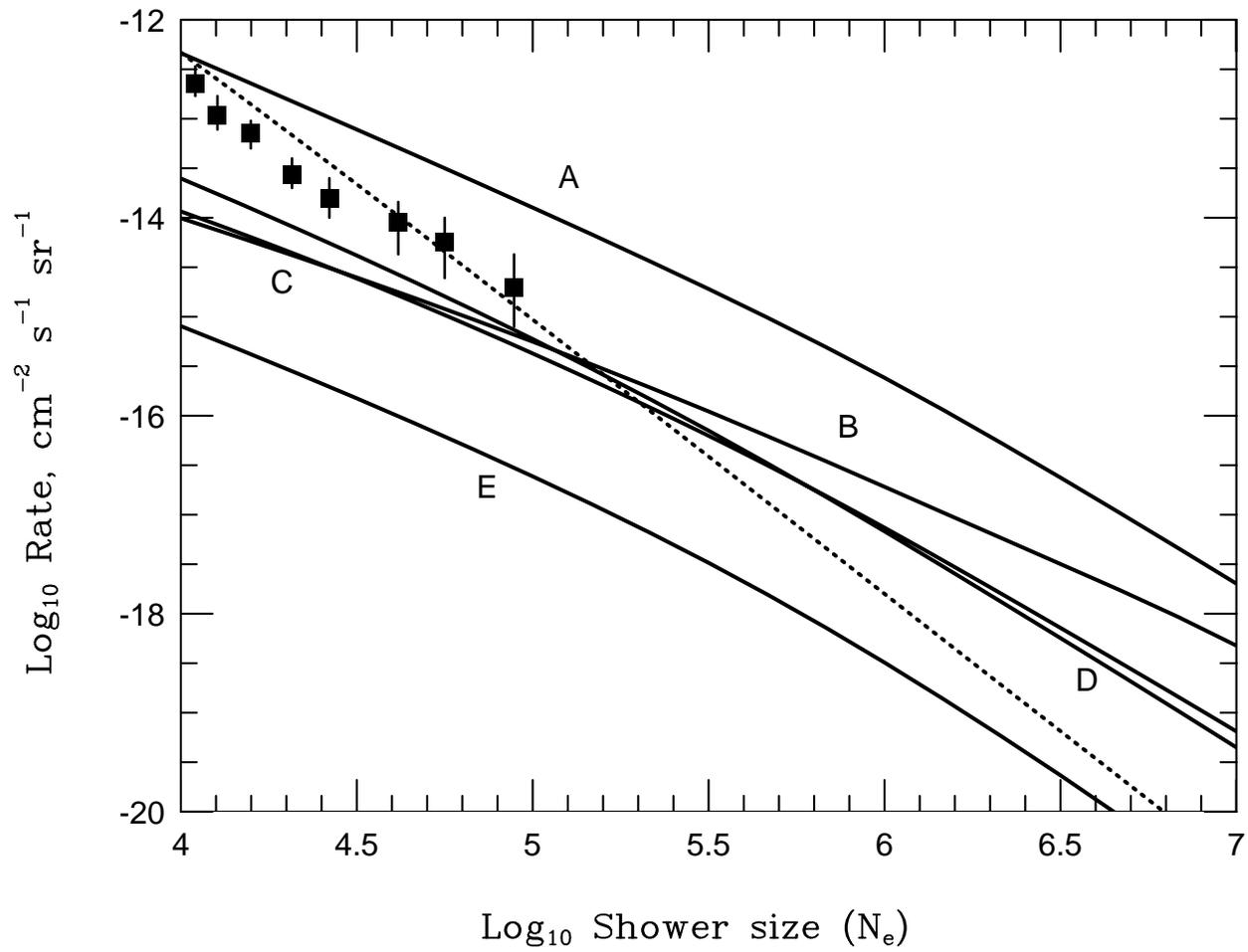
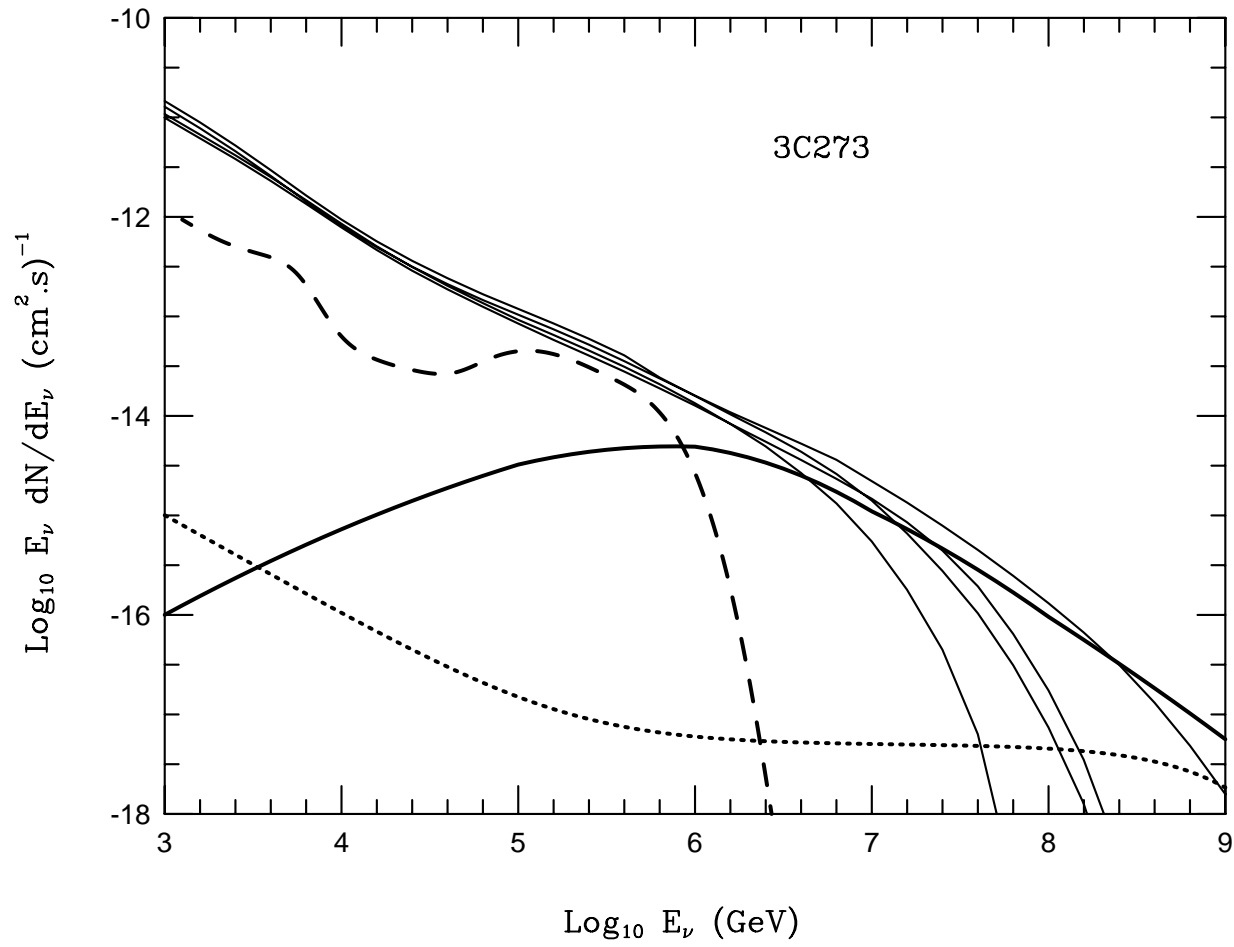
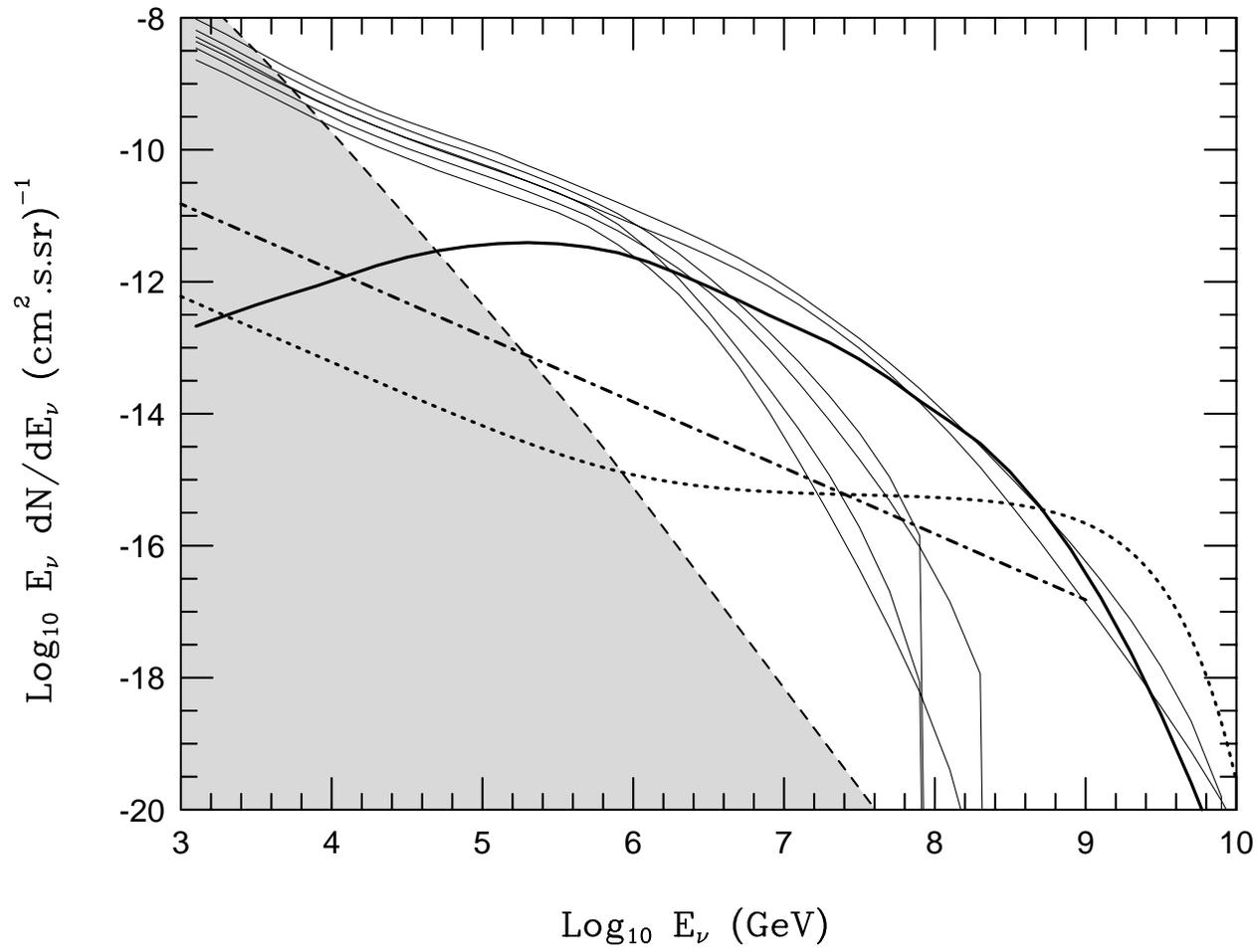
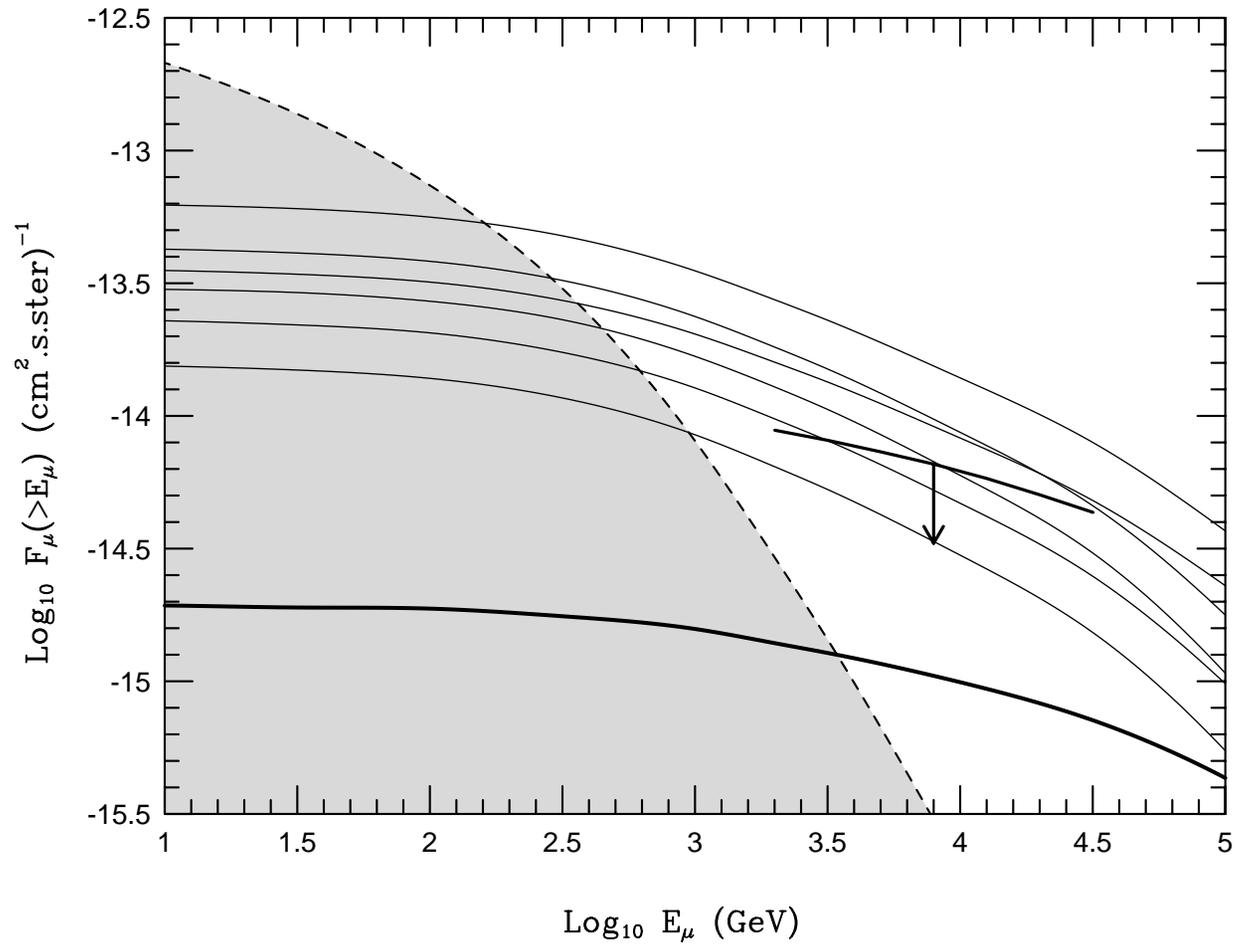
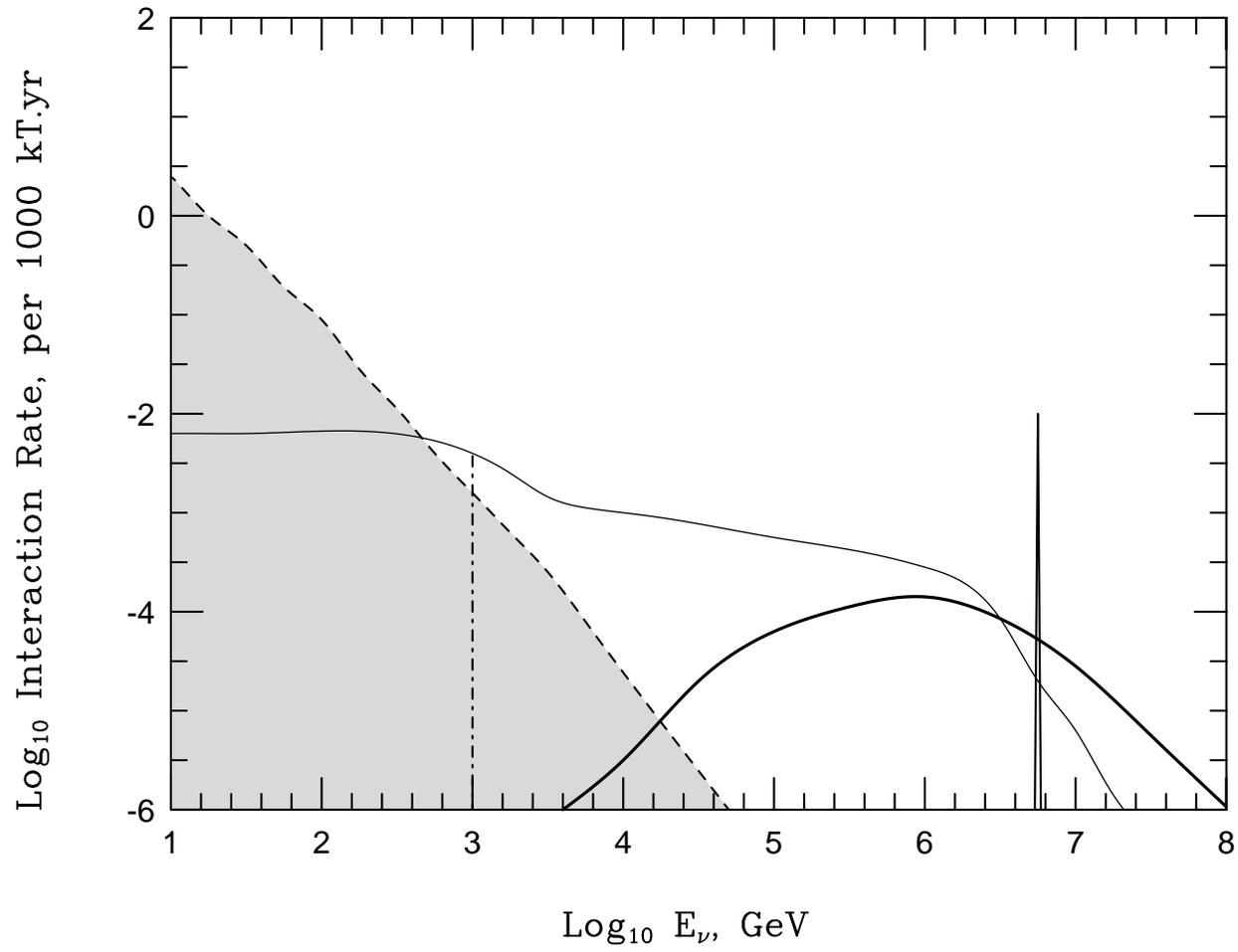
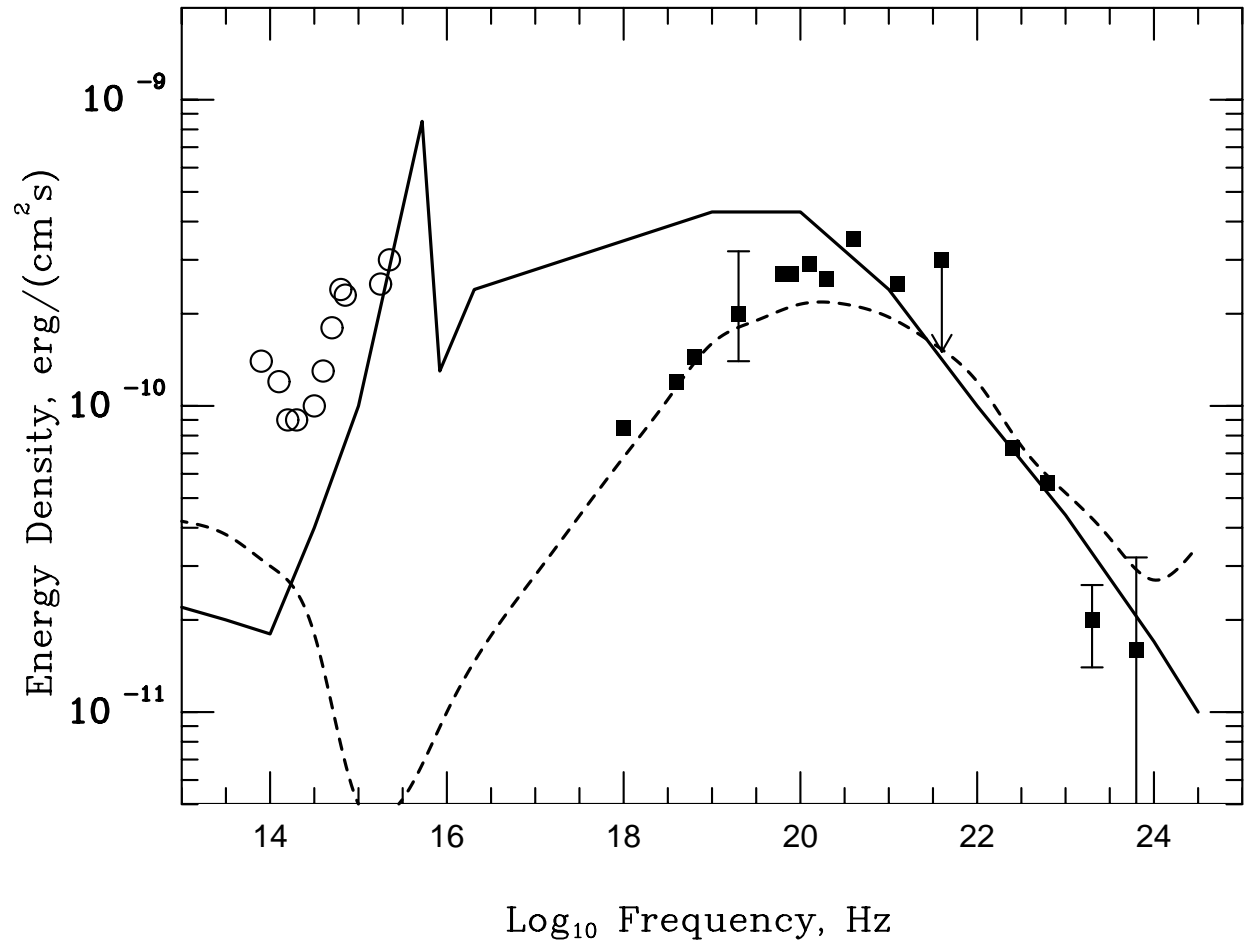
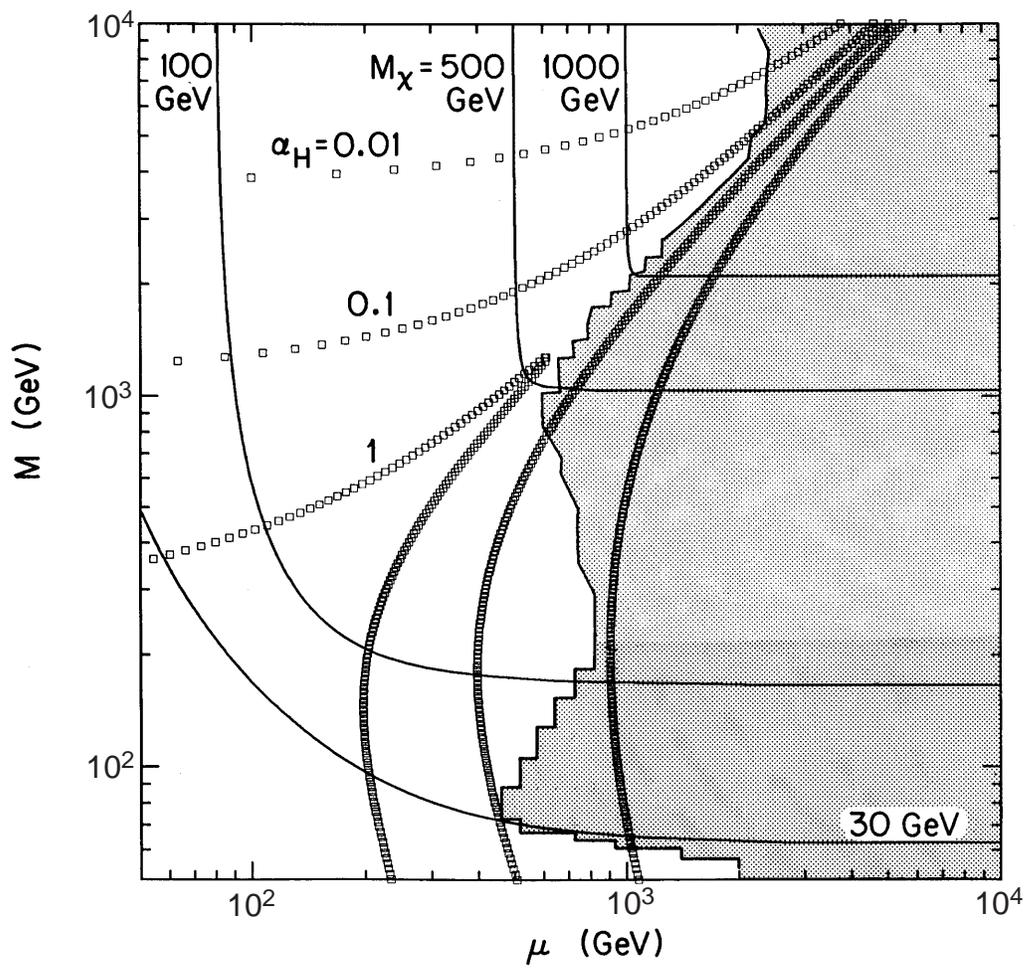
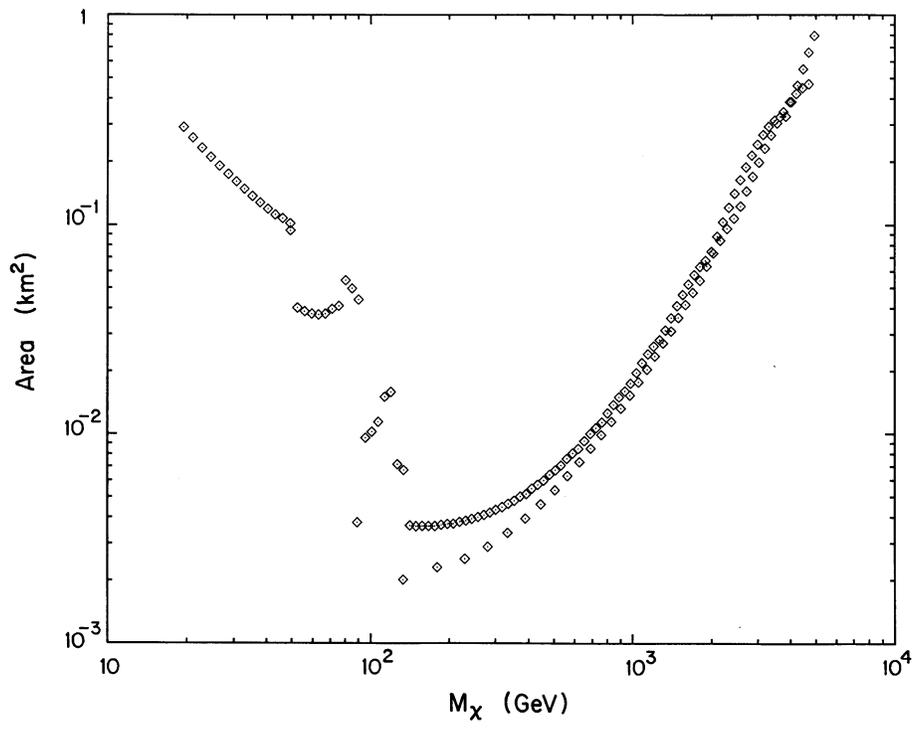
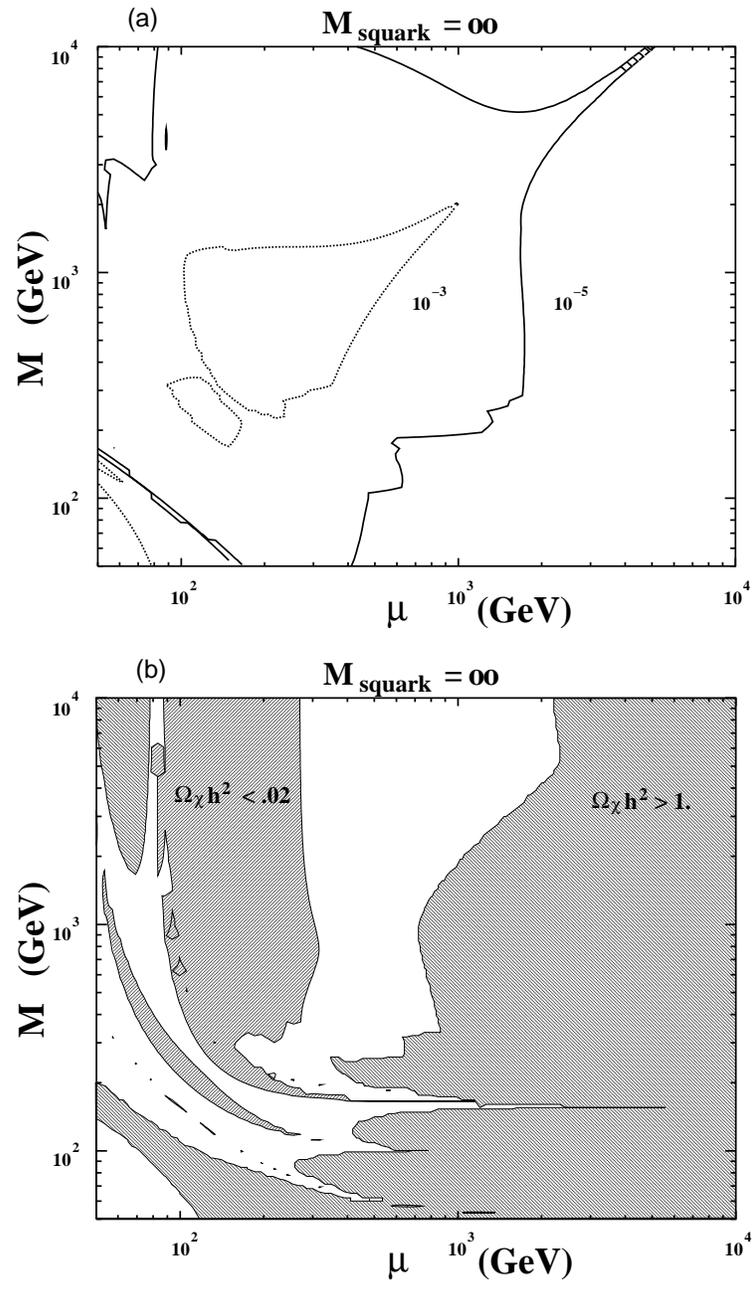
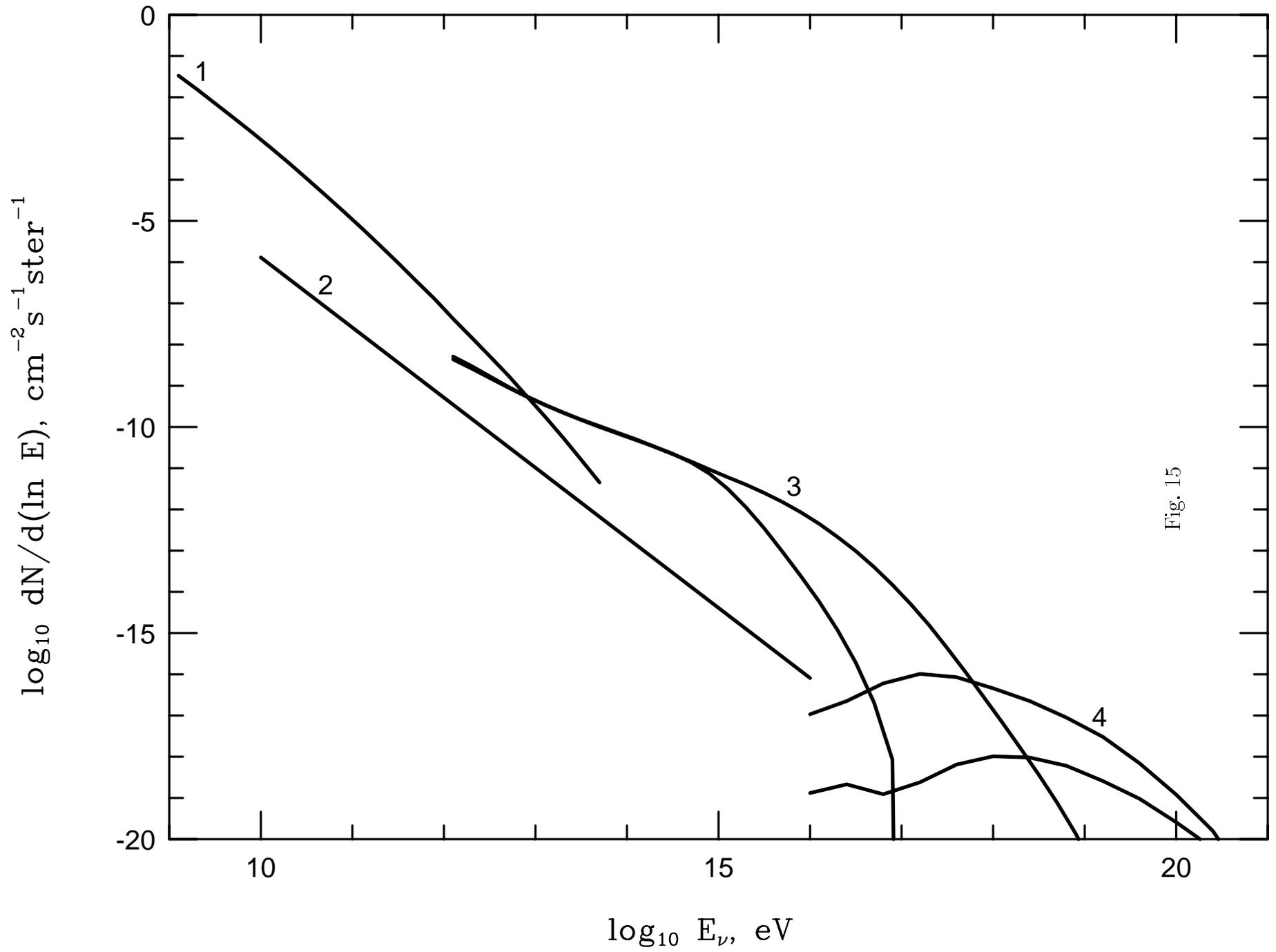
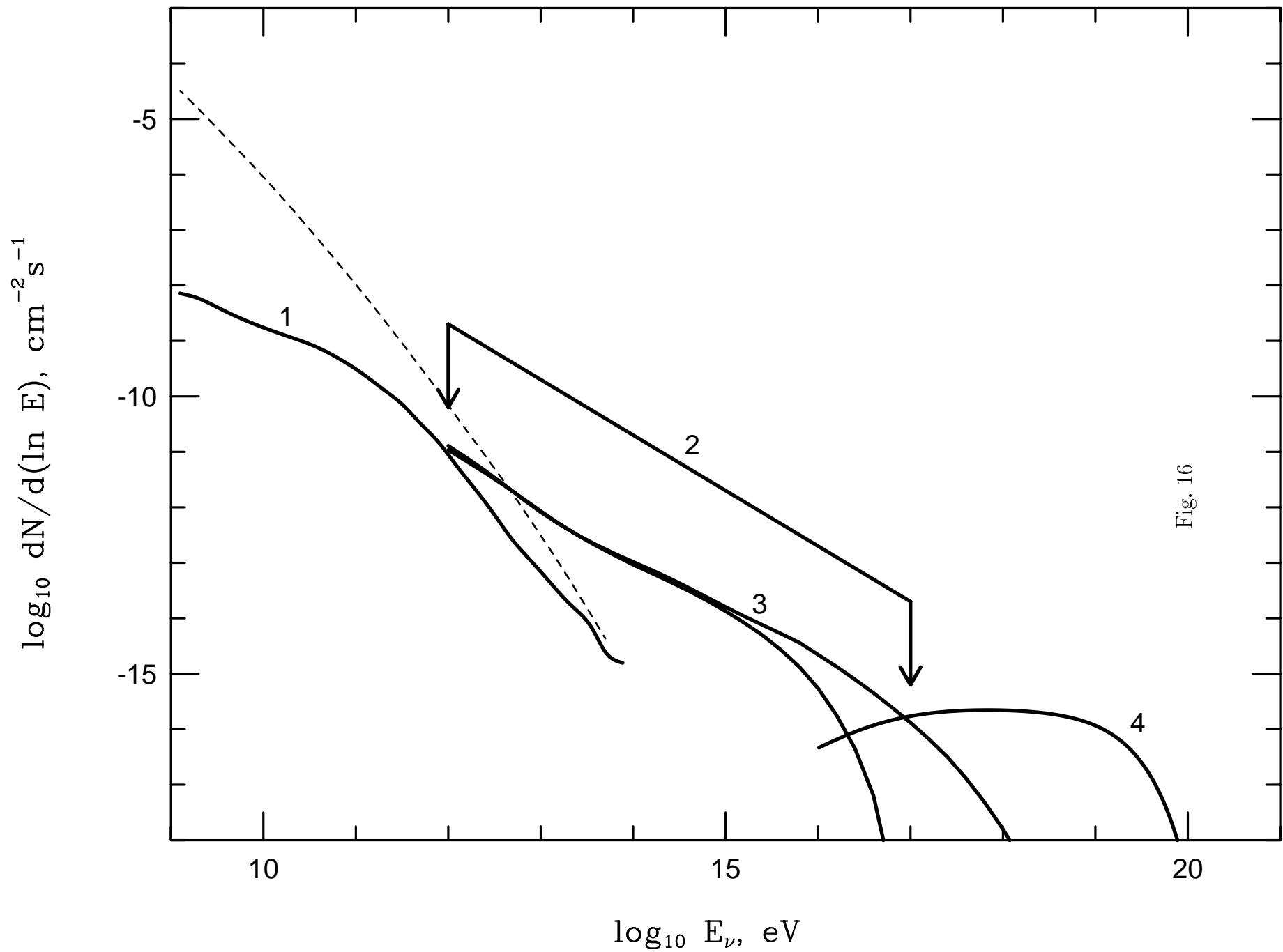
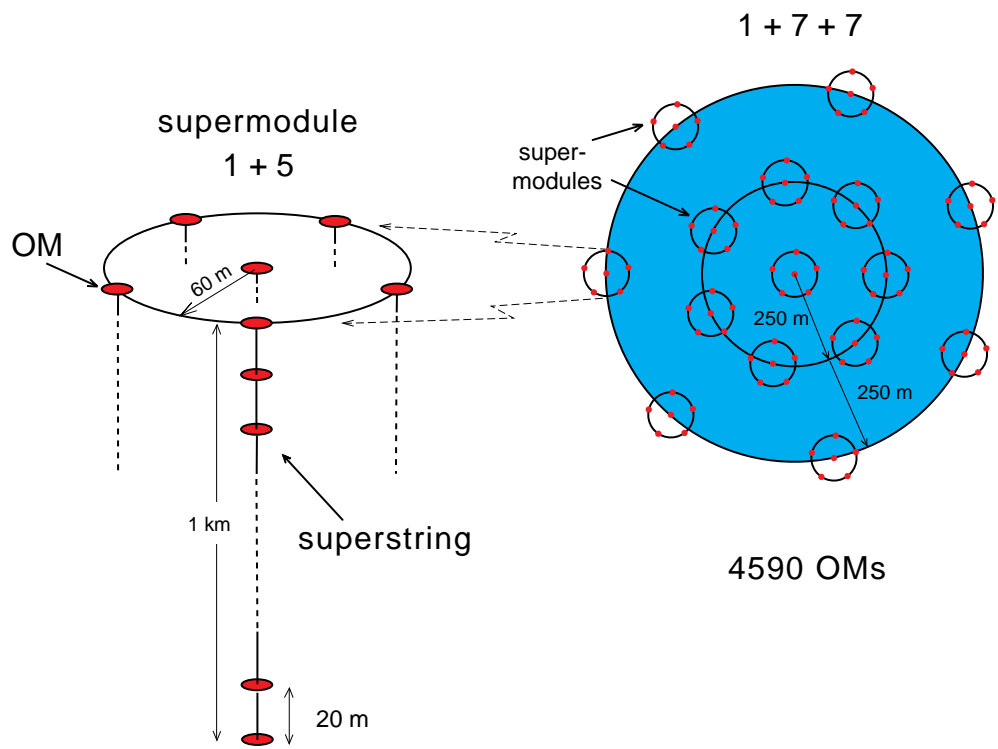
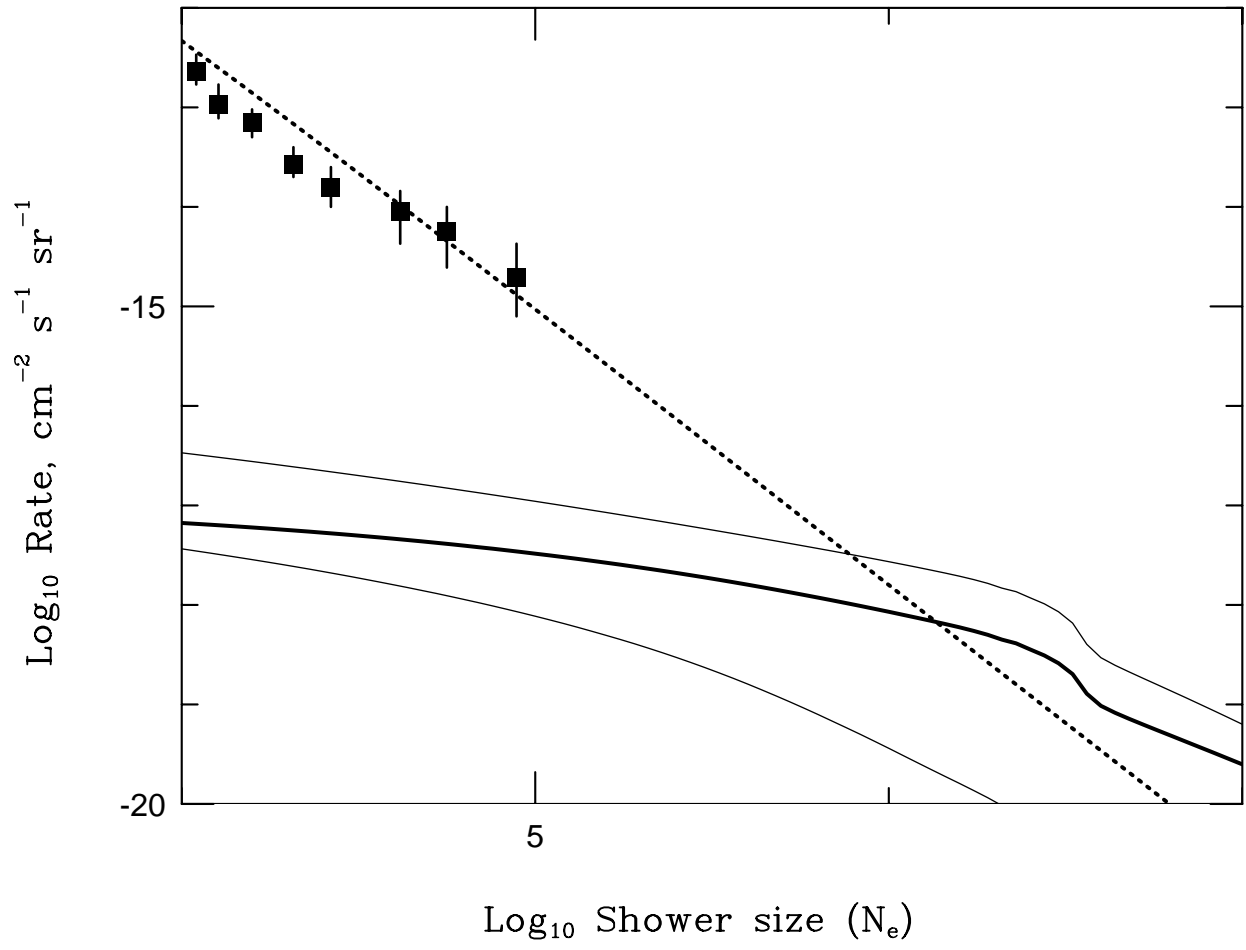
\end{document}